\documentclass[journal]{IEEEtran}

\usepackage{cite}
\ifCLASSINFOpdf
  \usepackage[pdftex]{graphicx}
\else
  \usepackage[dvips]{graphicx}
\fi

\usepackage[cmex10]{amsmath}
\usepackage{amsmath,amsfonts,amssymb}
\usepackage{mdwmath}
\usepackage{algorithm}
\usepackage{algorithmic}
\usepackage{array}
\usepackage{fixltx2e}
\usepackage{bm}
\usepackage{stfloats}
\usepackage{graphicx}
\usepackage{caption}
\usepackage{mdwtab}
\usepackage{subfig}
\usepackage{booktabs}
\usepackage{multirow}
\usepackage{etoolbox}
\usepackage{makecell}
\usepackage{xcolor}
\usepackage[bottom]{footmisc}
\usepackage{mdframed}
\usepackage{lettrine}
\usepackage{supertabular}
\usepackage{longtable}

\newcommand{\e}{\begin{equation}}
\newcommand{\ee}{\end{equation}}
\newcommand{\eqn}{\begin{eqnarray}}
\newcommand{\eeqn}{\end{eqnarray}}

\begin{document}

\title{A Comprehensive Survey of Machine Learning Applied to Radar Signal Processing}

\author{Ping Lang,~\IEEEmembership{Student Member,~IEEE}, Xiongjun Fu,~\IEEEmembership{Member,~IEEE}, \\
 Marco Martorella,~\IEEEmembership{Fellow,~IEEE}, Jian Dong, Rui Qin, Xianpeng Meng, and Min Xie

\vspace*{-5.0mm}
\thanks{This work was supported by the National Natural Science Foundation of China under Grant 61571043 and 111 Project of China under Grant B14010.
 {\it (Corresponding author: Xiongjun Fu)}}
\thanks{P. Lang, X. Fu are with School of Information and Electronics,
 Beijing Institute of Technology, Beijing 100081, China (E-mails: \{langping911220, fuxiongjun,
\}@bit.edu.cn).}
}

\markboth{Journal of \LaTeX\ Class Files,~Vol.~x, No.~x, Sep~2020}%
{Liao \MakeLowercase{\textit{et al.}}: Manuscript of IEEEtran.cls for IEEE Journals}

\maketitle

\begin{abstract}
 Modern radar systems have high requirements in terms of accuracy, robustness and real-time capability when operating on increasingly complex electromagnetic environments. Traditional radar signal processing (RSP) methods have shown some limitations when meeting such requirements, particularly in matters of target classification. With the rapid development of machine learning (ML), especially deep learning, radar researchers have started integrating these new methods when solving RSP-related problems. This paper aims at helping researchers and practitioners to better understand the application of ML techniques to RSP-related problems by providing a comprehensive, structured and reasoned literature overview of ML-based RSP techniques. This work is amply introduced by providing general elements of ML-based RSP and by stating the motivations behind them. The main applications of ML-based RSP are then analysed and structured based on the application field. This paper then concludes with a series of open questions and proposed research directions, in order to indicate current gaps and potential future solutions and trends.
\end{abstract}


\IEEEpeerreviewmaketitle

\section{Introduction}\label{S1}

 \lettrine[lines=2]{R}{ADAR} offers special advantages with respect to other types of sensors including all-day, all-weather operations, long detection distance and, depending on the frequency used, penetration. Moreover, radar can often be carried by a number of platforms, spanning from classic naval and airborne to more recent space-borne, UAVs, such as drones, and high-altitude platforms (HAPs). The ensemble of these characteristics can be exploited for military scenarios, such as target detection, tracking and recognition, and for civil scenarios, such as land use and classification, disaster assessment, urban and non-urban monitoring, making radar the perfect sensor for dual use applications  \cite{Stimson_Airborne_radar,Bassem_radar_system}. Radar signal processing (RSP) is one of the key aspects that characterize the radar field \cite{Richards_RSP} as its development allows for radar performances to be maximised and for several capabilities to be enabled, including the ability to operate in spectrally congested and contested scenarios and complex and dynamically changing environment \cite{Vakin_EW,EW_Spezio16}. Artificial Intelligence (AI) has pushed the research and development in many fields \cite{signal_information_processing}, including, among others, speech signal processing (SSP), computer vision (CV) and natural language processing (NLP). Such domains include logic programming, expert system, pattern recognition, machine learning (ML) and reinforcement learning \cite{Bishop_PRML}. Machine learning (ML), and especially deep learning (DL) \cite{Goodfellow_DL,LeCun_DL}, has achieved great breakthroughs thanks to large investments from a number of countries and through a pervasive cooperation of the scientific community. More specifically, ML-based RSP has been targeted by many to attempt to improve traditional RSP solutions and overcome their limitations. As a demonstration of the interest in this field, in the recent years, the Defense Advanced Research Projects Agency (DARPA) has launched many projects in this field, such as the radio frequency machine learning system (RFMLS) project \cite{RFMLS_1,RFMLS_2,RFMLS_3,RFMLS_4}, the behavior learning for adaptive electronic warfare (BLADE) project \cite{BLADE_1,BLADE_2}, and the adaptive radar countermeasures (ARC) project\cite{ARC}. In addition to DARPA's projects, there is ample support from the scientific literature, such as radar emitter recognition and classification \cite{emission_Dudczyk,Huang,Conning,Zhang}, radar image processing (e.g., synthetic aperture radar (SAR) image denoising \cite{Chierchia122,Zhao123,Wang124,Wang125,Mukherjee128}, data augmentation \cite{Cui101,Ma102,Wang103,Lv104,Yu105}, automatic target recognition (ATR) \cite{Cui152,Zhao163,Georgios164,Zhao165,Ding159,Furukawa160,Zhao161,Chen162,Zhang175}, target detection \cite{Metcalf302,Liu304}, also with specific emphasis on ship detection \cite{fater_RCNN_SAR,Fan224,Gao225,Kang227,An228}, anti-jamming \cite{DQN_anti_jamming_frequency_hopping}, optimal waveform design \cite{Wang296}, array antenna selection \cite{Elbir303}, and cognitive electronic warfare (CEW) \cite{You301}. These ML algorithms include traditional machine learning (e.g., support vector machines (SVMs), decision tree (DT), random forest (RF), boosting methods), and deep learning (e.g., deep belief networks (DBNs), autoencoders (AEs), convolutional neural networks (CNNs), recurrent neural networks (RNNs), generative adversarial networks (GANs)). This survey paper has comprehensively reviewed state-of-the-art of ML-based RSP algorithms, including traditional ML and DL.

  \vspace{-1.0mm}
\subsection{Motivation}\label{S1.1}

 Due to the large success of ML in many domains, the radar community has started applying ML-based algorithms to classic and new radar research domains to tackle traditional and new challenges from a novel prospective. Being ML a relatively new paradigm, the research results that have been obtained have not been systematically surveyed and analyzed. A thorough and reasoned review of new technologies is key for providing

 i) a solid basis for new researchers and practitioners who are approaching this field for the first time;

 ii) an important reference for more experienced researchers who are working in this field;

 iii) existing terms for comparison for newly developed ML-based algorithms;

 iv) means to identify gaps;

 v) a full understanding of strengths and limitations of ML-based approaches.

 \vspace{-1.0mm}
\subsection{Related works}\label{S1.2}

 This section will briefly survey some this topic-related review scientific literatures.

 \textbf{i) ML algorithms and applications} There are many review papers either about the development of ML algorithms, such as DL \cite{LeCun_DL,Hatcher_DL,Liu_DNN}, deep reinforcement learning (DRL) \cite{Arulkumaran_DR}, transfer learning (TL) \cite{Tan_tansfer}, GANs \cite{GANs}, developments of CNNs \cite{CNN_latest}, efficient processing technologies of DNN \cite{efficient_learning_DNN}, adversarial learning for DNN classification \cite{adversarial_learning_DNN_classification}, neural networks model compression and hardware acceleration \cite{model_compression_hardware_acceleration} or applications in special topics such as ML applied to medical image processing \cite{Litjens_med.,DL_ultrasound_imaging}, robotics \cite{Harry_robotics}, agriculture \cite{Kamilaris_agriculture}, sentiment analysis \cite{Zhang_sentiment}, object detection \cite{obj_detection_1,obj_detection_2}.

 As a most popular DNN model, CNN has been successfully applied in most of ML tasks. In \cite{CNN_latest}, the authors comprehensively investigated the state-of-the-art technologies about the development of CNN. This paper systematically introduced the CNN models from LeNet to latest networks such as GhostNet, including one-dimension (1D), two-dimension (2D), and multi-dimension (multi-D) convolutional models and their applications, such as 1D, 2D and multi-D models can be applied in time series prediction and signal identification, image processing, and human action recognition, X-ray, computation tomography (CT), respectively. Besides, some prospective trends have been proposed such as model compression\cite{model_compression_hardware_acceleration}, security, network architecture search\cite{NAS_survey}, and capsule neural network\cite{capsules_network}.

 TL aims to solve insufficient training data problem, which also used in RSP domain, such as radar emitter recognition \cite{Yang62}, micro-doppler for motion classification \cite{Seyfioglu279,Seyfio281}, SAR image processing with limited labeled data \cite{Ma153,Wang154,Huang178}. A TL-related review was developed in \cite{Tan_tansfer}, which categorized the TL techniques as four classes: instances-based, mapping-based, network-based, and adversarial-based, respectively.

 Object detection, as one of most important tasks of CV, is a fundamental and challengeable task, which not only concentrates on classifying different images but also tries to precisely estimate the concepts and locations of objects contained in each image \cite{obj_detection_1}. The authors in \cite{obj_detection_1,obj_detection_2} have studied the latest development of object detection in the past few years. These review papers have covered many aspects of object detection, including detection frameworks (such as R-CNN, Fast R-CNN, Faster R-CNN, Mask R-CNN, YOLOv1-v4), training strategy, evaluation metrics, and the analysis of some typical object detection examples, such as salient object detection, face detection and pedestrian detection.

 \begin{figure*}[!tp] 
\vspace{-2.0mm}
\begin{center}
\includegraphics[width = 1.4\columnwidth, keepaspectratio]{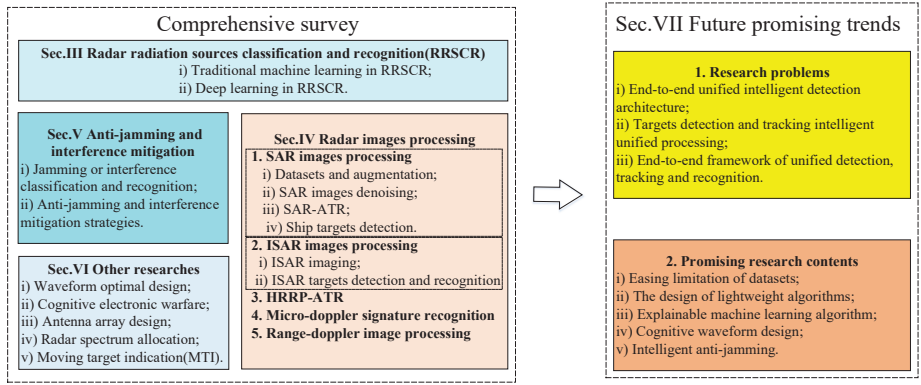}
\end{center}
\captionsetup{font = {footnotesize}, singlelinecheck = off, justification = raggedright, name = {Fig.}, labelsep = period}%
\vspace{-2.0mm}
\caption{The overview contents of this paper.}
\label{FIG1}
\end{figure*}

 \textbf{ii) Remote sensing} Besides, some review papers, focused on ML applied to remote sensing (RS) domain, have been published in \cite{DL_RS_1,DL_RS_2,DL_RS_3,ML_RS,DL_RS_4}. These survey papers investigated the state-of-the-art technologies of ML to solve the challenges in RS domain, such as RS image processing (e.g., hyperspectral image, SAR image, hyper resolution satellite image, 3D reconstruction), target recognition, scene understanding, object detection and segmentation.

 The challenges of using DL for RS data analysis were analyzed in \cite{DL_RS_1}, and then the recent advances in images classification, recognition, detection, multi-model data fusion, and 3D reconstruction were reviewed . The DL models mainly included AEs and CNNs. The authors in \cite{DL_RS_2} surveyed the recent developments of RS field with DL and provided a technique tutorial on the design of DL-based methods for processing the optical RS data, including image preprocessing, pixel-based classification, target recognition, and scene understanding. The comprehensive survey of state-of-the-art DL in RS research was developed in \cite{DL_RS_3}, which focused on theories, tools, challenges for the RS community, and specifically discussed unsolved challenges and opportunities, such as inadequate data sets, human understandable solutions for modeling physical phenomena. In \cite{DL_RS_4}, the authors systematically reviewed the DL in RS applications by meta-analysis method containing image fusion, registration, scene classification and object detection, semantic segmentation, and even accuracy assessment. The recent progress of RS image scene classification, especially DL-based methods was surveyed in \cite{scene_classify_RS}. In addition, a large-scale remote sensing image scene classification (RESISC) benchmark data set, termed ``NWPU-RESISC45" was proposed. The traditional ML algorithms applied to classification of RS research was also investigated in \cite{ML_RS}, including SVM, boosted DTs, RF, artificial neural network (ANN), K nearest neighbor (K-NN). The study aspects contained the selection of classifier, the requirements of training data, definition of parameters, feature space operation, model interpretability, and computation costs. Some key findings such as SVM, RF, and boosted DTs have higher accuracy for classification of remotely sensed data, compared to alternative machine classifiers such as a single DT and K-NN.

 A comprehensive state-of-the-art survey for SAR-ATR techniques was developed in \cite{Darymli146}, which was categorized to model-based, semi-model-based, and feature-based. These SAR-ATR techniques, however, were unilaterally based on pattern recognition or prior knowledge. The AE model and its variants applied to RS and SAR images interpretation was investigated in \cite{AEs_SAR_RS}, including original AE, sparse AE, denoising AE, convolutional AE, variational AE, and contrastive AE. The authors in \cite{vessel_detection_optical} surveyed temporal developments of optical satellite characteristics and connected these with vessel detection and classification after analyzed 119 selected literatures. Although there are some review papers about RS domain based on ML algorithms, as a subset of RS, the comprehensive survey of ML algorithms applied to RSP has not emerged so far.

 \textbf{iii) Multi-representation learning algorithms} There are also some other survey papers related the topic of this area, such as multi-view learning (MVL) \cite{multi-view}, multi-task learning (MTL) \cite{multi-task}.

 MVL and MTL have rapidly grown in ML and data mining in the past few years, which can obviously improve performance of model learning. In RSP domain, these related methods are popular in DL-based SAR-ATR, e.g., \cite{Pei174,Pei177,Pei180,Ren185,Zhao192,Ning186,Zhang189,Lv196}. Therefore, it is necessary to make a brief introduction about the review papers in MVL \cite{multi-view} and MTL \cite{multi-task}. MVL is concerned as the problem of learning representations (or features) of the multi-view data that facilitates extracting readily useful information when developing prediction models \cite{multi-view}.

 According to the state-of-the-art overview on MVL studied in \cite{multi-view}, multi-view representation alignment and multi-view representation fusion were two categories of MVL. The former aims to capture the relationships among multiple different views through feature alignment, including multi-modal topic learning, multi-view sparse coding, and multi-view latent space Markov networks. The latter seeks to fuse the separate features,  learned from multiple different views, into a single compact representation, including multi-modal AEs, multi-view CNNs, and multi-modal RNNs.

 MTL can be roughly defined as: a some task learning can improve generalization capability by shared representation between related tasks or optimize multi-loss function simultaneously. A comprehensive survey on MTL in deep neural networks (DNNs) was developed in \cite{multi-task}, which introduced (i) two common MTL methods in DL, i.e., hard and soft parameters sharing, (ii) MTL neural network models and non-NN models, such as block-sparse regularization, learning tasks relationship, and (iii) auxiliary tasks in order to reap the benefits of multi-task learning.

 \vspace{-1.0mm}
\subsection{Contributions and Organization}\label{S1.3}

 Motivated by the research community and our research interests, this article collects state-of-the-art achievements about ML-based RSP algorithms from public databases such as \emph{IEEE Xplore, Web of Science, and dblp}, most of which come from recent 5 years, i.e., from 2015 to 2020. We systematically analyze these findings on this research domain, to pave the access to promising and suitable directions for future research. Hopefully, this paper can help relative researchers and practitioners to quickly and effectively determine potential facts of the this topic by clearly knowing about key aspects and related body of research.

 In this consideration, we make mainly three contributions:

 (i) Based on a deep literatures analysis of more than 600 papers, we firstly provide an systematical overview of the existing approaches of ML-based RSP domain from different perspectives;

 (ii) We propose a comprehensive background regarding the main concepts, motivations, and implications of enabling intelligent algorithm in RSP;

 (iii) A profound discussion about the future promising research opportunities and potential trends in this field is proposed.

 Accordingly, the reminder of this review article is organized as follows. Section II briefly introduces the basic principles of typical ML algorithms; section III surveys the latest developments in radar radiation sources classification and recognition; section IV investigates state-of-the-art achievements in radar image processing; section V investigates the developments of anti-jamming and interference mitigation; other RSP-related research that does not fall in previous categories, such as waveform design, anti-interference, has been reviewed in section VI; section VII profoundly discusses open problems and possible promising research directions, in order to indicate current gaps and potential future solutions and trends. Finally, the conclusion of this article is drawn in section VIII. The overview contents of this paper is shown in Fig. 1.

\section{The Basic Principles of Typical Machine Learning Algorithms }\label{S2}

 ML has achieved great success in many domains, mainly related to three determining factors: data, model algorithm, and computation power. As a data-driven pattern, big data is the basic motivation for development of ML. Computation power is supported by hardware equipments to drive ML model training, such as graphical processing units (GPUs), tensor processing units (TPUs), Kunpeng 920 produced by Huawei corporation. This section will briefly introduce the RSP-related typical ML model algorithms.

  \vspace{-1.0mm}
\subsection{Traditional Machine Learning Models}\label{S2.1}

 \textbf{1) Support Vector Machines (SVMs)}.

 Support Vector Machines are the most popular ML algorithms for binary classification \cite{SVM}, especially high-efficiently in solving non-linear binary classification issues, through the projection of low dimensional feature space to a higher one with kernel function \cite{kernal_SVM} (e.g., polynomial kernels, radial basis function (RBF) kernel, Gaussian kernel). SVMs address the classification problem by finding an optimal hyperplane in the feature space to maximize the samples margin between the support vectors of two classes, as shown in Fig. 2. The optimal problem can be expressed as in Eq.(1), which is a convex optimization problem, and the sequence minimization optimization (SMO) \cite{SMO} can be used as an optimization algorithm. SVMs have been widely applied to radar emitter classification and recognition\cite{Jordanov25,zhang47,Yuan48,Xu49}.

\begin{figure}[h] 
\vspace{-2.0mm}
\begin{center}
\includegraphics[width=1.0\columnwidth, keepaspectratio]{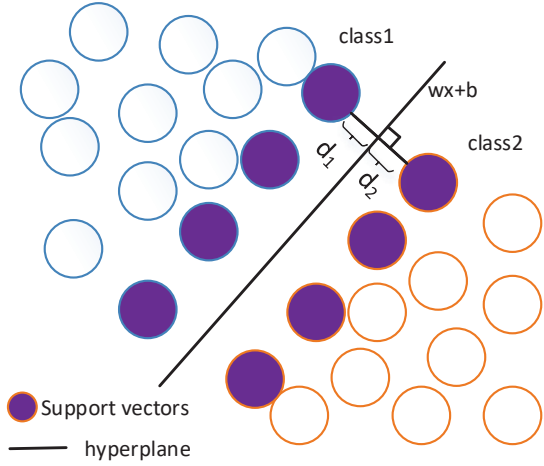}
\end{center}
\captionsetup{font = {footnotesize}, name = {Fig.}, singlelinecheck = off, justification = raggedright, labelsep = period}\
\vspace{-6.0mm}
\caption{The diagram of SVM,$\bm{d_{1}=d_{2}}$.}
\label{FIG3}
\vspace{-3.0mm}
\end{figure}

\begin{equation}\label{P_(v,h)} 
  \begin{aligned}
    &\min \frac{1}{2}{\left\| \bm{\omega}  \right\|^2} \\
    &s.t.\ {\rm{  }}{y^{(i)}}(\left\langle {\bm{\omega} ,{x^{i}}} \right\rangle  + \bm{b}) \ge 1\ (i = 1,2...m),
  \end{aligned}
\end{equation}
 where ${\bm{\omega}}$ and ${\bm{b}}$ are the hyperplane parameters, ${m}$ is the number of samples, ${x}$, ${y}$ are the samples and the labels, respectively. When the classes do not have an explicit classification hyperplane, i.e., inherently not separable. Soft-SVMs can be used to tackle this issue. This means that a small number of samples is allowed to fall into the wrong side. The objective of soft-SVMs adds a penalty term based on SVMs to restrict the slack term ${\bm{\varepsilon}}$, as follow:

 \begin{equation}\label{P_(v,h)} 
  \begin{aligned}
    &\min\ (\frac{1}{2}{\left\| \bm{\omega}  \right\|^2} + C\sum {{\varepsilon _i}} ) \\
    &s.t.\ {\rm{  }}{y^{(i)}}(\left\langle {\bm{\omega} ,{x^{i}}} \right\rangle  + \bm{b}) \ge 1 - {\varepsilon _i}\ (i = 1,2...m)\\
    &{\varepsilon _i} \ge \max \left\{ {0,\ 1 - {y^{(i)}}(\left\langle {\bm{\omega},{x^{i}}} \right\rangle  + \bm{b})} \right\},
    \end{aligned}
\end{equation}
 where ${C}$ is the penalty term.

\textbf{2) Decision Trees (DTs)}.

 Decision Trees are intuitively the simplest case of ML algorithm. They are suitable for addressing these situations where the labels of data are non-continuous. DTs adopt \emph{if-then} rules to split the input data according to features and suitable threshold values based on a binary tree structure \cite{DT}. The root nodes, middle nodes, and leaf nodes represent input data, features and threshold attributes and outputs, respectively. Each branch represents an output of the discrimination process. The loss function is usually implemented as a \emph{mean\ square\ error (MSE)} for regression and \emph{cross\ entropy (CE)} for classification. DTs typically use a limitation of the tree structure depth and pruning operations to address the overfitting problem. Although pruning will reduce the task accuracy to some extent, it generally improves the generalization. Information entropy-based ID3\cite{DT}, C4.5\cite{C4.5}, and Gini coefficient-based classification and regression tree (CART)\cite{CART} are usually the optimization algorithms that are implemented during the training process. DTs has been applied to radar emitter classification and recognition\cite{Matuszewski58}.

\textbf{3) Boosting Ensemble Learning}.

 The Ensemble Learning (EL)\cite{ensemble_learning} builds multi-classifier to jointly make prediction of inputs. The advantages of ensemble learning are as follow: (i) improving prediction accuracy with joint decision; (ii) can easily deal with either large or small datasets, i.e., large dataset can be divided into multiple subsets to build a multi-classifier, small dataset can be sampled to reform multiple datasets to establish a multi-classifier; and (iii) suitable to address the complex decision boundary problems, homologous and  heterogeneous datasets. EL can be categorized into two classes: \emph{bootstrap} (such as random forest) and \emph{boosting} (such as adaboost\cite{adaboost_SAR_ATR}, gradient boosting decision tree (GBDT)\cite{gradient_boosting}, extreme GBDT (XGBoost)\cite{Xgboost}). Gradient boosting methods\cite{gradient_boosting,Pavlyshenko64} were used as classification model in radar emitter recognition.

 \emph{\textbf{Random\ forest (RF)}} As one of the ensemble learning algorithms\cite{RFs}, RF is a bootstrap ensemble classifier, consisting of relatively independent multi-CART, to overcome the high prediction error problem with a single DT. Every sub-DT is a weak learning model as a part of an over learning task, trained by a random subset bootstrapped from the training dataset, and determined splitpoint with random features. The final prediction output is determined by \emph{voting} rules with all DTs. RF may reach the global optimum, instead of a local optimum as in the case of a single tree. The radar signals recognition based on RF models was proposed in \cite{Jordanov25} to obtain comparable performance.

 \emph{\textbf{Adaboost}} Adaboost, i.e., adaptive boosting, which firstly produce a set of hypothesis functions by repeatedly using basic learning algorithm based on multi-sampled training data. Then, these hypothesis functions are connected to ultimately form an ensemble learner via linear weighted vote rules \cite{adaboost_SAR_ATR}. An AdaBoost algorithm was employed as a classifier in \cite{Guo66} to complete the different types recognition of radar signals with 1D harmonic amplitude data sets.

 Given the hypothesis function ${{\bm{{\rm H}}} = \{h(x):{x} \to \bm{R}\}}$ and unknown data ${x}$, ${h(x)}$ donates weak learners or base learners, then the ultimate ensemble learner can be given by:

\begin{equation}\label{P_(v,h)} 
 F(x) = \sum\limits_{t = 1}^T {{\alpha _t}{h_t}(x)},
\end{equation}
 where ${\alpha _t}$ is the connection coefficients of ${t}$-th iteration, ${T}$ is the number of iteration. ${\bm{\alpha}  = [{\alpha _1},{\alpha _2},..,{\alpha _T}]}$ and ${h(x)}$ are optimally generated during the minimization of loss function ${C}$, as showed in Eq.(4). Initially, the weight of every sample is set equal to ${\frac{1}{N}}$, ${N}$ being the number of samples. When the sample is misclassified, it gets a larger weight in the following iterations, the base learner is forced to focus on these hard-to-classify cases in the subsequent training steps. This characterizes the adaptation of boosting methods.

\begin{equation}\label{P_(v,h)} 
 C = \frac{1}{N}\sum\limits_{n = 1}^N {\exp ( - {y_n}F({\bm{x}_n}))},
\end{equation}
 where ${{y_n} \subset \{  + 1, - 1\} }$  is the label of data ${{x}_n}$.

 \emph{\textbf{GBDT}} As a residual learning type, the prediction score of GBDT \cite{gradient_boosting} is determined by summing up the scores of a multi-CART regression tree, instead of a classification tree. In detail, adding a new tree structure to learn the residual (i.e., the gap between the prediction and the actual value) of previous tree at each iteration based on negative gradient learning, to iteratively approach the actual value. For a given dataset with ${n}$ samples and ${m}$ features ${D = \{ ({{x}_i},{y_i})\}\ (i = 1...n,{{x}_i} \in {{R}^m},{y_i} \in \bm{R})}$ which uses ${K}$ additive tree functions to predict the output (take a regression tree as an example)\cite{Xgboost}:

\begin{equation}\label{P_(v,h)} 
 {\hat{y}_i} = \phi ({\bm{x}_i}) = \sum\limits_{k = 1}^K {{f_k}({\bm{x}_i})} ,{f_k} \in \Gamma,
\end{equation}
 where ${\Gamma  = \{ f(x) = {\omega _{q(x)}}\}\ (q:{{R}^m} \to T,\bm{\omega}  \in {{R}^T})}$ is the space of regression tree. ${q}$ represents the structure of each tree that maps an example to the corresponding leaf index. ${T}$ is the number of leave nodes in the tree. Each ${f_k}$ corresponds to an independent tree structure ${q}$ and leaf weights $\bm{\omega}$. Let ${y_i}$ and ${\hat{y}_i}$ be the actual and prediction values, then we minimize the following loss function:

\begin{equation}\label{P_(v,h)} 
 L(\phi ) = \sum\limits_i {l({y_i},{\hat{y}_i})}\ (i = 1...n,\ {y_i} \in \bm{R}),
\end{equation}
 where ${l}$ is a loss function, usually the MSE.

 \emph{\textbf{XGBoost}} As an implementation of gradient tree boosting, XGBoost\cite{Xgboost}, an end-to-end scalable tree boosting system, is widely used in data mining. As one of the most popular ML model, it provides the state-of-the-art performance in many Kaggle competitions in recent years. For example, 17 solutions used XGBoost (eight solely used XGBoost, while others combined XGBoost with neural networks) among the 29 challengeable winning solutions at 2015 Kaggle competition\cite{Xgboost}. XGBoost was also used in the top-10 in the KDDCup 2015 by each award-winning team\cite{Xgboost}. In addition, the authors in \cite{Chen35} used weighted-XGBoost for Radar emitter classification. XGBoost's widespread scalability as its one of the most important factor of success, which can scale to billions of examples in distributed or memory-limited setting and have higher computation efficiency than existing popular solutions on a single machine. Compared to GBDT, XGBoost adds a penalty term (i.e., regularized term) in objective function to overcome overfitting, and introduces the first and second order gradient in objective based on Taylor expansion. The XGBoost minimizes the following objective,

\begin{equation}\label{P_(v,h)} 
  \begin{aligned}
    &L(\phi ) = \sum\limits_i {l({y_i},\hat{y}_{i})}  + \sum\limits_k {\Omega ({f_k})}, \\
    &\Omega (f) = \gamma T + \frac{1}{2}\lambda {\left\| \omega  \right\|^2},
  \end{aligned}
\end{equation}
 where ${\Omega}$ is the regularized term to penalize the complexity of the model, usually ${l_1\ norm}$ or ${l_2\ norm}$. The optimization algorithm is residual learning iteratively between the adjacent sub-model, and let ${\hat{y}_{i}^{t-1}}$ be as the prediction of ${i}$-th instance at ${(t-1)}$-th iteration, we minimize the following objective,

\begin{equation}\label{P_(v,h)} 
 {L^{(t)}} = \sum\limits_{i = 1}^n {l({y_i},\hat{y}_i^{t - 1} + {f_t}({x_i}))}  + \Omega ({f_t}),
\end{equation}
 where ${f_t}$ represents the residual between ${(t-1)}$-th and ${t}$-th iterations. Inspired with Taylor expansion (the second order expansion): ${f(x + \Delta{x}) \approx f(x) + {f^{'}}(x)\Delta{x} + \frac{1}{2}f ^{''}(x)\Delta{x^2}}$. The above equation can be rewritten as follow,

\begin{equation}\label{P_(v,h)} 
 {L^{(t)}} \approx \sum\limits_{i = 1}^n {[l({y_i},\hat{y}_i^{t - 1}) + {g_i}{f_t}({x_i}) + \frac{1}{2}{h_i}f_{_t}^2({x_i})]}  + \Omega ({f_t}),
\end{equation}
 where ${{g_i} = \partial {\hat{y}^{t - 1}}l({y_i},\hat{y}_i^{t - 1})}$ and ${{h_i} = \partial _{\hat{y}_i^{t - 1}}^2l({y_i},\hat{y}_i^{t - 1})}$ are the first and second order gradient statistics on the loss function, respectively. Removing the constant term ${[l({y_i},\hat{y}_i^{t - 1})]}$ to simplify the Eq.(7) by

 \begin{figure}[!tp] 
\vspace{-2.0mm}
\begin{center}
\includegraphics[width=1.0\columnwidth, keepaspectratio]{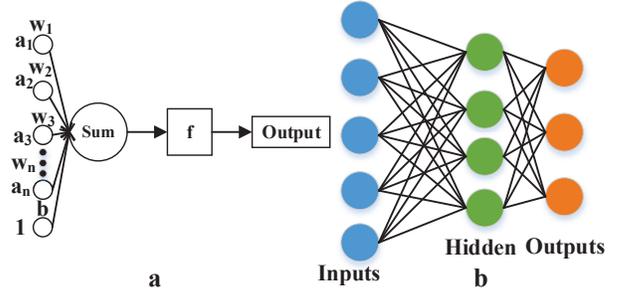}
\end{center}
\captionsetup{font = {footnotesize}, name = {Fig.}, singlelinecheck = off, justification = raggedright, labelsep = period}\
\vspace{-6.0mm}
\caption{The diagram of ANN,(a)single neutron,(b)artificial neural network with one-hidden layer.}
\label{FIG3}
\vspace{-3.0mm}
\end{figure}

\begin{equation}\label{P_(v,h)} 
 {L^{(t)}} = \sum\limits_{i = 1}^n {[{g_i}{f_t}({x_i}) + \frac{1}{2}{h_i}f_{_t}^2({x_i})]}  + \Omega ({f_t}).
\end{equation}

 Define ${C = \{ j\left| {q({x_i}) = j} \right.\}}$ as the set of leaf nodes ${j}$, Eq.(8) can be rewritten as

\begin{equation}\label{P_(v,h)} 
  \begin{aligned}
    &{L^{(t)}} = \sum\limits_{i = 1}^n {[{g_i}{f_t}({x_i}) + \frac{1}{2}{h_i}f_{_t}^2({x_i})]}  + \gamma T + \frac{1}{2}\lambda \sum\limits_{j = 1}^T {\omega _j^2} \\
    & = \sum\limits_{j = 1}^T {[(\sum\limits_{i \in C} {{g_i})} {\omega _j}{\rm{ + }}\frac{1}{2}(\sum\limits_{i \in C} {{h_i})} \omega _j^2] + \gamma T + \frac{1}{2}\lambda \sum\limits_{j = 1}^T {\omega _j^2} } \\
    &{\rm{ = }}\sum\limits_{j = 1}^T {[(\sum\limits_{i \in C} {{g_i}} ){\omega _j}{\rm{ + }}\frac{1}{2}(\sum\limits_{i \in C} {{h_i}}  + \lambda )\omega _j^2] + \gamma T}.
  \end{aligned}
\end{equation}

 When fixing a tree ${q(x)}$, the optimal score ${\omega _j^*}$ of leaf nodes ${j}$ is given by

\begin{equation}\label{P_(v,h)} 
 \omega _j^* =  - \frac{{\sum\limits_{i \in C} {{g_i}} }}{{\sum\limits_{i \in C} {{h_i}}  + \lambda }}.
\end{equation}

 The optimal objective at ${t}$-th iteration is given by

\begin{equation}\label{P_(v,h)} 
 {L^t}(q) =  - \frac{1}{2}\sum\limits_{j = 1}^T {\frac{{{{(\sum\limits_{i \in C} {{g_i}} )}^2}}}{{\sum\limits_{i \in C} {{h_i}}  + \lambda }} + \gamma T}.
\end{equation}

 Eq.(10) can be used as a scoring function to measure the quality of a tree structure ${q}$. A greedy algorithm is used to search for optimal tree structure ${C_L}$ and ${C_R}$ (${C = {C_L} \cup {C_R}}$) are sets of left and right nodes after being split. The reduction of loss after being split is given by

\begin{equation}\label{P_(v,h)} 
 {L_{split}} = \frac{1}{2}[\frac{{{{(\sum\limits_{i \in {C_R}} {{g_i}} )}^2}}}{{\sum\limits_{i \in {C_R}} {{h_i}}  + \lambda }} + \frac{{{{(\sum\limits_{i \in {C_L}} {{g_i}} )}^2}}}{{\sum\limits_{i \in {C_L}} {{h_i}}  + \lambda }} - \frac{{{{(\sum\limits_{i \in C} {{g_i}} )}^2}}}{{\sum\limits_{i \in C} {{h_i}}  + \lambda }}] - \gamma.
\end{equation}

\textbf{4) Artificial Neural Networks (ANNs)}.

 Inspired by human's brain neural network, ANN, a simplified mathematical analogue of human's neural network\cite{ANN}, is used to process the input information by the layer-wise style for regression and classification tasks, as shown in Fig. 3. A basic ANN has one input layer, one hidden layer, and output layer, each of which has many artificial neutrons. The number of neurons is determined by dimension of input data for input layer, number of classes for output layer, and alternative for hidden layer. All neurons in one layer are connected to all neurons in all adjacent layers (i.e., fully connection) with weights, bias, and non-linear activation function for every neuron. Obviously, more neurons in hidden layer or more hidden layers, will rapidly increase the ability of information processing because of improved power of feature extraction of data, which characterizes the deep learning algorithm (will be introduced in next subsection). The optimal training method is backpropagation algorithm to iteratively update the parameters of ANN. ANNs were widely used in radar emitter recognition\cite{Shieh52,Liu53,Yin54,Zhang55,Granger56}.

 \begin{figure}[h] 
\vspace{-2.0mm}
\begin{center}
\includegraphics[width=0.8\columnwidth, keepaspectratio]{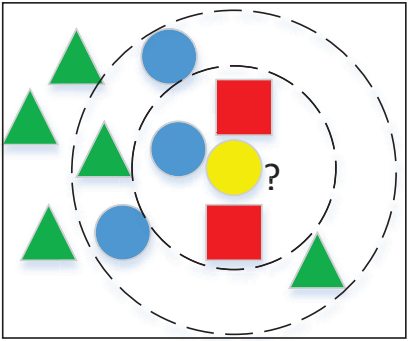}
\end{center}
\captionsetup{font = {footnotesize}, name = {Fig.}, singlelinecheck = off, justification = raggedright, labelsep = period}\
\vspace{-6.0mm}
\caption{The diagram of K-NN.}
\label{FIG3}
\vspace{-3.0mm}
\end{figure}

\textbf{5) K-Nearest Neighbors (K-NNs)}.

 As an instance-based learning style, K-NNs are not like other classifiers to explicitly train a classification model to classify unknown samples\cite{KNN}. Instead, samples have aforehand classes and features, and the class of unknown sample is determined by K nearest neighbors surrounding it, which is evaluated by the distances between feature spaces of aforehand samples and unknown samples (such as Euclidean distance, cosine distance, correlation, Manhattan distance). The unknown sample belongs to the class where the highest frequency in K nearest neighbor samples. For example, in Fig. 4, when K = 3, the color of yellow cycle classified to red, and the color of yellow cycle is classified to blue when K = 7. K is very significant for classification, which usually starts with K = 1, iteratively finding the smallest error with increment 1. K-NN is adopted to classify instantaneous transient signals based on radio frequency fingerprint extraction in \cite{Rehman36}.

  \vspace{-1.0mm}
\subsection{Deep Learning Models}\label{S2.2}
 DL models, also called DNN, consist of multi-layer ANN, i.e., input layers, multi-hidden layer, and output layer, which transform input data (e.g., images, sequences) to outputs (e.g., classes) with the high-level feature representation learning by multi-hidden layers.

 In 2006, Hinton has successfully achieved training of DBN with gradient decent backpropagation algorithm, and experiments results determined promissing performance in CV tasks \cite{Hinton_2006}. This breakthrough quickly draw insights from the industrial and academics. Especially, CNN-based AlexNet architecture has firstly won the human in the competition of ImageNet contest in 2012 \cite{Imagenet_2012}. DL has developed rapidly in many domains, such as speech recognition \cite{speech_recognition}, image processing \cite{Imagenet_2012,objection,Scene_Recognition,Semantic_Segmentation}, audio signal processing \cite{Acoustic_Modeling,music_recommendation}, video processing \cite{Video_analysis_1,Video_analysis_2}, and NLP \cite{NLP_1,NLP_2,NLP_3}. In the following years, many novel DL architectures and domain achievements have developed, including CNNs, RNNs, and GANs.

 The remainder of this section is contributed to briefly introduce several commonly used DL models in RSP, including unsupervised AEs, DBNs and GANs, and supervised CNNs, RNNs.

 \begin{figure}[h] 
\vspace{-2.0mm}
\begin{center}
\includegraphics[width=1.0\columnwidth, keepaspectratio]{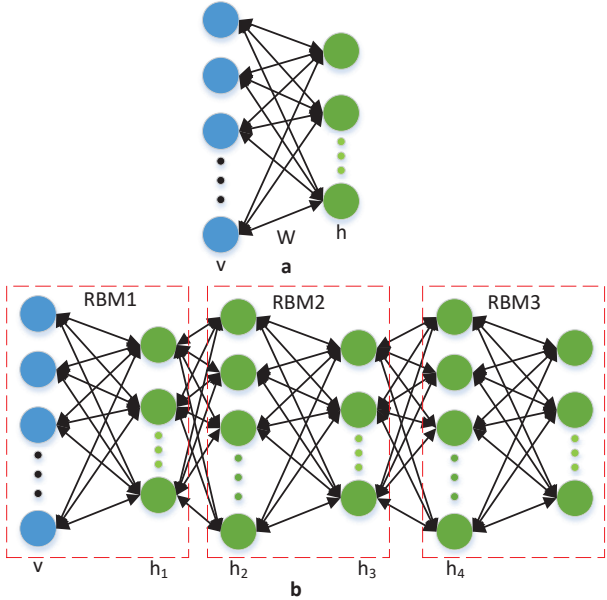}
\end{center}
\captionsetup{font = {footnotesize}, name = {Fig.}, singlelinecheck = off, justification = raggedright, labelsep = period}\
\vspace{-3.0mm}
\caption{The RBM and DBN model, (a) RBM,(b) DBN.}
\label{FIG2}
\vspace{-3.0mm}
\end{figure}

\textbf{1) Restricted Boltzmann Machines (RBMs) and Deep Belief Networks (DBNs)}.

 A restricted Boltzmann machine (RBM), composed by a visible layer ${x}$ and a hidden layer ${h}$, and  symmetric connections between these two layers represented by a weight matrix ${\bm{W}}$, is a generative stochastic undirectional neural network \cite{Acoustic_Modeling}. The joint distribution of visible and hidden units is defined by its energy function as follow \cite{RBM_1,RBM_2}
\begin{equation}\label{P_(v,h)} 
 \bm{P(v,h)} = \frac{1}{Z} e^{(-\bm{E(v,h)})} ,
\end{equation}
 where Z is the partition function. If the visible units are binary-value, ${\bm{E(v,h)}}$ can be defined as
\begin{equation}\label{E_(v,h)} 
 \bm{E(v,h)} = - \sum\limits_{i,j} {v_i}\bm{W_{ij}}{h_j} - \sum\limits_j {b_j}{h_j} - \sum\limits_i {c_i}{v_i} ,
\end{equation}
 where ${b_j}$ and ${c_i}$ are hidden unit bias and visible unit bias respectively. ${b,c,\bm{W}}$ are the parameters of RBM model.

 A DBN can be viewed as a stacked structure of multi-RBM model \cite{Hinton_2006,Acoustic_Modeling,DBN_1}, which is regarded as a generative probabilistic graphical model. DBN can break the limitation of RBM representation with a fast training algorithm \cite{Hinton_2006}. The RBM and DBN examples are shown in Fig. 5. In RSP domain, DBN has been used to radar emitter recognition and classification\cite{Jeong83,Liu84,Cao85}, HRRP-ATR\cite{Pan255,Peng256}, SAR-ATR\cite{Zhao161}.

\textbf{2) Autoencoders (AEs)}.

 AEs are basically unsupervised learning algorithms, which normally accomplish the tasks of data compression and dimensionality reduction in unsupervised manner. An AE model consists of three opponents: encoder, activation function, and decoder, as shown in Fig. 6 and Fig. 7\cite{AEs_SAR_RS}.

 \begin{figure}[h] 
\vspace{-2.0mm}
\begin{center}
\includegraphics[width=1.0\columnwidth, keepaspectratio]{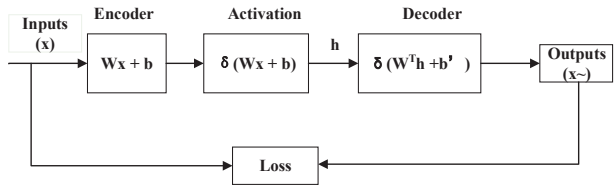}
\end{center}
\captionsetup{font = {footnotesize}, name = {Fig.}, singlelinecheck = off, justification = raggedright, labelsep = period}\
\vspace{-3.0mm}
\caption{The general autoencoder model.}
\label{FIG2}
\vspace{-3.0mm}
\end{figure}

\begin{figure}[h] 
\vspace{-2.0mm}
\begin{center}
\includegraphics[width=1.0\columnwidth, keepaspectratio]{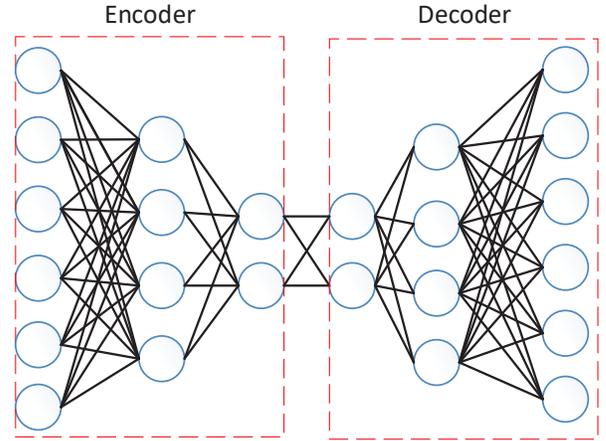}
\end{center}
\captionsetup{font = {footnotesize}, name = {Fig.}, singlelinecheck = off, justification = raggedright, labelsep = period}\
\vspace{-3.0mm}
\caption{The fully connection neural network model of Autoencoder.}
\label{FIG2}
\vspace{-3.0mm}
\end{figure}

\begin{figure*}[!tp] 
\vspace{-2.0mm}
\begin{center}
\includegraphics[width = 2.0\columnwidth, keepaspectratio]{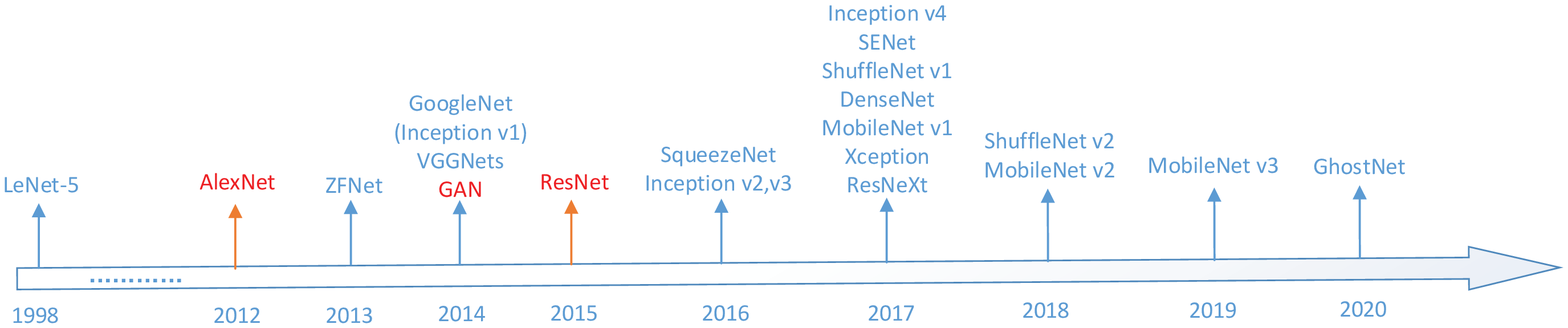}
\end{center}
\captionsetup{font = {footnotesize}, singlelinecheck = off, justification = raggedright, name = {Fig.}, labelsep = period}%
\vspace{-2.0mm}
\caption{The classic CNN models \cite{CNN_latest}.}
\label{FIG4}
\end{figure*}

 Encoder ${f}$ can be regarded as a linear feed-forward filter of input ${x}$ determined by weight matrix $\bm{W}$ and bias $\bm{b}$, i.e., ${f = \bm{W}x + \bm{b}}$.

 Activation function ${\sigma}$ performs a non-linear mapping that transforms the ${f}$ into latent representation ${h}$ of input ${x}$ at the range of ${[0,1]}$, i.e., ${h = \sigma(\bm{W}x + \bm{b})}$.

 The decoder ${g}$ is a reverse linear filter to produce the reconstruction $\widetilde{{x}}$ of the input ${x}$, i.e., ${\widetilde{{x}} = g(\bm{W}^{T}h + \bm{b}^{'})}$

 A loss function ${L}$ is used to measure how close the AE can reconstruct the output $\widetilde{{x}}$, i.e.,  ${L(\widetilde{x},x)}$. The training processing is minimizing the loss between  $\widetilde{{x}}$ and ${x}$, i.e., ${min(L(\widetilde{x},x))}$.

 \emph{\textbf{Sparse\ autoencoder (SAE)}} In order to accelerate the training of AE model, SAE characterizes by adding sparsity constraints to the hidden layers, and only activating the neurons whose outputs are close to 1. Therefore, the only small amount of parameters was needed to learn greatly reduce training time.

 \emph{\textbf{Denoising\ autoencoder (DeAE)}}  To increase the robustness of AE with small various input data, DeAE has been proposed in \cite{denoise_AE}. Before entering into the input layer, the original input ${x}$ is corrupted as ${x^{'}}$. Binary noisy and Gaussian noise are usually the two corruption methods.

 \emph{\textbf{Variational\ autoencoder (VAE)}}  Different from original AE model, VAE\cite{VAE_1,VAE_2} is a probabilistic generative model. The latent representation ${h}$ of inputs is not directly learned by encoder, but being encoded learning by encoder to generate a desired latent probabilistic distribution at condition of probabilistic constraint. Generally, this constraint is standard normal distribution, i.e., N(0,1). In phase of decoding, sampling from the latent distribution representation ${h}$, the decoder generates the output. Therefore, the VAE has two loss function: one for encoder to evaluate the similarity between generated distribution by encoder and standard distribution, the other is for measuring how close between the original input and the output data. The idea of generator of GANs is also from the VAE. please refer to related literature\cite{AEs_SAR_RS} for other AEs, such as, contractive autoencoder \cite{contractive_AE}, convolutional autoencoder \cite{conv_AE}.

\textbf{3) Convolutional Neural Networks (CNNs)}.

 Inspired by animal's visional neural information processing system, CNNs are extensively applied to many research domains \cite{CNN_latest}, including CV, NLP, speech recognition. Specialized convolution layer and pooling layer, CNNs can quickly extract latent features of data by shared convolutional kernel and downsampling with pooling operation, which characterizes with the positive properties of partiality, identity, and invariance. Up to now, many famous CNNs architectures have emerged, (including one-dimension, two-dimension, and multi-dimension, the relative diagrams showed in Fig. 9), such as LeNet-5\cite{LeNet}, AlexNet\cite{AlexNet}, VGGNet\cite{VGGNet}, GoogleNetv1-v4\cite{GoogleNet_v1,GoogleNet_v2,GoogleNet_v3,GoogleNet_v4}, ResNet\cite{ResNet}, MobileNetv1-v3\cite{MobileNet_v1,MobileNet_v2,MobileNet_v3}, ShuffleNetv1-v2\cite{ShuffleNet_v1,ShuffleNet_v2}, and the latest GhostNet\cite{GhostNet}. The developments of classic CNNs models are shown in Fig. 8. The increasing depth of model is a main fashion at the starting time of DL, e.g, from the starting with a 5-layer (two convolution-pooling layers and three fully connection layers) of LeNet in 1998 to hundreds of layers of ResNet in 2015\cite{ResNet}. In recent years, the lightweight models design, i.e., small volume of parameters, is increasingly popular, e.g., ShuffleNet, MobileNet, EfficientNet\cite{EfficientNet}.

 LeNet-5 has been successfully applied to handwritten digits recognition \cite{LeNet}, which is equipped with two convolution-pooling layers (convolution kernel: ${3*3}$, and ${5*5}$) and three fully connection layers, but without activation function. This structure pattern was widely used in most of CNNs models. A 8-layer AlexNet has firstly won the championship in ImageNet large-scale visual recognition challenge (ILSVRC) in 2012\cite{AlexNet}, which quickly expands the intensive research interest of deep learning from the industrials and academics. This competition verified that the deeper the model is, the better performance will be. The structure of AlexNet has 5-convolution-pooling layers (convolution kernel: ${3*3}$, ${5*5}$, and ${11*11}$), 3-fully connection layers, others containing Relu activation function, dropout. To increase the depth of model, VGGNet of has been proposed for in ILSVRC 2014 \cite{VGGNet}, won the second place, including VGG-11, VGG-13, VGG-16, VGG-19. Its convolutional kernels are all ${3*3}$, instead of ${11*11}$ and ${5*5}$.

  \begin{figure}[h] 
\vspace{-2.0mm}
\begin{center}
\includegraphics[width=1.0\columnwidth, keepaspectratio]{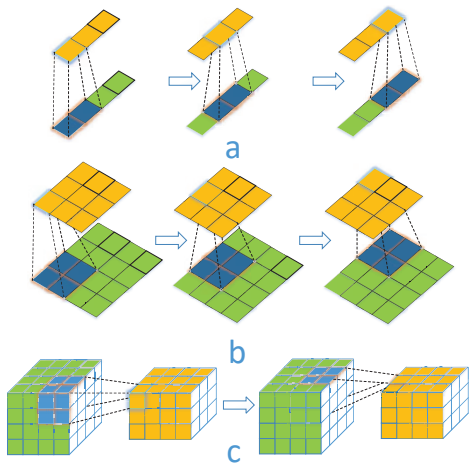}
\end{center}
\captionsetup{font = {footnotesize}, name = {Fig.}, singlelinecheck = off, justification = raggedright, labelsep = period}\
\vspace{-6.0mm}
\caption{The diagram of convolution operation: (a) 1D convolution; (b) 2D convolution; and (c) 3D convolution.}
\label{FIG3}
\vspace{-3.0mm}
\end{figure}

  The large volume of parameters of deep model, however, leads to low computation efficiency during training process. Combined with multi-parallel filters in the same layer (i.e., Inception module), a 22-layer GoogleNet has won the champion in the competition of ILSVRC 2014\cite{GoogleNet_v1}, whose number of parameters is 12 times less than AlexNet, but has higher performance. GoogleNet firstly verified that the deep model can work well by increasing the width of model, not just depth, including GoogleNetv1-v4\cite{GoogleNet_v1,GoogleNet_v2,GoogleNet_v3,GoogleNet_v4}. With the increasing of layers, the problem of difficult training of model is more and more obvious, i.e., gradient exploding and vanishing. To address this issue, in \cite{ResNet}, the authors proposed a 34-layer of ResNet, which won the ILSVRC 2015 as a champion model. The excellent design of ResNet is the skip connection operation of input to directly output, not through the hidden layer. In this way, the model just learns the residual part between the ultimate output and original input, which can keep the gradient existing in a suitable range during all training process to efficiently train more deeper networks. ResNet makes extreme deep network possible, such as, ResNet-152.

 Although ResNet can improve the computation efficiency, a large volume of parameters remains a challenge for optimally training the model in some practical applications, because of the insufficient computing power and low efficient performance. In recent years, the lightweight DL models have become the main research direction, including the design of lightweight model (such as MobileNets(v1-v3)\cite{MobileNet_v1,MobileNet_v2,MobileNet_v3}), ShuffleNets(v1-v2)\cite{ShuffleNet_v1,ShuffleNet_v2}), EfficientNet\cite{EfficientNet}), model compression and hardware acceleration technique\cite{model_compression_hardware_acceleration}. For example, MobileNets was proposed by Google corporation to embed in portable devices, such as mobile phones.

 To solve the problem of redundant features extraction of existing CNNs, Ghost module was proposed in \cite{GhostNet}, which can be embedded in existing CNNs models to construct a high computation efficiency model, i.e., GhostNet, to achieve state-of-the-art performance results in DL tasks.

\textbf{4) Recurrent Neural Networks (RNNs)}.

 Different from the CNNs, RNNs, inspired by the memory function of animal-based information processing system, are used to solve the problem of data prediction with the temporal memory series. In other words, the current output results are related to previous data sequences. The memory unit, as the basic module of RNNs, is shown as in Fig. 10. This unit includes one-layer fully connected neutral network, two input: state ${s}$ (i.e., the memory of previous unit has ${m}$ dimensions) and data feature ${x}$ (${n}$ dimensions), and output as the state input of next memory unit. One-layer RNN consists of multi-memory units sequently connected, as shown in Fig. 11. The number of memory units is determined by the length of data series (${x^{(0)},x^{(1)},...,x^{(R-1)}}$, ${R}$ is the length of input sequence). The output of last memory unit is the ultimate results of RNN learning. All memory units share identical parameters in the same layer of RNN: weights ${\bm{W}}$ of ${m+n}$ dimensions, bias ${\bm{b}}$ of ${m}$ dimensions. Multiple one-layer RNN stacks to form multi-layer RNN.

\begin{figure}[h] 
\vspace{-2.0mm}
\begin{center}
\includegraphics[width=0.7\columnwidth, keepaspectratio]{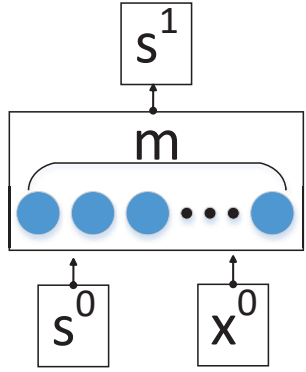}
\end{center}
\captionsetup{font = {footnotesize}, name = {Fig.}, singlelinecheck = off, justification = raggedright, labelsep = period}\
\vspace{-6.0mm}
\caption{The memory unit in RNNs.}
\label{FIG3}
\vspace{-3.0mm}
\end{figure}

 \begin{figure}[h] 
\vspace{-2.0mm}
\begin{center}
\includegraphics[width=1.0\columnwidth, keepaspectratio]{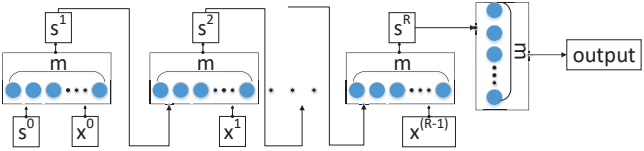}
\end{center}
\captionsetup{font = {footnotesize}, name = {Fig.}, singlelinecheck = off, justification = raggedright, labelsep = period}\
\vspace{-6.0mm}
\caption{The one layer RNN.}
\label{FIG3}
\vspace{-3.0mm}
\end{figure}

 Although the RNNs architecture can achieve the function of memory, the gradient vanishing issue is obvious with the increasing length of time series during the training process. To address this problem, long short-term memory (LSTM) architecture is proposed in \cite{LSTM}. Compared to original memory unit, a LSTM module has two states: \emph{long-term\ memory} unit (${C}$) and \emph{short-term\ memory} unit (${h}$), both are ${m}$ dimensions. ${C}$ can selectively memorize valuable information of long temporal series, which efficiently transmit the early information to current unit. LSTM consists of four gate units: \emph{forget}, \emph{memory}, \emph{information} and \emph{output}, respectively. Each gate unit includes a fully connection neural network layer with ${m}$ neutrons, and the output of each gate is short time memory ${h}$ and data features. The activation functions are \emph{sigmoid}, except for \emph{information} gate is \emph{tanh} function, since the output of \emph{sigmoid} ranges from 0 to 1, contributing to the functions of \emph{forget}, \emph{memory}, and \emph{output} dramatically. The structure of LSTM is shown in Fig. 12 and the main relationship is shown as the follow,

  \begin{figure}[h] 
\vspace{-2.0mm}
\begin{center}
\includegraphics[width=1.0\columnwidth, keepaspectratio]{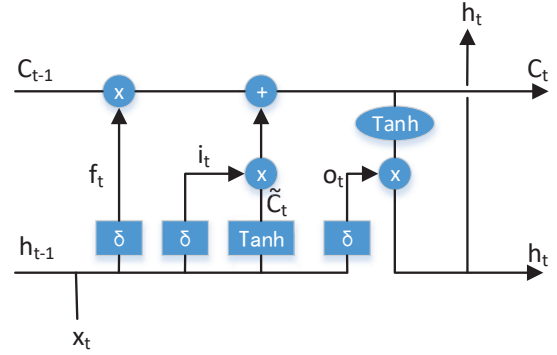}
\end{center}
\captionsetup{font = {footnotesize}, name = {Fig.}, singlelinecheck = off, justification = raggedright, labelsep = period}\
\vspace{-6.0mm}
\caption{The LSTM module.}
\label{FIG3}
\vspace{-3.0mm}
\end{figure}

 Firstly, the ${forget}$ gate unit determines which kind of information should be discarded from the input

\begin{equation}\label{channel} 
 {f}_{t} = {sigmoid}(\bm{W}_{f}[{h}_{t-1}\  {x}_{t}]) + \bm{b}_{f}.
\end{equation}

 The following is the ${memory}$ and ${information}$ gate units to determine the input of new information, i.e, ${input}$ gate unit

\begin{equation}\label{channel} 
 {i}_{t} = {sigmoid}(\bm{W}_{i}[{h}_{t-1}\  {x}_{t}]) + \bm{b}_{i}.
\end{equation}

\begin{equation}\label{channel} 
 \widetilde{C}_{{t}} = {Tanh}(\bm{W}_{c}[{h}_{t-1}\  {x}_{t}]) + \bm{b}_{c}.
\end{equation}

 The new long time memory (${C}$) is acquired by

\begin{equation}\label{channel} 
 {C}_{{t}} = {C}_{t-1} * {f}_{{t}} + {i}_{t} * \widetilde{C}_{{t}}.
\end{equation}

 The ${output}$ gate unit is

\begin{equation}\label{channel} 
 \widetilde{o}_{t} = sigmoid(\bm{W}_{o}[{h}_{t-1}\  {x}_{t}]) + \bm{b}_{o}.
\end{equation}

 Lastly, the short time memory (${h}$) is given by

\begin{equation}\label{channel} 
 {h}_{t} = {o}_{t} * {Tanh}(\widetilde{C}_{{t}}).
\end{equation}

 \begin{figure}[h] 
\vspace{-2.0mm}
\begin{center}
\includegraphics[width=1.0\columnwidth, keepaspectratio]{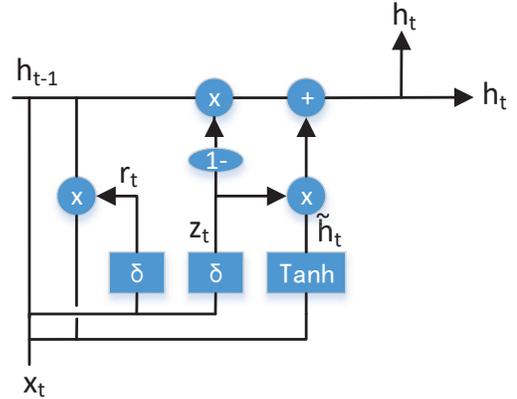}
\end{center}
\captionsetup{font = {footnotesize}, name = {Fig.}, singlelinecheck = off, justification = raggedright, labelsep = period}\
\vspace{-6.0mm}
\caption{The GRU module.}
\label{FIG3}
\vspace{-3.0mm}
\end{figure}

 Accordingly, LSTM architecture can solve the gradient vanishing issue, thanking to the ${long\ time\ memory}$ unit (${c}$) and ${forget}$ gate unit. ${forget}$ gate discards much non-valuable redundant information and ${c}$ can preserve valuable information with large numerical value, therefore, the gradient will not become smaller after layer-by-layer gradient decent training, and avoid emerging gradient vanishing to some extent. As a variant of LSTM, gated recurrent unit (GRU), which combines ${forget}$ gate with ${input}$ gate (i.e., ${memory}$ and ${information}$ gates aforementioned) as a single ${update}$ gate, is more simple than LSTM \cite{GRU}. The GRU module is shown in Fig. 13. The relationships of variables are shown as following,

\begin{equation}\label{channel} 
 {r}_{{t}} = {sigmoid}(\bm{W}_{r}[{h}_{t-1}\  {x}_{t}]) + \bm{b}_{r},
\end{equation}

\begin{equation}\label{channel} 
 {z}_{{t}} = {sigmoid}(\bm{W}_{z}[{h}_{t-1}\  {x}_{t}]) + \bm{b}_{z},
\end{equation}

\begin{equation}\label{channel} 
 \widetilde{h}_{{t}} = {Tanh}(\bm{W}_{h}[{h}_{t-1} * {r}_{{t}}\  {x}_{t}]) + \bm{b}_{h},
\end{equation}

\begin{equation}\label{channel} 
 {h}_{{t}} = ({1-z_{t}}) * {h}_{t-1} + {z}_{t} * \widetilde{h}_{t}.
\end{equation}

 Although the performance of GRU is similar to LSTM, the structure of GRU is simpler than that of LSTM. The amount of parameters of GRU is only one third of LSTM. Therefore, GRU converges fast  and does not cause overfitting.

\textbf{5) Generative Adversarial Networks (GANs)}.

 Similar to VAE, GANs are also unsupervised generative models, which consist of ${Generator\ G}$ and ${Discriminator\ D}$ \cite{GAN,GAN_overview_1,GAN_overview_2}, as shown in Fig. 14. The input of ${G}$ is usually noise with standard normal distribution, i.e., ${N(0,1)}$, to generate a new sample (e.g., image) as output. The ${D}$ is an two-class classifier, to discriminate whether the generated sample is true or not. So the inputs are new generated sample and true sample, and the output is the probability of classification. The loss function of ${D}$ has two parts: ${loss}_{1}$, determined by true sample and true labels, and ${loss}_{2}$ is for generated sample and its label. The ${D}$ makes correct discrimination between generated and true samples by minimizing the $({loss}_{1} + {loss}_{2})$. The ${G}$ has just one loss function ${loss}$ determined by generated sample and true label, to try to trick the ${D}$. The loss function of GAN is as follow

\begin{equation}\label{channel} 
 \begin{aligned}
   \mathop {\min }\limits_G \mathop {\max }\limits_D V(G,D) = {E_{x\sim{p_{data}}(x)}}[\log D(x)] + \\ {E_{z\sim{p_z}(z)}}[\log (1 - D(G(z)))],
 \end{aligned}
\end{equation}
 where ${{p_{data}}(x)}$, ${z}$, ${p_z(z)}$, and ${G(z)}$  represent true distribution of sample, noise signal, distribution of noise signal, and generated new sample, respectively. The distribution of ${G(z)}$ is ${p_G(x)}$. ${D(x)}$ and ${1-D(G(z))}$ denote the loss of discriminator and generator respectively.

 GAN firstly trains discriminator to maximize the expectation of discrimination, which tries to correctly discriminate the true and generated samples. Then, fix the parameters of generator to minimize the divergence (i.e., Jensen-Shannon (JS) divergence\cite{GAN}) between the true and generated samples. In other words, the purpose of this phase is making the distribution of the generated sample close to distribution of true sample as close as possible. So the discriminator is used to measure the gap between the generated and true distribution, instead of directly computing the generated distribution of generator ${p_(G(x))}$. The training process will not stop until the discriminative probability of true and generated sample is equal, i.e., ${0.5}$.

 Although supervised learning representation with CNNs has developed many achievements in CV domain, the labeled datasets remains a great challenge. GANs have been demonstrated huge potentials in unsupervised learning, which bridges the gap between supervised learning and unsupervised learning with deep CNNs architecture. Deep convolutional GANs (DCGANs) were proposed in \cite{DC_GAN}. However, GANs suffer from training instability, and it is difficult to adjust the discriminator to an optimal state.

 \begin{figure}[!tp] 
\vspace{-2.0mm}
\begin{center}
\includegraphics[width=1.0\columnwidth, keepaspectratio]{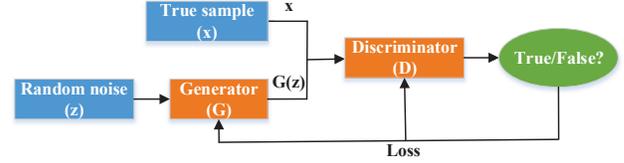}
\end{center}
\captionsetup{font = {footnotesize}, name = {Fig.}, singlelinecheck = off, justification = raggedright, labelsep = period}\
\vspace{-6.0mm}
\caption{The diagram of GAN.}
\label{FIG3}
\vspace{-3.0mm}
\end{figure}

 To address this issue, the authors proposed Wasserstein GAN (WGAN) model in \cite{WGAN,Improve_WGAN} to make the process of training easier by using a different formulation of the training objective that does not suffer from the gradient vanishing problem. WGAN replaces JS divergence in original GAN model with Wasserstein distance as objective loss function, which transforms the binary classification into regression model to fit Wasserstein distance. The discriminator of WGAN must satisfy the space of 1-Lipschitz functions, which enforces through weight clipping. The objective of WGAN is as follow\cite{Improve_WGAN}

 \begin{equation}\label{channel} 
 \mathop {\min }\limits_G \mathop {\max }\limits_{D \in \Omega } W({p_r},{p_g}) = {E_{x\sim{p_r}(x)}}[D(x)] - {E_{z\sim{p_g}(z)}}[D(G(z))],
\end{equation}
 where ${\Omega}$ is the set of 1-Lipschitz functions,  ${{p_g}(z)}$ is the distribution of generator, and ${{p_r}(x)}$ is the true distribution of sample.

 Moreover, there are also some other GANs like condition GAN\cite{conditional_GAN}, cycle GAN\cite{cycle_GAN}, conditional cycle GAN\cite{conditional_cycleGAN}, InfoGAN\cite{InfoGAN}.

\textbf{6) Reinforcement Learning (RL)}.

 Reinforcement learning (RL) system is also an unsupervised learning framework concerning on iteratively making optimal tactical actions act on environment to obtain maximum total amount of rewards\cite{Arulkumaran_DR}. RL is a Markov decision process (MDP) with the interactions between the artificial agent and complex and uncertain environment regarding the sets of states and actions. The exploration-exploitation trade-off is a typical training processing of RL. The former is to explore the whole space to aggregate more information while the latter is to exploit the information with more value at the conditions of current information. As the usual RL algorithm, Q-learning (also action value function) aims to obtain a Q function to model the action-reward relationship. Bellman equation is used to calculate the reward in Q learning. The neural network is often used to model the Q function in deep Q network (DQN).

\section{Radar Radiation Sources Classification and Recognition}\label{S3}

 Electronic warfare (EW) is one of the crucial aspects of modern warfare \cite{EW_Spezio16}. EW receivers are passive systems that receive emission from various platforms that operate in the relative vicinity. The received signals are typically analysed \cite{Ataa23} to obtain valuable information about characteristics and intentions of various elements that are presented in the battlefield. A significant example in modern military warfare is represented by radar radiation sources classification and recognition (RRSCR) \cite{K_means_Yang4,Roe11}, which is one of the tasks that are associated to electronic support measures and electronic signal intelligence systems (ESM/ELINT) \cite{histogram_Zak2,knowledge_J3}. The former (ESM) focuses on classifying different radar types, such as military or civil radar, surveillance or fire control radar, whereas the latter further concerns the identification of individual radar emitter parameters of the same classification, also called specific emitter identification (SEI) \cite{emission_Dudczyk,Huang,Conning,Zhang}. Such operations are based on radio frequency distinct native attribute (RF-DNA) fingerprint features analysis methods \cite{RF_DNA}, such as pulse repetition interval (PRI) modulations analysis, intra-pulse analysis. For example, kernel canonical correlation analysis \cite{kernel_Shi} and nonlinear dynamical characteristics analysis \cite{Huang} have been used to recognize radar emitters. In addition, analysis of out-of-band radiation and fractals theory were reported in \cite{Dudczyk,Dudczyk1}. These radar radiation sources (RRSs) include signal carrier frequency (SCF), linear frequency modulation (LFM), non-LFM, sinusoidal frequency modulation (SFM), even quadratic frequency modulation (EQFM), binary frequency-shift keying (2FSK), 4FSK, dual linear frequency modulation (DLFM), mono-pulse (MP), multiple linear frequency modulation (MLFM), binary phase-shift keying (BPSK), Frank, LFM-BPSK and 2FSKBPSK \cite{Qu,multi_branch_ACSE_radar_signal_recognition}. In this section, RRSCR include the classification and recognition of radar signal automatic modulations (such as intra-pulse modulations, PRI modulations), radar emitter types, radar waveforms, and jamming or interference, as shown in Fig. 15. Examples of time-frequency samples of RRSs are shown in Fig. 16.

 \begin{figure}[!tp] 
\vspace{-2.0mm}
\begin{center}
\includegraphics[width=0.7\columnwidth, keepaspectratio]{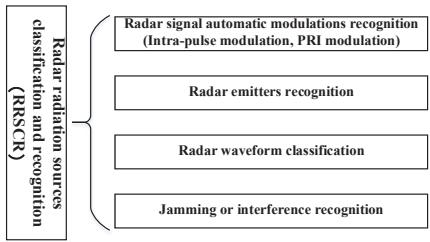}
\end{center}
\captionsetup{font = {footnotesize}, name = {Fig.}, singlelinecheck = off, justification = raggedright, labelsep = period}\
\vspace{-6.0mm}
\caption{The contents of RRSCR.}
\label{FIG3}
\vspace{-3.0mm}
\end{figure}

 \begin{figure}[h] 
\vspace{-2.0mm}
\begin{center}
\includegraphics[width=1.0\columnwidth, keepaspectratio]{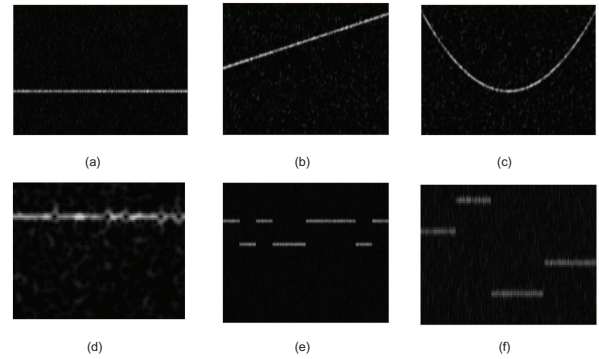}
\end{center}
\captionsetup{font = {footnotesize}, name = {Fig.}, singlelinecheck = off, justification = raggedright, labelsep = period}\
\vspace{-6.0mm}
\caption{The time-frequency images of RRSs. (a)SCF,(b)LFM,(c)non-LFM,(d)BPSK,(e)2FSK,(f)4FSK.}
\label{FIG3}
\vspace{-3.0mm}
\end{figure}

 RRSCR mainly concerns the following four aspects:

 i) denoising and deinterleaving (or separation) of collected pulse streams;

 ii) improving accuracy of recognition in low SNR scenarios, in conditions of missing and spurious data and in real-time;

 iii) boosting robustness and generalization of algorithms;

 iv) identification of unknown radiation sources.

 The methods of RRSCR mainly have three classes:

 i) knowledge based;

 ii) statistical modeling based;

 iii) ML based.

 The knowledge-based methods depend on the prior radar knowledge summarized from the collected raw data by radar experts to achieve RESCR-related works. A novel knowledge-related radar emitter database was built by relational modeling in \cite{emitter_database}. In\cite{knowledge_J3}, the authors proposed radar signal knowledge representation determined by rules with semantic networks. The authors also analyzed signal parameters, feature extraction using linear Karhunen-Loeve transformation and applied knowledge-based techniques to recognize the intercepted radar signals \cite{Matuszewski}. Concerning traditional statistical modeling methods, an autocorrelation spectrum analysis was applied to \cite{Wang16} for modulation recognition of multi-input and multi-output (MIMO) radar signals. In\cite{Nguyen17}, a joint sparse and low-rank recovery approach was proposed for radio frequency identification (RFI), i.e., radar signal separation. In addition, a feature vector analysis based on a fuzzy ARTMAP classifier for SEI was developed in \cite{Conning19}, a wavelet-based sparse signal representation technique was defined for signal separation of helicopter radar returns in \cite{Helicopter20}, and an entropy-based theoretical approach for radar signal classification was developed in \cite{Li21}.

 The increasingly growing complexity of electromagnetic environment demonstrates severe challenges for RRSCR, such as the increasingly violent electronic confrontation and the emergence of new types of radar signals generally degrade the recognition performance of statistic modeling techniques, especially at low signal noise ratio (SNR) scenario. Although these aforementioned technologies can improve performances, they are not sufficient to face these challenges. Knowledge-based methods spend considerable time to extract signal features. Conventional statistical modeling methods depend on statistical features of the collected data. However, this operation pattern do not have competitive performance.

  \begin{figure}[h] 
\vspace{-2.0mm}
\begin{center}
\includegraphics[width=1.0\columnwidth, keepaspectratio]{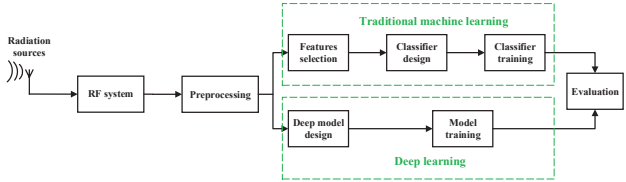}
\end{center}
\captionsetup{font = {footnotesize}, name = {Fig.}, singlelinecheck = off, justification = raggedright, labelsep = period}\
\vspace{-6.0mm}
\caption{The pipeline of RRSCR.}
\label{FIG3}
\vspace{-3.0mm}
\end{figure}

\begin{table*} 
 \centering
 \caption{The traditional ML algorithms in RRSCR}
 \begin{center}
  \begin{tabular}{m{0.3\textwidth}<{\centering}m{0.3\textwidth}<{\centering}m{0.3\textwidth}<{\centering}} 
   \toprule
   \textbf{Features} & \textbf{Models} & \textbf{Accuracy} \\
   \midrule
   PWDs\cite{Jordanov25,Shieh30,Granger31,Matuszewski34};entropy theory\cite{Li21}; spectrum features\cite{Anjaneyulu26,waveform32,intra_pulse_sparse_representation}; wavelet packets\cite{Azimisadjadi27}; dynamic parameters searching\cite{Yin28}; rough sets\cite{Zhang29};  energy envelope\cite{Rehman36}; time-frequency analysis\cite{Zeng40,Zeng41,Mingqiu42,Konopko43}; autocorrelation images\cite{Rigling44,Wang45,waveform_recognition_auto_image,Wang16}; CWTFD\cite{waveform_recognition_CWD,LPI_recognition_CWD,radar_signal_recognition_CWD,radar_emitter_recognition_CWD}; PCA\cite{Yu46}, ambiguity function images\cite{radar_emitter_recognition_AF} & SVMs\cite{Jordanov25,zhang47,Yuan48,Xu49, Liu51}; ANNs\cite{Shieh52,Liu53,Yin54,Zhang55,Granger56,Wong57,Jordanov25,Li21,Anjaneyulu26,Petrov37}; DT\cite{Matuszewski58,LPI_recognition_DT,radar_signal_recognition_DT}; RF\cite{Jordanov25}; Adaboost\cite{radar_signal_recognition_adaboost}; clustering\cite{Li21,radar_emitter_recognition_clustering,radar_emitter_recognition_online_clustering,non_cooperative_radar_emitter_recognition_online_clustering,radar_emitter_identification_K-means}; K-NN\cite{RF_fingerprint_KNN,radar_signal_recognition_KNN,radar_emitter_recognition_KNN,radar_emitter_identification_KNN}; weighted-Xgboost\cite{Chen35}; HMMs\cite{Kim60} & 84\% (-5 dB)\cite{Li21}; 97.3\% (-6 dB)\cite{radar_emitter_recognition_CWD}\\
   \bottomrule
  \end{tabular}
 \end{center}
\end{table*}

\begin{table*} 
  \centering
  \caption{The DL algorithms in RRSCR}
  \begin{tabular}{m{0.3\textwidth}<{\centering}m{0.3\textwidth}<{\centering}m{0.3\textwidth}<{\centering}}
   \toprule
   \textbf{Features} & \textbf{Models} & \textbf{Accuracy} \\
   \midrule
   IQ 1D time sequences\cite{Sun67,Zhang72,Chongqing82,radar_emitter_recognition_IQ}; STFT\cite{Chongqing69,Wang77,Kong78,radar_pulse_detection_STFT,intra_pulse_identification_STFT,radar_emitter_identification_energy_cumulant_STFT,waveform_identification_resnet}; CWTFD\cite{Ming71,Zhang72,Time_frequency73,multi_branch_ACSE_radar_signal_recognition,radar_emitter_identification_CWD_CNN,waveform_recognition_CNN_SVM}; amplitude-phase shift\cite{Selim68}; CTFD\cite{Qu75,Wang76,intra_pulse_identification_CTFD};  bivariate image with FST\cite{FST_bivariate_image}; bispectrum\cite{Li86}; autocorrelation features\cite{PRI_modulation_recognition,ACSE_PRI_autocorrelation,multi_branch_ACSE_radar_signal_recognition}; ambiguity function images\cite{automatic_modulation_classification_AF,radar_waveform_classification_AF}; fusion features\cite{Intra_Pulse74,automatic_modulation_classification_CNN} & CNNs\cite{Sun67,Selim68,Wang77,Ming71,Zhang72,Time_frequency73,Qu75,Intra_Pulse74,Kong78,Ye79,Zhou81,Wang76,GoogleNet_v1,Li86,Chongqing82,radar_signal_modulation_recognition_CNN_transfer_learning}; RNNs\cite{Liu70,radar_emitter_classification_RNN,PRI_modulation_recognition_RNN,waveform_identification_GRU}; DBNs\cite{Liu84,Cao85,radar_emitter_identification_energy_cumulant_STFT,radar_emitter_recognition_DBN_LR}; AEs\cite{Wang76}; SENet\cite{Chongqing69,PRI_modulation_recognition}; ACSENet\cite{ACSE_PRI_autocorrelation,multi_branch_ACSE_radar_signal_recognition}; CDAE + CNN\cite{Wang76,waveform_recognition_CDAE,waveform_recognition_DCDAE}; CNN + DQN\cite{intra_pulse_identification_CTFD}; CNN + LSTM\cite{intra_pulse_identification_CNN_LSTM,waveform_recognition_CNN_LSTM}; CNN + TPOT\cite{LPI_waveform_recognition_CNN_TPOT}; CNN + SVM\cite{waveform_recognition_CNN_SVM} & 94.5\% (-2 dB)\cite{Zhang72}; more than 95\% (-9 dB)\cite{Wang76}; 93.7\% (-2 dB)\cite{Ming71}; over 96.1\% (-6 dB)\cite{Qu75}; 96\% (-2 dB)\cite{waveform_identification_resnet}; more than 90\% (-6 dB)\cite{intra_pulse_identification_CNN_LSTM}; more than 94\% (-6 dB)\cite{intra_pulse_identification_CTFD}; 95.4\% (-7 dB)\cite{waveform_recognition_CDAE}; 94.42\% (-4 dB)\cite{LPI_waveform_recognition_CNN_TPOT}; 97.58\% (-6 dB)\cite{radar_signal_modulation_recognition_CNN_transfer_learning}.\\
   \bottomrule
  \end{tabular}
\end{table*}

 In recent years, because of the high-efficiency of ML algorithms and the rapid development of novel RSP technology, ML-based methods have been successfully applied to RRSCR to face some critical challenges. To better understand these research developments and grasp future research directions in this domain, we provide a comprehensive survey on ML-related RRSCR in this section. This is roughly divided into two parts: one concerning traditional ML algorithms and the other is DL-based methods. A concise summary of some examples of the existing algorithms is shown in Table I and Table II for traditional ML and DL-based algorithms,  respectively. A generic pipeline of ML-based methods is also shown in Fig. 17, to represent a visual framework of ML algorithms.

  \vspace{-1.0mm}
\subsection{Preprocessing}\label{S3.1}

 Data preprocessing is the first step, which processes collected raw data (i.e., sequence data) to prepare for the following  classification or recognition tasks, including denoising, deinterleaving\cite{Liu24}, data missing processing\cite{Jordanov25,Petrov37}, unbalanced dataset\cite{Chen35,Sun38}, noise and outliers, features encoding and transformation\cite{Jordanov25,Sun38,Matuszewski39}, and data scaling. we will introduce the denoising, deinterleaving, and features transformation.

 Because of complex electromagnetic environment, amount of interleaving radio signals are hard to classify and recognize directly in short time, so deinterleaving is the first step. Multi-parameter cluster deinterleaving methods are usually adopted for deinterleaving the pulse streams (such as pulse repetition interval (PRI) deinterleaving methods\cite{PRI_deinterleaving}, time of arrival (TOA) deinterleaving methods\cite{TOA_deinterleaving}). Some novel methods have emerged based on ML algorithms in recent years. Parameter clustering technology was proposed to deinterleave the receptive radar pulses based on Hopfield-Kamgar\cite{Kamgar22} and  Fuzzy ART neural network\cite{Ataa23}. To solve the deinterleaving problems of pulse streams, a group of forward/backward prediction RNNs was established in\cite{Liu24} to understand the current context of the pulses and predict features of upcoming pulses. The cluster and SVM classifier were employed to interleave mixed signals with similar pulse parameters in \cite{deinterleaving_cluster_SVM}. In\cite{MLP_deinterleaving}, MLP structure was used to deinterleave the radar pulse train. As for denoising aspects, RNNs was used for denoising the pulse train in\cite{Liu24}. AEs are also used to address pulse denoising problem by extracting features from TOA sequences\cite{AEs_denoising}.

 As for features transformation, the one-dimension and two-dimension features are usually the inputs of DNN models. The former are encoded IQ time sequences\cite{Sun67,Zhang72, Chongqing82,radar_emitter_recognition_IQ}, and the latter usually are time-frequency distribution (TFD) images, which are produced by short time fourier transformation (STFT)\cite{Chongqing69}, Choi-Williams time-frequency distribution (CWTFD)\cite{Ming71,Zhang72}, and Cohen's time-frequency distribution (CTFD) image\cite{Qu75,Wang76}. In addition, there are some other two-dimension feature images, such as amplitude-phase shift image\cite{Selim68}, the spectrogram of the time domain waveform based on STFT\cite{Kong78}, bispectrum of signals\cite{Li86}, ambiguity function images\cite{automatic_modulation_classification_AF,radar_waveform_classification_AF}, and autocorrelation function (ACF) features\cite{PRI_modulation_recognition,ACSE_PRI_autocorrelation,multi_branch_ACSE_radar_signal_recognition}.

  \vspace{-1.0mm}
\subsection{Traditional Machine Learning in RRSCR}\label{S3.2}

 Traditional ML algorithms based in RRSCR usually includes features selection, classifier design, classifier training and evaluation. Two-phase method of feature extraction and classification based on common machine learning algorithm, is a typical pattern in RRSCR reported in many literatures. There are many classifier models applied to RRSCR, such as supervised learning methods: ANN\cite{ANN}, SVMs\cite{SVM,kernal_SVM}, decision DT\cite{DT}, RF\cite{RFs}, as shown in Table.I.

 The accuracy rate of the traditional two-step methods is mainly determined by the feature extraction algorithm. Artificial feature extraction regarding specific types of radar signals, however, depend mostly on the experience of the experts. Compared to two-step method, DL-based methods can develop feature extraction automatically and potentially learn the latent features of data, so it has higher accuracy. Challenges on generalization, big dataset, and optimal training algorithm, however, are main problems for DL-based methods.

 Feature extraction is used to extract signal features from the preprocessed data for classification model training and recognition\cite{Matuszewski39}. These features include pulse description words (PDWs) of radar signal\cite{Jordanov25,Shieh30,Granger31,Matuszewski34}, information entropy theory\cite{Li21}, high order spectra\cite{Anjaneyulu26}, wavelet packets\cite{Azimisadjadi27}, dynamic parameters searching\cite{Yin28}, rough sets\cite{Zhang29}, acousto-optic spectrums\cite{waveform32}, energy envelope\cite{Rehman36}, time-frequency analysis\cite{Zeng40,Zeng41,Mingqiu42,Konopko43}, autocorrelative functions\cite{Rigling44,Wang45}, principal component analysis (PCA) method\cite{Yu46}. Parameters of signal, however, is time-variable, which can lead to uncertainty of signal. Vector neural network was reported in \cite{Liu33} to deal with the uncertainty of parameters.

\textbf{1) SVMs classifiers}.

 With the typical advantage of efficiently using kernel function to deal with non-linear binary classification, SVMs are mainstream of ML methods applied to RRSCR\cite{Jordanov25,zhang47,Yuan48,Xu49}, which maximizes the distance or margin between the support vectors of classes to search for an optimal hyperplane in feature space of samples.

 In\cite{Jordanov25}, SVM was used in radar signal classification and source identification based on the PDWs of radar, including continuous, discrete and grouped radar data signal train pulse sources. To simplify SVM structure and improve recognition accuracy, SVM with binary tree architecture was proposed in\cite{zhang47}, a roughly pre-classification method was used before SVM with resemblance coefficient classifier. Transient energy trajectory-based SVM method was proposed in \cite{Yuan48} for specific emitter identification with robustness to Gaussian-noise, which used PCA to deduce dimensions of features space. To address the non-linear classification, there are lots of researches on kernel-SVM in RRSCR with different kernel functions. However, optimal kernel function is basically relative to excellent performances in stability and accuracy. In\cite{Xu49}, the authors developed the comprehensive estimation method for choosing optimal kernel functions of SVM for radar signal classifier, which used separability, stability and parameter numbers as evaluation indexes.

 To identify the radar emitter sources with high accuracy rate at low SNR scenario, a SVM classifier based on the scale-invariant feature transform (SIFT) in position and scale features was employed in\cite{Liu51}. The SIFT scale and position features of the time-frequency image were extracted based on the Gaussian difference pyramid. The extracted noise feature points were suppressed based on the scale features. Finally, SVM was used for the automatic identification of radiation sources based on the SIFT position features.

 However, SVM classifier does not good at learning new knowledge in real-time. Hull vector and Parzen window density estimation\cite{Zhu50} were reported for online learning of radar emitter recognition.

\textbf{2) Artificial Neural Networks (ANNs) classifiers}.

 This part will review ANNs-based methods popularly applied to RRSCR, only considering superficial layer NNs, which have not more than 3 hidden layers, including vector neural network\cite{Shieh52,Liu53}, SPDS-neural network\cite{Yin54}, radial basis function neural network (RBF-NN)\cite{Zhang55}, and fusion neural network\cite{Granger56}. The DNNs-based related works will be surveyed in later section.

 To guarantee the accuracy rate of approximately 100\% in exacting one-dimensional parameter, a modified back propagation (BP) neural network was proposed in\cite{Yin54} for radar emitter recognition with uncomplicated data and enough train time. RBF-NN was developed in\cite{Zhang55} to classify radar emitter signals. The decision rules of RBF-NN, to determine signal types, are extracted from rough sets and the cluster center of RBF-NN by rough K-means cluster method. A what-and-where neural network architecture was developed for recognizing and tracking multiple radar emitters in\cite{Granger56}.

 The multi-layer perceptron (MLP) achieved more than 99\% recognition rate at SNR ${\ge}$ 0 dB, only six features were selected by genetic algorithm (GA) algorithm\cite{Wong57}. One hidden layer with a full connected layer as classifier model for classification of 11 classes radar signals was developed in\cite{Jordanov25}. Neural network classifier based on three types of entropy features to achieve 100\% recognition rate at high SNR, and 84\% at -5dB for 4-class signals in\cite{Li21}. An ANN was trained to detect and identify the low probability intercept (LPI) radar signal whose type was unknown at the received SNR of -3 dB in\cite{Anjaneyulu26}. Different from one-hidden layer feedforward neural network topologies were developed in\cite{Petrov37} to classify 2-class and 11-class civil and military radar emitters, and achieved accuracy rate of 82\%, 84\%, and 67\% for civil, military, and other classes, respectively.

 In addition, other classification learning models were also researched in RRSCR, such as DT\cite{Matuszewski58}, RF\cite{Jordanov25}, weighted-XGBoost\cite{Chen35}, Hidden Markov Models (HMMs)\cite{Kim60}, Adaboost\cite{radar_signal_recognition_adaboost}, clustering\cite{Li21,radar_emitter_recognition_clustering,radar_emitter_recognition_online_clustering,non_cooperative_radar_emitter_recognition_online_clustering,radar_emitter_identification_K-means}, K-NN\cite{RF_fingerprint_KNN,radar_signal_recognition_KNN,radar_emitter_recognition_KNN,radar_emitter_identification_KNN}.

 The SVM classifier is suitable for binary classification, whose labels of classes are continuous distributed and have the strict boundary. However, when the label of class is nonlinear, the SVM do not function. It is convenient to split the feature space into unique subsets, containing similar radar functions and parameters, to reduce recognition time and simplify classifier design. For this type of classification task, the DT classifier is working. In \cite{Matuszewski58}, a DT model was developed for a classification system, containing 15,000 signal vectors from 125 radars about different applications. To address the high error rate of single DT, a RF model was employed in \cite{Jordanov25} to recognize radar signals with better performance, compared to NN and SVM classifiers.

 Relevant vector machine (RVM) model-based methods has been also applied to radar emitters classification\cite{Yang61,Yang62}. A hybrid method of rough k-Means classifier combining with three-dimensional distribution feature was proposed in \cite{Yang61}. The robust RVM was developed in \cite{Yang62}. In addition, gradient boost\cite{gradient_boosting,Pavlyshenko64} was used as classification model, K-NN as a classifier to classify instantaneous transient signals based on RF fingerprint extraction\cite{Rehman36}. As for unknown radar signals recognition, class probability output network (CPON) was proposed in \cite{Kim65} for classification of trained and untrained radar emitters signal types.

 When classifing intercepted radar signals, there exists a data deviation in practical application. Weighted-xgboost algorithm was applied to \cite{Chen35} to address this problem. Compared to existing methods, this novel method achieved 98.3\% accuracy rate, while the SVM, RVM\cite{Yang62}, the gradient boost methods, and DBN obtained 89.1\%, 79.6\%, 91.9\%, 95.4\%, respectively.

 In\cite{Guo66}, AdaBoost algorithm was developed as a classifier based on fast correlation-based filter solution (FCBF) feature selection, to complete the recognition of different types of radar signals with 1D harmonic amplitude datasets. These datasets were decomposed by frequency-domain analysis from radar time-domain signals. The simulation results showed that this method was more effective than the SVM algorithm in accuracy rate and stability.

    \vspace{-1.0mm}
\subsection{Deep learning in RRSCR}\label{S3.3}

 Compared to statistics-based analysis methods, traditional ML-based have developed many achievements in RRSCR introduced in section B, which can improve the classification and recognition performance dramatically. However, the weakness of standard 2-phase-method is hard to further extract latent features by domain experts to facilitate classification model training, because of the limitation of expert knowledge and lots of time costs in general.

 Nowadays with the advantages of deeply automatic feature extraction, radar experts exploit apply DL in RRSCR to improve the classification performance based on DNN models. In general, the 1D and 2D features are as the inputs of DNN models aforementioned. Since the CNNs have excellent performance and have been applied widely to image classification and recognition. In this section, we mainly make a comprehensive survey on radar signals classification based on CNNs architecture. In addition, RNNs\cite{Liu70}, DBNs\cite{Liu84,Cao85}, and AEs\cite{Wang76} are also briefly investigated.

 A novel unidimensional convolutional neural network (U-CNN) was proposed in \cite{Sun67} to classify radar emitters, which is based on encoded high dimension sequences as extracted features. The U-CNN has three independent convolution parts followed by a fully-connected part. Three encoded sequences:{${RF_i}$}, {${PRI_i}$}, {${PW_i}$} act as inputs of the corresponding convolution parts. Experiments on a large radar emitter classification (REC) dataset, including 67 types of radars and 227,843 samples, demonstrated that U-CNN can achieve the highest accuracy rate and competitive computation cost for classification, compared with other classifier models, such as NN, SVM, DT.

 A CNN model with five convolution-maxpooling layers, two fully connection layers, and one softmax output layer, was proposed in\cite{Selim68} to classify radar bands from mixed radar signals. Experiments results showed that amplitude-phase shift property as inputs of CNN achieved 99.6\% of accuracy rate, compared to that of  98.6\%. when spectrograms as inputs. Sequeeze-and-excitation network (SENet) was proposed in\cite{Chongqing69} to identify five kinds of radar signals, each of which has 4,000 training samples. This novel model achieved accuracy rate of 99\% with time, frequency, and TFD images as the inputs. Combining with autocorrelation functions, SENet was also used in \cite{PRI_modulation_recognition} to recognize PRI modulations. Moreover, in \cite{ACSE_PRI_autocorrelation}, asymmetric convolutional squeeze-and-excitation network (ACSENet) and autocorrelation features were proposed for PRI modulations. Also, in \cite{multi_branch_ACSE_radar_signal_recognition}, multi-branch ACSENet and multi-dimension features based on SVM fusion strategy were developed for multiple radar signal modulation recognition. Similarly, a CNN model was employed in\cite{Wang77} based on multiple zero-means scaling denoised TFD images of radar emitter intra-pulse modulated signals.

 A cognitive CNN model was proposed in \cite{Ming71} to recognize 8 kinds of radar waveforms based on CWTFD-based TFD images. More than probability of successful recognition (PSR) of 90\% was achieved when the SNR was -2 dB. To improve the accuracy rate, an automatic radar waveform recognition system was exploited in\cite{Zhang72} to detect, track and locate the LPI radars. This novel system achieved overall PSR of 94.5\% at an SNR of -2 dB by a hybrid classifier. The model includes two relatively independent subsidiary networks, mainly CNN and Elman neural network (ENN) as auxiliary.

 In\cite{Time_frequency73}, the authors proposed a deep CNN based automatic detection algorithm for recognizing radar emitter signals, which leveraged on the structure estimation power of deep CNN and the CWTFD-based TFD images as inputs of model. This architecture had competitive performance compared with BP and SVM models. Combining CNN model with the new kernel function, CTFD as the inputs of model for identifying 12 kinds of modulation signals to achieve more than PSR of 96.1\% at the SNR of -6 dB\cite{Qu75}.

 To make full use of the features of inputs, a feature fusion strategy based on CNN architecture was proposed in\cite{Intra_Pulse74} to classify intra-pulse modulation of radar signals with fused frequency and phase features. Two independent CNNs learned frequency and phase related inputs respectively, and then followed by feature fusion layer to fuse the individual outputs as ultimate output. Similarly, two different neural networks were developed in \cite{Kong78} with spectrogram of the time domain waveform by STFT for radar emitter recognition.

 In order to accelerate feature learning of CNN, a PCA based CNN architecture was proposed in\cite{Ye79} to reduce dimensionality of TFD images. After feature extraction with CNN, random vector functional link (RVFL) was employed in\cite{Zhou81} to promote feature learning ability, and picked out the maximum of RVFL as identification results of signals.

 In general, TFD images remove noise by preprocessing process before them are as inputs of CNN, such as binarization and wiener filtering\cite{Intra_Pulse74,Qu75,Wang77}. Although this preprocessing pattern can reduce the impact of noise, it may cause a loss of information details contained in images to some extent. To address this problem, an end-to-end DL method was developed in\cite{Wang76} to recognize 12 classes of intra-pulse modulation signals based on convolutional denoising autoencoder (CDAE) and deep CNN with CTFD-based TFD images. CDAE was used to denoise and repair TFIs, and Inception\cite{GoogleNet_v1} based deep CNN was used for identification. The simulations showed that the proposed approach had good noise immunity and generalization and achieved PSR of more than 95\% at SNR of -9 dB for twelve kinds of modulation signals classification. An end-to-end RNN architecture was proposed in\cite{Liu70} for classification, denoising and deinterleaving of pulse streams. This structure used RNNs to extract long term patterns from previous collected streams by supervised learning and understand the current contexts of pulses to predict features of upcoming pulses.

 Pulse repetition interval (PRI) is a vital feature parameter of radar emitter signals. It is possible to recognize radar emitter only based on PRI of signals. Due to the high ratio of lost and spurious pulses in modern complex electromagnetic environments, however, PRI modulations are more difficult to separate and recognize. To address this issue, A CNN model was proposed in\cite{Li86} to recognize the PRIs modulations of radar signals. Simulation results showed that the recognition accuracy is 96.1\% with 50\% lost pulses and 20\% spurious pulses in simulation scenario.

 A more efficient threat library was generated in\cite{Jeong83} for radar signal classification based on DBN model, consisted of independent RBMs of frequency, pulse repetition interval, pulse width respectively, and a RBM fused the pervious results again. The experiments results showed more than 6\% performance improvement over the existing system. To accurately address the complex electromagnetic environment and various signal styles, a robust novel method based on the energy accumulation of STFT and reinforced DBN was developed in\cite{Liu84} to recognize radar emitter intra-pulse signals at a low SNR. Deep network based hierarchical extreme learning machine (H-ELM) was explored in\cite{Cao85} for radar emitter signal representation and classification with high order spectrum. After extracting the bispectrum of radar signals, the SAE in H-ELM was employed for feature learning and classification.

 Although DL has high accuracy and generalization for RRSCR, its black-box property makes it difficult to apply in practical applications, such as military and medical applications. To alleviate this issue, a novel method was presented in\cite{Zhang80} based on tree-based pipeline optimization tool (TPOT) and local interpretable model-agnostic explanations (LIME). The experimental results showed that the proposed method can not only efficiently optimize the ML pipeline for different datasets, but determine the types of indistinguishable radar signals in the dataset according to the interpretability.

 In summary, this subsection has done a comprehensive survey on the RRSCR based on ML algorithms, including the classification and recognition of radar signal modulations, LPI waveform, and radar emitters. The ML algorithms include traditional ML and DL, such as SVM, DT, adaboost, CNN, RNN, AE, DBN. The features include statistic, 1D, 2D, and fusion features.

\section{Radar Images Processing}\label{S4}

 Active radar imaging is an important tool for detection and recognition of targets as well for the analysis of natural and man-made scenes. Radar images in a broader sense include unidimensional high-resolution range profiles (HRRP) \cite{Li247,Yuan248,Li250,Li261}, two-dimensional SAR and ISAR images \cite{Chierchia122,Zhao123,Wang124,Mukherjee128}, micro-doppler images \cite{Patel288,Ritchie289,Rahman290,Thayaparan292,Shi291} and range-doppler images \cite{Jokanovi293,Wang294,Li295}. Several ML-based techniques have been developed for radar image processing, particularly for what concerns Synthetic Aperture Radar (SAR) and Inverse Synthetic Aperture Radar (ISAR). This section will review the scientific literatures that focus on radar image processing based on ML technology, including image preprocessing (e.g., denoising), feature extraction and classification.

   \vspace{-1.0mm}
\subsection{SAR Images Processing}\label{S4.1}

 Operating conditions of all weather, day-and-night and high-resolution imaging, synthetic aperture radar (SAR) is a popular research domain on remote sensing domain in military and civil applications. SAR is an active remote sensor, i.e., it carries its own illumination and does not depend on sunlight like optical imaging. With the rapid development of military and science technology, various types of SAR sensors have been appeared, which can be roughly divided into three main categories based on the carrier platform: satellite-borne SAR, airborne SAR, and ground-based SAR. Different SAR sensors can have different configured properties, even though within the same category, such as carrier frequency/wavelength, imaging mode (e.g., stripmap SAR, spotlight SAR, scanSAR, inverse SAR, bistatic SAR and interferometric SAR (InSAR)), polarization (e.g., horizontal (H) or vertical (V) polarization), resolutions in range and azimuth directions, antenna dimensions, synthetic aperture, and focusing algorithm (e.g., range doppler algorithm, chirp scaling algorithm, and the SPECAN algorithm).

 Focused SAR image is the 2D high resolution image, i.e., range and azimuth directions. At range direction, SAR transmits LFM waveform with huge product of pulse width and bandwidth, and obtains high resolution of the range direction by adopting pulse compression technology; as for azimuth, a long synthetic aperture, formed along the trajectory of relative motion between detected target and radar platform, to store the magnitude and phase of successive radar echoes to guarantee the high resolution at azimuth direction. Therefore, one of the vital conditions of forming SAR image is that there should exist relative motion between target and radar platform.

 The multiple configurations of SAR potentially characterize the distinctiveness of SAR imagery, which vastly contributes to classification and recognition of targets. Compared to optical counterparts, SAR images have distinctive characteristics including i) an invariant target size with the various distance between the SAR sensor and the target, ii) the imaging scene information is determined by the magnitude and phase of the radar backscatter (i.e., for a single-channel SAR and multi-channel SAR), iii) high sensitivity to the changes of target's postures and configurations such as the shadowing effect, the interaction of the target's backscatter with the environment (e.g., clutter, adjacent targets, etc.), projection of the 3-D scene (including the target) onto a slant plane (i.e., SAR's line of sight (LOS)), and the multiplicative noise (known as speckle) due to the constructive and destructive interference of the coherent returns scattered by small reflectors within each resolution cell \cite{Wang88}, and iv) SAR imagery can easily observe the hidden targets with the well penetration of suitable wavelength of electromagnetic wave.

 SAR imagery processing methodologies includes denoising, classification and recognition, detection and segmentation. In recent years, with the rapid development of ML in image processing, ML-based,  especially DL, has applied to SAR image processing widely and successfully (such as\cite{Cui101,Ma102,Chierchia122,Zhao192,Wang147,Zhang243}). In this section, we make a comprehensive survey for SAR image processing techniques based on DL algorithms, such as CNN, DBN, SAE.

\textbf{1) Datasets and Augmentation}.

 \emph{\textbf{Datasets}} Dataset is one of the important factors for the success of DL, including training datasets, validation datasets, and testing datasets, respectively. The collection of data and the building of formatted datasets are challengeable tasks, generally requiring huge human and economic costs. Since especially military backgrounds, the public big SAR datasets are not easily collected, compared with general CV datasets, such as ImageNet, COCO, CFAR-10, which depends on big data easily collected from the Internet. Luckily, with the cooperation and endeavor of radar community, there are still some public SAR datasets in military and civil application for target classification and recognition, detection, and segmentation. These targets include military vehicles, farmland, urban streets, and ships. Such as moving and stationary target acquisition and recognition (MSTAR)\cite{Keydel87,Wang88,MSTAR95} , TerraSAR-X high resolution imagery\cite{Bai148,Xing96}, San Francisco \cite{Zhou90}, Flevoland\cite{Zhou90,Feng91,Zhang92,Sterling93,html94}. Ship datasets includes SSDD\cite{Li97}, SAR-Ship-Dataset\cite{Wang98}, AIR-SARShip-1.0\cite{Sun99}, HRSID\cite{HRSID_ship_detection_dataset}.

 MSTAR is a typical widely applied as a baseline SAR imagery dataset, including 10 classes of ground targets. The dataset consists of X-band SAR images with 0.3 m * 0.3 m resolution of multiple targets, which includes BMP2 (infantry combat vehicle), BTR70 (armored personnel carrier), T72 (main tank), etc. All images are size of 128 * 128. The samples are as shown in Fig. 18.

\begin{figure}[h] 
\vspace{-2.0mm}
\begin{center}
\includegraphics[width=1.0\columnwidth, keepaspectratio]{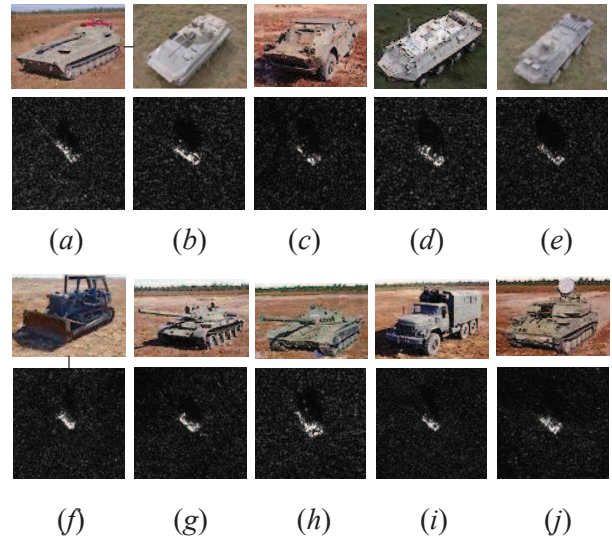}
\end{center}
\captionsetup{font = {footnotesize}, name = {Fig.}, singlelinecheck = off, justification = raggedright, labelsep = period}\
\vspace{-6.0mm}
\caption{The MSTAR data samples, optical images (top) and their corresponding SAR images (bottom). (a)2S1, (b)BMP2, (c)BRDM2, (d)BTR60, (e)BTR70, (f)D7, (g)T62, (h)T72, (i)ZIL131, (j)ZSU234.}
\label{FIG3}
\vspace{-3.0mm}
\end{figure}

 SSDD dataset\cite{Li97} includes 1,160 images and 2,456 ships totally, which follows a similar construction procedure as PASCAL VOC \cite{Everingham110}. SAR-Ship-Dataset\cite{Wang98} constructed with 102 Chinese Gaofen-3 images and 108 Sentinel-1 images. It consists of 43,819 ship chips of 256 pixels in both range and azimuth directions. These ships mainly have distinct scales and backgrounds. It can be used to develop object detectors for multi-scale and small object detection.

 AIR-SARShip-1.0\cite{Sun99} firstly released 31 images, scale of 3,000 * 3,000 pixels, the resolution of SAR images is 1 m and 3 m, imaging pattern including spotlight mode, stripemap mode, and single polarization mode. The landscapes including port, island, the sea surface with different sea conditions. The targets have almost thousands of ships with ten classes, including transport ship, oil ship, fisher boat, and so on.

 High resolution SAR images dataset (HRSID)\cite{HRSID_ship_detection_dataset} is used for ship detection, semantic segmentation, and instance segmentation tasks in high-resolution SAR images. This dataset contains a total of 5,604 high-resolution SAR images and 16,951 ship instances. ISSID draws on the construction process of the Microsoft common objects in context (COCO) dataset, including SAR images with different resolutions, polarizations, sea conditions, sea areas, and coastal ports. This is a benchmark dataset for researchers to evaluate their approaches. The resolution of ISSID is 0.5 m, 1 m, and 3 m.

 \emph{\textbf{Data augmentation}} Although there exist some public available datasets, the number of labeled samples is relatively small, which do not always satisfy the requirements of DL algorithm. Therefore, the SAR targets recognition and classification can be regarded as small samples recognition problem. To address the deficiency samples of datasets, many researchers have proposed novel methods to augment the dataset, such as GANs\cite{Cui101,Ma102}, or design novel efficient model to learn with limited labeled data, such as TL based methods\cite{Ma102,Wang103,Huang106}.

 Wasserstein GAN, with a gradient penalty (WGAN-GP), was proposed to generate new samples based on existing MSTAR data in\cite{Cui101}, which can improve the recognition rate from 79\% to 91.6\%, from 57.48\% to 79.59\%, for three-class and ten-class recognition problem, respectively, compared to original MSTAR. In \cite{Ma102}, the authors proposed least squares generative adversarial networks (LSGANs) combined with TL for data augmentation. Different from\cite{Cui101,Ma102}, some image processing methods were utilized in\cite{Wang103} i.e., manual-extracting sub-images, adding noise, filtering, and flipping, to produce new samples based on the original data. In\cite{Yan107}, the authors generate noisy samples at different SNRs, multiresolution representations, and partially occluded images with the original images, to enhance the robustness of CNN at various extended operating conditions (EOCs). In addition, three types of data augmentation based on MSTAR were developed in\cite{Ding108,Ding159}, i.e., translation of target, adding random speckle noise to the samples, and posture synthesis. Image reconstruction with sparse representation was proposed in \cite{Lv104,Junya155,Ding176} for data augmentation based on attributed scattering centers (ASCs).

 A accuracy-translation map based on domain-specific data augmentation method was developed in\cite{Furukawa160}, which can achieve a state-of-the-art classification accuracy rate of 99.6\% on MSTAR dataset. In\cite{Pei174}, the authors used a flexible mean to generate adequate multi-view SAR data with limited raw data. An electromagnetic simulation approach was proposed in\cite{Pei180} as an alternative to generate enough bistatic SAR images for network training. In\cite{Coman200}, amplitude and phase information of SAR image was also used to generate multichannel images as the inputs of CNN model to alleviate the over-fitting during the training phase.

 Except for data augmentation methods, high efficient classification model design is also adopted to alleviate the small samples challenge. In\cite{Yu105}, the authors proposed a new deep feature fusion framework to fuse the feature vectors, which were extracted from different layers of the model based on Gabor features and information of raw SAR images. A TL based method was employed in \cite{Huang106} to transfer knowledge learned from sufficient unlabeled SAR scene images to labeled SAR target data. A-ConvNet was proposed in \cite{Wang109} to vastly reduce the number of free parameters and the degree of overfitting with small datasets. The novel architecture replaced the fully-connected layers with sparsely-connected convolutional layers, which obtained average accuracy of 99.1\% on classification of 10-class targets on MSTAR dataset.

\textbf{2) SAR Images Denoising}.

 As a coherent imaging modality, SAR images are often contaminated by the multiplicative noise known as speckle, which severely degrades the processing and interpretation of SAR images. It is hard to balance  performances between speckle noise reduction and detail preservation. In general, traditional despecking methods\cite{Hemani126,Martino127}, (such as multi-look processing\cite{Oliver112}, filtering \cite{Baraldi113,Frost114,Giampaolo131,Zhou133}, blocking matching 3D (BM3D)\cite{Dabov115,Nicolas129}, wavelet-based\cite{Parrilli116,Achim117,Argenti118,Xie119,Giampaolo131}, separated component-based\cite{Patel120}), transform the multiplicative noise into additive noise by logarithm operation of observed data. These methods, however, can introduce more or less bias into denoised image. In addition, the local processing of these methods fails to preserve sharply useful features, e.g., edges, texture, detailed information, and often contains artifacts\cite{Argenti121}. Another problem is that most of traditional methods require statistics modeling. To address the problems of SAR image despeckling aforementioned, and inspired by the advantages of DL algorithm. DL-based algorithm has been applied to this field, especially CNN-based model algorithm.

 \emph{\textbf{CNN-based supervised methods}} A residual learning strategy with residual CNN model was firstly employed in\cite{Chierchia122} for SAR imagery despeckling, which achieved better performance on man-synthetic and real SAR data and guaranteed a faster convergence in the presence of limited training data, compared to state-of-the-art techniques. A probability transition CNN (PTCNN) was proposed in \cite{Zhao123} to increase noise-robustness and generalization for patch-level SAR image classification with noisy labels. The authors in\cite{Wang124} developed an end-to-end learning architecture (i.e., image despeckling convolutional neural network (ID-CNN)) to automatically remove speckle from noisy SAR images. In particular, this architecture contained a component-wise division-residual layer with skip-connection to estimate the denoised image. Similarly, in\cite{Zhang141}, the authors proposed a SAR dilated residual network (SAR-DRN) to learn a non-linear end-to-end mapping between the noisy and clean SAR images. DRN could both enlarge the receptive field while maintaining the filter size and layer depth with a lightweight structure to conduct image details and reducing the gradient vanishing problem. In addition, combined with ensemble learning method, the authors proposed a despecking CNN architecture in \cite{despeckling_CNN_ensemble}.

 To deal with the random noisy SAR imagery, despeckling and classification coupled CNNs (DCC-CNNs) was proposed in\cite{Wang125}, to classify ground targets in SAR images with strong and varying speckle. DCC-CNNs contained a despeckling subnetwork to firstly mitigate speckle noise, and a classification subnetwork for noise robustness learning of target information, which could achieve more than 82\% of overall classification accuracy rate for ten ground target classes at various speckle noise levels. A novel method to directly train modified U-Net\cite{Ronneberger136} with given speckled images was developed in\cite{Ravani135}. An extra residual connection in each convolution-block of U-Net, and the operations of replacing the transposed convolution with parameter free binary linear interpolation were also introduced. A DNN based approach was proposed in\cite{Foucher143} for speckle filtering, which based on DNN's application in super-resolution reconstruction, iteratively improved the first low resolution filtering results by recovering lost image structures.

 To overcome the problem of collecting a large number of speckle-free SAR images, a CNNs-based Gaussian denoiser was developed in\cite{Yang134}, which was based on multi-channel logarithm and Gaussian denoising (MuLoG) approaches. The TL-based pre-trained CNN models, trained by datasets with additive white Gaussian noise (AWGN), was also directly employed to process SAR speckle.

 Thanks to the excellent ability of exploiting image self-similarity with nonlocal methods, a CNN-powered nonlocal despeckling method was investigated in\cite{Cozzolino137} to improve nonlocal despeckling performance\cite{Vitale138} on man-synthetic and real SAR data. The trained CNN was used to discover useful relationships among target and predictor pixels and were converted into the weights of plain nonlocal means filtering. In\cite{Liu142}, the authors proposed CNN model combined with guided filtering based fusion algorithm for SAR image denoising. Five denoised images were firstly obtained via a seven-layer CNN denoiser acts on an noisy SAR image, then a final denoised image is acquired by integrating five denoised images with a guided filtering-based fusion algorithm.

 However, the DL model remains very sensitive to the inputs. To address the non-invariant denoising capability of DL-based methods, a novel automatical two-component DL network with texture level map (TLM) of images was proposed in \cite{Gu140} to achieve satisfactory denoising results and strong robustness for SAR imagery invariant denoising capability. Texture estimation subnetwork produced the TLM of images. Noise removal subnetwork learned a spatially variable mapping between the noise and clean images with the help of TLM.

 \emph{\textbf{DNN-based unsupervised methods}} Except for CNN-based supervised learning model methods, the DNN-based unsupervised methods are also developed in SAR images denoising. To solve the notorious problem of gradients vanishing and accelerate training convergence, a AE model was employed in\cite{Gu144} to denoise multisource SAR images, which adopted residual learning strategy by skip-connection operation. AE-CNN architecture was developed in\cite{Mukherjee128} for InSAR images denoising in the absence of clean ground truth images. This method can reduce artefact in estimated coherence through intelligent preprocessing of training data. To solve the trade-off of speckle suppression and information preservation, a CNN-based unsupervised learning solution scheme was proposed in \cite{Sergio130,Giampaolo131}. Taking into account of both spatial and statistical properties of noise, this model could suppress the noise while simultaneously preserve spatial information details by a novel cost function. In addition, a MLP model was elaborated in\cite{Tang145} for SAR image despeckling by using a time series of SAR images. The greatest advantage of MLP was that once the despeckling parameters were determined, they can be used to process not only new images in the same area, but also images in completely different locations.

\textbf{3) SAR Automatic Target Recognition (SAR-ATR)}.

 Originated from the military, the goal of ATR is to infer or predict the classes of detected targets via acquired sensory data with computer processing technology. Today, ATR technology is significantly applied to both military and civil domains, such as valuable military target recognition (e.g., missile, airplane, ship), human pose, gait, and action recognition. Due to the unique characteristics of SAR images aforementioned, it is difficult to easily interpret the SAR imagery with the common ATR system. Research on SAR-ATR system has increasingly absorbed attention from the researchers around the RSP community. The problems of SAR-ATR research domain based on ML algorithms mainly focus on improving performances in the following four aspects: i) accuracy with DNNs model, ii) generalization with limited labeled data\cite{Zhao192}, iii) robustness with speckle denoising, scale-variance and adversarial samples attack, and iv) real-time or alleviating computation cost at practical strict situations. Furthermore, interpretability of deep models are also studied. SAR-ATR has two main categories based on ML: traditional ML based methods, such as SVM\cite{Cui152,Zhao163,Georgios164,Zhao165}, genetic programming \cite{Stanhope166}, boosting\cite{Sun167}, Markov random field (MRF)\cite{Perciano168,Weisenseel169,Clausi170,Ruohong171}, ELM\cite{Gu187}, and DNN based methods, such as CNNs\cite{Ding159,Furukawa160}, DBNs\cite{Zhao161}, SAE\cite{Chen162}, RNNs\cite{Zhang175}. Besides, three classes of SAR-ATR methods have been categorized in a surveyed paper\cite{Darymli146}, i.e., feature-based, semi-model-based, and model-based, respectively. This section presents an understanding survey for SAR-ATR based on model-based DL algorithms, which is roughly categorized into four classes based on research aspects aforementioned. These state-of-the-art algorithms including basic CNNs, fusion models, highway model, multi-view, multi-task learning networks models, RNNs based spatial SAR image sequences learning, AEs, and DBNs.

 \emph{\textbf{i) Boosting Accuracy with DL Model}}

 \emph{\textbf{General DL models}} Similarly, CNNs are also widely applied to SAR-ATR. To understand the relationship between the convolution layers and feature extraction capability, a weighted kernel CNN (WKCNN) was presented in\cite{Tian194}. By modeling the interdependence between different kernels, this model integrated a weighted kernel module (WKM) into the common CNN architecture to improve the feature extraction capability of the convolutional layer. The CNN models were designed for MSTAR data\cite{Ievgen173} and polarimetric Flevoland SAR dataset (15 classes)\cite{Zhou149} classification, which achieved recognition accuracy of 99.5\% and 92.46\% respectively. In\cite{Liu203}, the authors proposed a dual channel feature mapping CNN (DCFM-CNN) for SAR-ATR, which achieved a average recognition accuracy of 99.45\% on MSTAR. In order to extract spatial discriminative features of SAR images, a DCNN was proposed to extract gray level-gradient co-occurrence matrix and Gabor features in \cite{scale_transformation_denoising,restrain_noise} for SAR image classification. A novel neighborhood preserved DNN (NPDNN) was proposed in \cite{spatial_relation_PolSAR_classification} to exploit the spatial relation between pixels by a jointly weighting strategy for PolSAR image classification. An convolution kernel of the fire module based effective max-fire CNN model, called MF-SarNet, was constructed in \cite{MF_SARNet_SAR_ATR} for effective SAR-ATR tasks.

 Combing DNN and traditional ML algorithm is also investigated. A unsupervised discriminative learning method based on AE and SVM models was proposed in \cite{patch_sorted_AE_SVM}, called patch-sorted deep neural network (PSDNN), which firstly adopted sorted patches based on patch-sorted strategy to optimize CNN model training for extracting the high-level spatial and structural features of SAR images, and a SVM classifier as the final classification task. The combination of CNN and SVM was developed in \cite{CNN_SVM_SAR_ATR, DCNN_SVM_SAR_ATR}. A modified stacked convolutional denoising auto-encoder (MSCDAE) was proposed in \cite{AEs_SVM_SAR_recognition} to extract hierarchical features for complex SAR target recognition, and SVM as final object classification with features extracted by MSCDAE model. To enhance the learning of target features, a novel deep learning algorithm based on a DCNN trained with an improved cost function, and combined with a SVM was proposed in \cite{DCNN_improved_cost_function_SVM} for SAR image target classification. A TL based pre-trained CNN was employed to extract learned features in combination with a classical SVM for SAR images target classification in \cite{TL_CNN_SVM}. Deep kernel learning method was employed in \cite{deep_kernel_learning_SVM_SAR_recognition} for SAR image target recognition, which optimized layer by layer with the parameters of SVM and a gradient descent algorithm. A novel oil spill identification method was proposed in \cite{oil_spillage_detection_CNN_SVM} based on CNN, PCA, and RBF-SVM, which could improve the accuracy of oil spill detection, reduce the false alarm rate, and effectively distinguish an oil spill from a biogenic slick. To take the advantage of manifold learning with modeling core variables of the target, and separate different data's manifold as much as possible, the authors proposed nonlinear manifold learning integrated with FCN for PolSAR image classification in \cite{nonlinear_manifold_learning_FCN_SAR_classify}.

 In \cite{ensemble_learning_CNN_PolSAR}, the authors proposed an ensemble transfer learning framework  to incorporate manifold polarimetric decompositions into a DCNN to jointly extract the spatial and polarimetric information of PolSAR image for classification. In order to effectively classify single-frequency and multi-frequency PolSAR data, the authors proposed a single-hidden layer optimized Wishart network (OWN) and extended OWN, respectively in \cite{single_layer_efficently_classify}, which outperformed DL-based architecture involving multiple hidden layers. To exploit the spatial information between pixels on PolSAR images and preserve the local structure of data, a new DNN based on sparse filtering and manifold regularization (DSMR) was proposed for feature extraction and classification of PolSAR data in \cite{sparse_manifold_regularized_CNN}. In \cite{DCNN_expert_knowledge_PolSAR_classify}, the authors made full use of existing expert knowledge to construct a novel deep learning architecture for deep polarimetric feature extraction, and a superpixel map was used to integrate contextual information. This model consisted of multiple polarimetric algebra operations, polarimetric target decomposition methods, and CNN to extract deep polarimetric features.

 A 20-layers (with 3 dense block and 2 transition layers) DenseNet was built to implement polarimetric SAR image classification in \cite{Dense_CNN_SAR_recognition}. Inception v3 model was adopted in \cite{Inception_v3_SAR_classiifcation} to develop an efficient and accurate method to detect and classify key geophysical phenomena signature among the whole sentinel-1 wave mode SAR dataset. In \cite{UNet_SAR_classification}, the authors proposed an end-to-end framework for the dense, pixel-wise classification of GF-3 dual-pol SAR imagery with convolutional highway unit network (U-net) to extract hierarchical contextual image features. To concern about the estimation of depression angle and azimuth angle of targets in SAR-ATR tasks, the authors proposed a new CNN architecture with spatial pyramid pooling(SPP) in \cite{spatial_pyramid_pool_CNN_SAR_ATR}, which could build high hierarchy of features map by dividing the convolved feature maps from finer to coarser levels to aggregate local features of SAR images.

 To address the redundant parameters and the negligence of channel-wise information flow, group squeeze excitation sparsely connected CNN was developed in\cite{Huang195}. Group squeeze excitation performed dynamic channel-wise feature recalibration with less parameters, and sparsely connected CNN demonstrated the concatenation of feature maps from different layers. This model achieved accuracy rate of 99.79\% on MSTAR, outperformed the most common skip connection models, such as ResNet and densely connected CNN. TL-based pre-trained ResNet50 and VGGNet model were employed in\cite{Ma153,Wang154} for SAR image classification. Pre-trained models can deeply extract multiscale features of data samples in short training time, and convolutional predictor were added after the pre-trained model for the target classification. The experiment results showed that the pre-trained model achieved accuracy rate of 98.95\% in\cite{Ma153} and higher performance with the suitable data augmentation technology than other methods in\cite{Wang154}. TL was developed in\cite{Shang157} to overcome the problem of difficulty in convergence.

 Instead of directly outputting the class of SAR image with the DNN, a class center metric based method with CNN model was proposed in\cite{Qi182}. This method used CNN to extract features from SAR images to calculate class center of each class under the new features representation. Then, the class of test sample was identified by the minimum distance between the center of class and learned features space of test sample. Similarly, a DNN model was employed in\cite{Kazemi202} to directly classify targets with slow-time and fast-time sampled signals. The decision-making strategy of classification is determined by the distance between the optimized sets of vectors and  classes. Each of class represented a new sample.

 As for complex SAR imagery, the complex-value CNN (CV-CNN) architecture was proposed in \cite{Xu150,Zhang151}. All components of CNNs were extended to the complex domain. CV-CNN achieved accuracy rate of 95\% on Flevoland dataset\cite{Xu150}, and 96\% with enough samples in\cite{Zhang151}. Moreover, a deep FCN was also employed in \cite{FCN_CV_SAR_classification} that used real-valued weight kernels to perform pixel-wise classification of complex-valued images. A CV-CAE was proposed in \cite{CV_CAE_complex_PolSAR_classify} for complex PolSAR images classification.  In order to sufficiently extract physical scattering signatures from PolSAR and explore the potentials of different polarization modes on this task, a contrastive-regulated CNN was proposed in the complex domain, attempting to learn a physically interpretable deep learning model directly from the original backscattered data in \cite{contrastive_regulated_CNN_PolSAR_classify}. A novel deep learning framework, deep SAR-Net, was constructed in \cite{SAR_Net_PolSAR_classification} to take complex-valued SAR images into consideration to learn both spatial texture information and backscattering patterns of objects on the ground.

 To exploit the performance of generative models in SAR-ATR based on unsupervised learning, an SAE model with feature fusion strategy was adopted in\cite{Kang188} for SAR target recognition. The local and global features of 23 baselines and three patch local binary pattern (TPLBP) features were extracted from the SAR image, which achieved an classification accuracy rate of 95.43\% on MSTAR. A single-layer CNN model combined with features extraction by SAE was developed in\cite{Chen162}, which achieved accuracy rate of 90.1\% and 84.7\% for 3-class and 10-class targets classification on MSTAR.  A novel framework for PolSAR classification based on multilayer projective dictionary pair learning (MPDPL) and SAE was proposed in \cite{MPDPL_SAE_PolSAR_classify}. To learn more discriminative features of SAR images, an ensemble learning based discriminant DBN (DisDBN) was proposed in\cite{Zhao161} to learn high-level discriminant features of SAR images for classification. Some weak classifiers were trained by several subsets of SAR image patches to generate the projection vectors, which were then input into DBN to learn discriminative features for classification. An unsupervised deep generative network-poisson gamma belief network (PGBN) was proposed to extract multi-layer feature from SAR images data for targets classification tasks in \cite{deep_Bayesian_GNN_SAR_recognition}. An unsupervised PolSAR image classification method using deep embedding network-SAEs was built in \cite{unsupervised_SAEs_clustering_SAR_classification}, which used SVD method to obtain low-dimensional manifold features as the inputs of SAEs, and the clustering algorithm determined the final unsupervised classification results. As for In-SAR data, a DBN was used to model data in \cite{DBN_InSAR_classification} for classification, which could fully explore the correlation between intensity and the coherence map in space and time domain, and extract its effective features. Inspired by DL and probability mixture models, a generalized gamma deep belief network (g-DBN) was proposed for SAR image statistical modeling and land-cover classification in \cite{gB_DBN_mixed_model_SAR_classify}. Firstly, a generalized Gamma-Bernoulli RBM (gB-RBM) was developed to capture high-order statistical characterizes from SAR images. Then a g-DBN was constructed to learn high-level representation of different SAR land-covers. Finally, a discriminative network was used to classify. In addition, the deep RNN based model was adopted in \cite{deep_RNN_agricultural_classify} for agricultural classification using multitemporal SAR Sentinel-1.

 \emph{\textbf{Multi-aspect fused learning methods}} Multi-aspect fused learning methods are very popular in improving the accuracy of SAR-ATR tasks such as multi-view, multi-task, multi-scale, multi-dimension. To fully extract features of images, a CNN based fusion framework was proposed in\cite{Yu156}, including  a preprocessing module initialized with Gabor filters, an improved CNN and a feature fusion module. This model could achieve an average accuracy rate of 99\% on MSTAR, even obtained a high recognition accuracy on limited data and noisy data. The authors developed concurrent and hierarchy target learning architecture in\cite{Touafria193}. Three CNN models simultaneously extracted features of SAR images in two scenarios, final classification was finished by different combination and fusion approaches based on extracted features. Based on multi-view learning manner, the authors proposed a multi-input DCNN for bistatic SAR-ATR system in\cite{Pei180}. A multi-stream CNN (MS-CNN) was proposed in\cite{Zhao192} for SAR-ATR by leveraging SAR images from multiple views. A Fourier feature fusion framework derived from kernel approximation based on random Fourier features, to unravel the highly nonlinear relationship between images and classes, which fused information from multi-view of the same target in different aspects.

 To capture the spatial and spectral information of a SAR target simultaneously with kernel sparse representation (KSR) technology, multi-task kernel sparse representation framework was developed in\cite{Ning186} for SAR target classification. SAR target recognition was formulated as a joint covariate selection problem across a group of related tasks. A multi-task weight optimization scheme was developed to compensate for the heterogeneity of the multi-scale features and enhance the recognition performance. A two-stage multi-task learning representation method was also proposed in\cite{Zhang189}. After finding an effective subset of training samples and constructing a new dictionary by multi-feature joint sparse representation learning as the first stage, the authors utilized multi-task collaborative representation to perform target images classification based on the new dictionary in second stage. A multi-level deep features-based multi-task learning algorithm was developed in\cite{Lv196} for SAR-ATR. This architecture employed joint sparse representation as the basic classifier and achieved an recognition rate of 99.38\% on MSTAR under standard operating conditions (SOCs).

 A mixed framework based on multimodal, multidiscipline, and data fusion strategy was proposed in\cite{Odysseas201} for SAR-ATR. An adaptive elastic net optimization method was applied to balance the advantages of ${l_1-norm}$ and ${l_2-norm}$ optimization on scene SAR imagery by a clustered AlexNet with sparse coding. The clustered AlexNet with a multiclass SVM classification scheme was proposed to bridge the visual-SAR modality gap. This framework achieved 99.33\% and 99.86\% for the three and ten-class problems on MSTAR, respectively. A SAR and infrared (IR) sensors based multistage fusion stream strategy with dissimilarity regularization using CNN architecture was developed in \cite{multistage_fusion_SAR_IR_recognition} to improve the performance of SAR target recognition. In order to make full use of phase information of PolSAR images and extract more robust discriminative features with multidirection, multiscale, and multiresolution properties, a complex Contourlet CNN was proposed in \cite{complex_contourlet_CNN_PolSAR_classify}.

 However, most of DL based SAR-ATR methods present a limitation that each learning process only handles static scattering information with prepared SAR image, while missing the space-varying information. To involve space-varying scattering information to improve the accuracy rate of recognition, a novel multi-aspect-aware method was proposed in\cite{Zhang175} to learn space-varying scattering information through the bidirectional LSTM model. The Gabor filter and three-patch local binary patterns were progressively implemented to extract comprehensive spatial features of multi-aspect space-varying image sequences. After dimensionality reduction with MLP, a bidirectional LSTM learned the multi-aspect features to achieve target recognition. This method achieved accuracy rate of 99.9\% on MSATR data.

 To fully exploit the characteristics of continuous SAR imaging instead of utilizing single image for recognition, a bidirectional convolution-recurrent network (BCRN) was developed in\cite{Bai183} for SAR image sequence classification. Spatial features of each image were extracted through DCNNs without the fully connected layer, and then sequence features were learned by bidirectional LSTM networks to obtain the classification results. In order to exploit the spatial and temporal features contained in the SAR image sequence simultaneously, a spatial-temporal ensemble convolutional network (STEC-Net) was proposed for a sequence SAR target classification in \cite{sparse_temporal_ensemble_CNN}, which achieved a higher accuracy rate (99.93\%) in the MSTAR dataset and exhibited robustness to depression angle, configuration, and version variants. A SAR sequence image target recognition network based on two-dimensional (2D) temporal convolution was proposed in \cite{2D_temporal_convolution_SAR_recognition}, including three stages: feature extraction, sequence modeling and classification. To using rotation information of PolSAR image for improving classification performance, the authors built a convolutional LSTM (ConvLSTM) along a sequence of polarization coherent matrices in rotation domain for PolSAR image classification in \cite{convolutional_LSTM_rotation_information_SAR_recognition}.

 In order to automatically and precisely extract water and shadow areas in SAR imagery, the authors proposed multi-resolution dense encoder and decoder (MRDED) network framework in \cite{water_shadow_classify_dense_network}, which integrated CNN, ResNet, DenseNet, global convolutional network (GCN), and ConvLSTM. MRDED outperformed by reaching 80.12\% in pixel accuracy (PA) and 73.88\% in intersection of union (IoU) for water, 88\% in PA and 77.11\% in IoU for shadow, and 95.16\% in PA and 90.49\% in IoU for background classification, respectively. A feature recalibration network with multi-scale spatial features (FRN-MSF) was built in \cite{multi_spatial_features_SAR_scene_classify}, which achieved high accuracy in SAR-based scene classification. FRN was used to learn multi-scale high-level spatial features of SAR images, which integrated the depthwise separable convolution (DSC), SE-Net block and CNN.

 In order to make full use of pose angle information and intensity information of SAR data for boosting target recognition performance, a CNN-based SAR target recognition network with pose angle marginalization learning, called SPAM-Net was proposed in \cite{pose_angle_information_CNN_SAR_recognition} that marginalized the conditional probabilities of SAR targets over their pose angles to precisely estimate the true class probabilities. A combination of multi-source and multi-temporal remote sensing imagery was proposed in \cite{crop_classify_SAR_optical_CNN} for crops classification, which used CNN and visual geometry group (VGG) to classify crops based on the different numbers of input bands composed by optical and SAR data. Deep bimodal autoencoders were proposed for classification of fusing SAR and multispectral images in \cite{AEs_multi_modal_fusing_SAR_multispectral}, which was trained to discover both independencies of each modality and correlations across the modalities. Combining polarimetric information and spatial information, a dual-branch DCNN (dual-DCNN) was proposed to realize the classification of PolSAR images in \cite{spatial_polarimetric_information_fusing_PolSAR_classify}. The first branch was used to extract the polarization features from the 6-channel real matrix, which are derived from the complex coherency matrix. The other was utilized to extract the spatial features of a Pauli RGB (Red Green Blue) image. These extracted features were first combined into a fully connected layer sharing the polarization and spatial property. Then, the softmax classifier was employed to classify these features.

 \emph{\textbf{ii) Enhancing Generalization with Limited Labeled Data}}

 \emph{\textbf{Data augmentation based technology}} To eliminate the overfitting of small dataset in SAR-ATR, a CNN model with feature extractor and softmax classifier, combining with data augmentation technique, was proposed in\cite{Ding159}. An improved DQN method for PolSAR image classification was proposed in \cite{DQN_PolSAR_image_classification}, which could generate amounts of valid data by interacting with the agent using the ${\varepsilon}$-greedy strategy. A multi-view DL framework was proposed in\cite{Pei174} for SAR-ATR with limited data, which introduced a unique parallel deep CNN topology to generate multi-view data as inputs of model. The distinct multi-view features were fused in different layers progressively. A Gabor-DCNN was proposed to overcome the overfitting problem due to limited data in \cite{Gabor_DCNN_SAR_recognition}. Multi-scale and multi-direction-based Gabor features and a DCNN model were used for data augmentation and for SAR image target recognition, respectively. A novel adversarial AE was proposed to improve the orientation generalization ability for SAR-ATR tasks in \cite{adversarial_AEs_generalization_SAR_classify}, which learned a code-image-code cyclic network by adversarial training for the purpose of generating new samples at different azimuth angles. A new dual-channel CNN was developed in \cite{dual_CNN_PolSAR_classify} for PolSAR image classification when labeled samples were small, which firstly used a neighborhood minimum spanning tree to enlarge the labeled sample set and then extracted spatial features by DC-CNN model.

 A DNN-based semi-supervised method was proposed in \cite{class_probability_vector_PolSAR_classify} to tackle the PolSAR image classification when labeled samples was limited. The class probability vectors were used to evaluate the unlabeled samples to construct an augmented training dataset. The feature augmentation and ensemble learning strategies were proposed in \cite{adaboost_rotation_forest_SAR_limited_data} to address the limited samples issue in SAR-ATR tasks. The cascaded features from optimally selected convolutional layers were concatenated to provide more comprehensive representation for the recognition. The adaboost rotation forest was introduced to replace the original softmax layer to realize a more accurate limited sample-based recognition task with cascaded features. In \cite{superpixel_classification_PolSAR}, a superpixel restrained DNN-based multiple decisions strategy, including nonlocal decision and local decision, was developed to select credible testing samples. The final classification map was determined by the deep network, which was updated by the extended training set.

 \emph{\textbf{Fine-grained DNN structure design-based technology }} In\cite{Xu150,Shang157}, the authors used convolutional layer to replace full connection layer and proposed deep memory CNNs (M-Net) to overcome overfitting caused by small samples data, which achieved accuracy rate of more than 99\% on MSTAR. Aiming to improve the classification performance with greatly reduced annotation cost, the authors proposed an active DL approach for minimally-supervised PolSAR image classification\cite{active_DL_PolSAR_classify}, which integrated active learning and fine-tuning CNN into a principled framework. A microarchitecture called CompressUnit-based deeper CNN was proposed in \cite{CompressUnit_deeper_CNN_SAR_classify}. Compared with the fewest parameters-based networks for SAR image classification, this architecture was deeper with only about 10\% of parameters. An efficient transferred max-slice CNN with L2-regularization term was proposed in \cite{transferred_MS_CNN_SAR_recognition} for SAR-ATR, which could enrich the features and recognize the targets with superior performance with small samples. An asymmetric parallel convolution module was constructed in \cite{asymmetric_parallel_CNN_SAR_recognition} to avoid severe overfitting. In \cite{compact_CNN_PolSAR_classify}, the authors developed a systematic approach, based on sliding-window classification with compact and adaptive CNNs, to overcome drawbacks of limited labelled data whilst achieving state-of-the-art performance levels for SAR land use/cover classification.

 TL methods are significantly used in DNN design to solve the problems caused by limited data. A TL-based algorithm was proposed in\cite{Huang178} to transfer knowledge, learned from sufficient unlabeled SAR scene images, to labeled SAR target data. The proposed CNN architecture consisted of a classification pathway and a reconstruction pathway (i.e., stacked convolutional auto-encoders), together with a feedback bypass additionally. A large number of unlabeled SAR scene images were used to train the reconstruction pathway at first. Then, these pre-trained convolutional layers were reused to transfer knowledge to SAR target classification tasks, combining with reconstruction loss introduced by feedback bypass.

 TL strategy was used to effectively transfer the prior knowledge of the optical, non-optical, hybrid optical and non-optical domains to the SAR target recognition tasks in \cite{small_samples_SAR_ATR}. The approach of transferring knowledge from electro-optical domains to SAR domains was developed in \cite{optical_SAR_knowledge_transfer_SAR_classification} to eliminate the need for huge labeled data in the SAR classification. This method learned a shared domain-invariant embedding by cross-domain knowledge transfer pattern. The embedding was discriminative for both related electro-optical and SAR tasks, while the latent data distributions of both domains remained similar. Two TL strategies, based on FCN and U-net architecture, were proposed in \cite{TL_robustness} for high-resolution PolSAR image classification with only 50 image patches. The distinct pretraining datasets were also applied to different scenarios. To adapt deep CNN model for PolSAR target detection and classification with limited training samples while keeping better generalization performance, expert knowledge of target scattering mechanism interpretation and polarimetric feature mining were incorporated into CNN to assist the model training and improve the final application performance\cite{expert_knowledge_CNN_PolSAR_recognition}.

 The semi-supervised and unsupervised learning methods are also the significant technologies to alleviate the overfitting with small labeled data. A semi-supervised TL method based on GAN was presented in \cite{GANs_TL_SAR_semi_supervised} to address the insufficient labeled SAR data. Firstly, A GAN was trained by various unlabeled samples to learn generic features of SAR images. Subsequently, the learned parameters were readopted to initialize the target network to transfer the generic knowledge to specific SAR target recognition task. Lastly, the target network was fine-tuned by using both the labeled and unlabeled training samples with a semi-supervised loss function. In \cite{adversarial_learning_multi_band_SAR_classify}, an unsupervised multi-level domains adaptation method based on adversarial learning was proposed to solve the problem of time-consuming for multi-band labeled SAR images classification. A semi-supervised recognition method combining GAN with CNN was proposed in \cite{adversarial_learning_semi_supervised_SAR_classify}. A dynamic adjustable multi-discriminator GAN architecture was used to generate unlabeled images together with original labeled images as inputs of CNN. In order to alleviate the time-consuming problems of obtaining the labels of radar images, a semi-supervised learning method based on the standard DCGANs was presented in \cite{DCGANs_semi_supervise_SAR_recognition}. Two discriminators sharing the same generators for joint training.

 To alleviate the burden of manual labeling, a CNN-based unsupervised domain adaptation model was proposed in\cite{unsupervised_domain_adaption_SAR} to learn the domain-invariant features between SAR images and optical aerial images for SAR image retrieving. An unsupervised learning method to achieve SAR object classification with no labeled data was introduced in \cite{unsupervised_clustering_SAR_object_classify}. This approach regared object clustering as a recurrent process, in which data samples were gradually clustered together according to their similarity, and feature representations of them were obtained simultaneously. To address the problem of insufficient labelled training, an unsupervised DL model was implemented in the encoding-decoding architecture to learn feature maps at different scale and combine them together to generate feature vectors for SAR object classification in \cite{multi_scale_CAE_SAR_object_recognition}.

 In \cite{WAEs_generation_SAR_images}, the authors employed an extension of Wasserstein AE as deep generative model for SAR image generation to achieve SAR image target recognition with high accuracy. A novel generative-based DNN framework was proposed in \cite{zero_shot_learning_SAR_images} for zero-shot learning of SAR-ATR. A generative deconvolutional neural network was referred to as a generator to learn a faithful hierarchical representation of known targets, while automatically constructing a continuous SAR target feature space spanned by orientation-invariant features and orientation angle. In \cite{conv_biLSTM_few_shot_SAR_classification}, the authors proposed a new few-shot SAR-ATR method based on conv-biLSTM prototypical networks. A conv-biLSTM network was trained to map SAR images into a new feature space where it was easy for classification. Then, a classifier based on Euclidean distance was utilized to obtain the recognition results. A virtual adversarial regularization term was introduced in a neural nonlocal stacked SAEs architecture to regularize the network for keeping the network from being overfitting \cite{nonlocal_AEs_VAR_PolSAR_classify}. A multilayer AE, combining with Euclidean distance as a supervised constraint, to be used in\cite{euclidean_distance_constrain_AE_SAR_recognition} for SAR-ATR tasks with the limited training images.

 A new deep network in the form of a restricted three-branch denoising auto-encoder (DAE) was proposed in \cite{DAE_triplet_restriction_SAR_object_classify} to take the full advantage of limited training samples for SAR object classification. In this model, a modified triplet restriction, that combined the semi-hard triplet loss with the intra-class distance penalty, was devised to learn discriminative features with a small intra-class divergence and a large inter-class divergence. In order to solve overfitting problem, the authors introduced a dual-input Siamese CNN into the small samples oriented SAR target recognition in \cite{Siamese_network_small_samples_SAR_recognition}. The recognition accuracy rate of this method outperformed the SVM, A-ConvNet, and 18-layers ResNet by 31\%, 13\%,  and 16\%, respectively, in the experiment of 15 training samples and 195 testing data. A novel method of target classification of SAR imagery based on the target pixel grayscale decline with a graph CNN was introduced in \cite{graph_CNN_SAR_classify}, which transformed the raw SAR image from Euclidean data to graph-structured data by a graph structure and these transformed data were as the inputs of graph CNN model. To balance the anti-overfitting and features extraction abilities with small training samples for SAR targets images classification, the authors proposed a novel hinge loss (HL)-based CAE semi-greedy network in \cite{CAE-HL-CNN}, i.e., CAE-HL-CNN. Compared with existing state-of-the-art network, the CAE-HL-CNN had best performances in classification accuracy and computation costs with the SOC and EOC MSTAR datasets.

 \emph{\textbf{iii) Improving Robustness of Recognition Algorithms}}

 The speckle noise, clutter, scale-variance of inputs, and adversarial samples can severely cause unstability of DNN algorithm in SAR-ATR. Therefore, the robustness improvement of DNN algorithms is very vital. A new multi-view sparse representation classification algorithm based on joint supervised dictionary and classifier learning was developed in\cite{Ren185} for SAR image classification. During training peocess, classification error was back propagated to the dictionary learning procedure to optimize dictionary atoms. In this way, the representation capability of the sparse model was enhanced. This new architecture was more robust for depression variation, configuration variants, view number, dictionary size, and noise corruption, compared to other state-of-the-art methods, such as SVM.

 SAR-ATR is performed on either global or local features of acquired SAR images. The global features can be easily extracted and classified with high efficiency. However, they lack of reasoning capability thus can hardly work well under the EOCs. The local features are usually more difficult to extract and classify, but they can provide reasoning capability for target recognition. To make full use of global and local features of SAR-ATR at the EOCs, a hierarchical fusion scheme of the global and local features was proposed in\cite{Ding179} to jointly achieve high efficiency and robustness in ATR system. The global random projection features can be extracted and classified by sparse representation-based classification mthod effectively. The physical related local descriptors, i.e., ASCs, were employed for local reasoning to handle various EOCs like noise corruption, resolution variance, and partial occlusion.

 To improve robustness of model for noise and invariance of models, a multiple feature-based CNN model was employed in\cite{Cho181} to recognize the SAR image in an end-to-end learning way. The strong features more effected by noise and smoothed features less influenced were aggregated into a single column vector to build complementary relationships for recognition by a full connection network. As for target rotation behavior recognition, a rotation awareness based self-supervised DL model was proposed in\cite{Zhang184}. This model suggested that more attention should be paid on rotation-equivariant and label-invariant features. To explore the property of translation-invariance of CNNs, the authors verified that ResNet could  achieve translation-invariance with aligned SAR images\cite{Furukawa160}, even ResNet do not adopt data augmentation. A scale-invariant framework based on CNN was proposed in \cite{scale_invariant_CNN} to improve the robustness of model with respect to scale and resolution variations in dataset. This architecture developed an uniform representation method to enlarge the feature space for the variants of data by concatenating scale-variant and scale-invariant features.

 Luminance information of SAR images was used to form the target's profile in \cite{luminance_analysis_SAR_recognition} to significantly reduce the influence of speckle noise on CNN model. A scale transformation layer was embedded in deep convolutional autoencoder model to reduce the influence of noise in \cite{scale_transformation_denoising}. To restrain the influence of speckle noise and enhance the locally invariant and robustness of the encoding representation, the operations of contractive restriction and graph-cut-based spatial regularization in DSCNN were adopted in \cite{restrain_noise}. A SRDNN-based SAE was proposed to capture superpixel correlative features to reduce speckle noises in \cite{superpixel_classification_PolSAR}. A speckle-noise-invariant CNN was developed in \cite{noise_invariant_CNN}, which employed regularization term to improve Lee sigma filter performance, i.e., minimizing feature variations caused by speckle noise.

 A TL-based top-2 smooth loss function with cost-sensitive parameters was introduced to tackle the problems of label noise and imbalanced classes in \cite{TL_robustness}. A CNN-based recognition method of synthetic SAR dataset with complex background was proposed in \cite{complex_backgroud_SAR}. As for noise and signal phase errors, the authors proposed a advanced DL based adversarial training method to mitigate these influence in \cite{advanced_technology_noise_phase_errors}. A point-wise discriminative auto-encoder was proposed in \cite{noise_clutter_robustness} to extract noise and clutter robust features from the target area of SAR images. In order to alleviate the speckle influence on the scattering measurements of individual pixels in PolSAR images, local spatial information was introduced into stacked sparse autoencoder to learn the deep spatial sparse features automatically in \cite{local_spatial_information_SSAE_denoising}.

 Moreover, The DL-based SAR target recognition algorithms are potentially vulnerable to adversarial examples \cite{SAR_recognition_adversarial_attacks}. In \cite{advanced_technology_noise_phase_errors}, the authors involved a adversarial training technology to ensure the robustness of DL algorithm under the attacks of adversarial samples. HySARNet, as a hybrid ML model, was proposed in \cite{hySARNet_robustness} to determine the robustness of model when faced variations in graze angle, resolution, and additive noise in SAR-ATR tasks. A wavelet kernel sparse deep coding network under unbalanced dataset was proposed in \cite{deep_AEs_unbalanced_PolSAR_data_classify} for unbalanced PolSAR classification.

 The issue of different characters of heterogeneous SAR images will lead to poor performances of TL algorithm in SAR image classification. To address this problem, a semi-supervised model named as deep joint distribution adaptation networks was proposed in \cite{joint_distribution_adaptation_Net_TL_SAR_classifiy} for TL model, which learning from a source SAR images to similar target SAR images. In order to increase the stability of GANs model training in SAR targets recognition, the authors proposed a new semi-supervised GANs with multiple generators and a classifier in \cite{semi_supervised_GANs_multi_generators_SAR_recognition}. Multiple generators were employed to keep stability of training.

 \emph{\textbf{iv) Promoting the Real-Time or Reducing Computation Costs}}

 A CNN-based framework consisted of SqueezeNet network and a modified wide residual network was developed in \cite{Bai148} to build real-time damage mapping for classifying different damaged regions on the SAR image. A  direct ATR method was employed in\cite{Huang195} for large-scene SAR-ATR task, which directly recognized targets on large-scene SAR images by encapsulating all of the computation in a single DCNN. Experiments on MSTAR and large-scene SAR images (with resolution 1478 * 1784) showed this model outperformed other methods, such as CFAR+SVM, region-based CNN, and YOLOv2\cite{Redmon218}. The PCANet was employed in\cite{Qi197} for SAR-ATR to achieve more than 99\% accuracy rate on MSTAR. A-convNet was proposed in\cite{Wang109} to achieve an average accuracy rate of 99.1\% on MSTAR. A novel stacked deep convolutional highway unit network was proposed in\cite{Zhao172} for SAR imagery classification, which achieved accuracy rate of 99\% with all MSTAR data, and still reached 94.97\% when the training data was reduced to 30\%.

 The complex multi-view processing of images, however, can cause huge computation costs for multi-view learning method. To address this problem, a optimal target viewpoints selection based multi-view ATR algorithm was developed in\cite{Pei177}. This algorithm used two-channel CNNs as multi-view classifiers, which was based on ensemble learning\cite{ensemble_learning}. A direct graph structure-based single source shortest path search algorithm was also adopted to represent the tradeoff between the recognition performance and flight distance of SAR platform. A heterogeneous CNN-based ensemble learning method was employed in\cite{Xue199} to construct noncomplete connection scheme and multiple filters stacked.

 A lightweight CNN model was designed in\cite{Shao190} to recognize the SAR images. The channel attention by-pass and spatial attention by-pass were introduced to enhance the feature extraction ability. Depthwise separable convolution was used to reduce the computation costs and heighten the recognition efficiency. In addition, a new weighted distance measured loss function was introduced to weaken the adverse effects of data imbalance on accuracy rate of minority class. This architecture has better performance than ResNet, A-ConvNet\cite{Wang109,Chen191}.

 A one-layer based novel incremental Wishart broad learning system was specifically designed in \cite{incremental_learning_PolSAR_image_classification} to achieve PolSAR image classification, which could effectively transfer essential Wishart distribution and other types of polarimetric decomposition and spatial features to establish feature map and enhancement nodes in just one layer without DL structures. Therefore, the training consumption could be decreased significantly. Similarly, a superpixel-driven optimized Wishart network was introduced in \cite{fast_Wishart_network} for fast PolSAR image classification. In \cite{computation_cost_tricks_DNN}, the authors applied some tricks (such as BN, drop-out strategy) and concatenated ReLU to reduce computation cost of DL algorithm. A spatial-anchor graph based fast semi-supervised classification algorithm for PolSAR image was introduced in \cite{computation_cost_spatial_anchor_graph}.

 In \cite{pixel_grayscale_graph_CNN}, the authors proposed a novel method based on target pixel grayscale decline by a graph representation network to accelerate the training time and achieve classification accuracy rate of 100\%. In order to speed up computation and improve classification accuracy, a classification method of full-polarization SAR images, based on DL with shallow features,  was proposed in \cite{shallow_features_computation_cost}. Aiming to solve the problems of energy consumption, so as to deploy the DL model on embedded devices conveniently and train the model in real-time, a custom AI streaming architecture was employed in \cite{real_time_embedded_device} for SAR maritime target detection. A more flexible structure as the new implementation of CNN classifier was proposed in \cite{flexible_structure_CNN}, which had less free parameters to reduce training time. An atrous-inception module-based lightweight CNN was proposed in \cite{small_samples_SAR_ATR}, which combined both atrous convolution and inception module to obtain rich global receptive fields, while strictly controlling the parameter amount and realizing lightweight network architecture. In \cite{high_speed_SAR_classify}, apache spark clustering framework was presented for classification of high-speed denoised SAR image patches. An asymmetric parallel convolution module was constructed in \cite{parallel_CNN_alleviate_computation_cost} to alleviate the computation cost. In order to alleviate the trade-off between real-time and high performance, the authors proposed a semi-random DNN to exploit random fixed weights for real-time training with comparable accuracy of general CNNs in \cite{semi_random_real_time}.

 To tackle the issues of low memory resources and low calculation speed in SAR sensors, the authors proposed a a micro CNN for real-time SAR recognition system in \cite{micro_CNN_knowledge_distillation}, which only had two layers, compressed from a 18-layer DCNN by a novel knowledge distillation algorithm, i.e., gradual distillation. Compared with the DCNN, the memory footprint of the proposed model was compressed 177 times, and the computation costs was 12.8 times less. In order to deploy a real-time SAR platform, three strategies of network compression and acceleration were developed in \cite{NN_acceleration_three_strategies} to decrease computing and memory resource dependencies while maintaining a competitive accuracy. Firstly, weight-based network pruning and adaptive architecture squeezing method were proposed to reduce the consumption of storage and computation time of inference and training process of DL model. Then weight quantization and coding were employed to compress the network storage space. In addition, a fast approach for pruning convolutional layers was proposed to reduce the number of multiplication by exploiting the sparsity of the inputs and weights.

 At present, most of neutral network-based classification methods need to expand the dataset by data augmentation technology or design the light-weighted network model to improve their classification performance. However, optimal training and generalization are two main challenges for DNN model. Instead of DNN model, a novel deep forest model was constructed in\cite{Zhang198} by multi-grained cascade forest (gcForest) to classify 10-class targets on MSTAR. This was the first attempt to classify SAR targets using the non-neural network model. Compared with DNN-based methods, gcForest had better performances in calculation scale, training time, and interpretability.

\textbf{4) Ship Targets Detection based on SAR Images}.

 In section 3) we make a comprehensive survey on SAR-ATR based on DL algorithm. From the overview in published literatures, the SAR-ATR is a very important research domain widely involved in military and civil applications. The SAR-based ship targets detection (STD), one of the important research aspects in maritime surveillance (such as marine transportation safety), is an another significant research direction for SAR image processing. Of course, optical imagery-based ship detection and classification is also a hot research direction, please refer to \cite{vessel_detection_optical}. The SAR images of the STD usually have a large scale, which contains many different scale ship targets. The goal of STD is detection and recognition of each target on the SAR image.

 Traditional STD approaches include constant false alarm rates (CFAR) based on the distributions of sea clutter \cite{Gao205,Farrouki206}, extracted features manually based on ML algorithm \cite{Huang207,Xing208,Ma209,Xie210}, dictionary-based sparse representation, SVM, template matching, K-NN, Bayes, saliency object detection models. Traditional methods, however, intensively depend on the statistics modeling and the experts' feature extraction ability, which degrades the detection performances of SAR imagery to some extend.

 In recent years, DL-based methods have produced many great achievements in objects detection domain. These DL algorithms can be roughly categorized into two classes: two-stage methods and one-stage methods. The former firstly generates positive region proposals to discard the most of negative samples, then performs the candidate regions classification, such as region convolutional neural networks (R-CNN) \cite{Girshick211}, fast R-CNN \cite{Girshick212}, faster R-CNN \cite{Ren213}, mask R-CNN \cite{He214}, cascade R-CNN \cite{Cai215}, feature pyramid networks (FPNs) \cite{Lin216}. The latter directly detects the objects by obtaining objects' coordinate values and the class probability, which considers both accuracy and computation costs, such as You Only Look Once (YOLOs: v1-v4, poly-v3) \cite{Redmon217,Redmon218,Redmon219,YOLOv4,ploy-YOLOv3}, single shot multiBox detector (SSD) \cite{Liu220}, RetinaNet \cite{Lin221}. The two-stage methods have higher accuracy, but slower training than one-stage methods.

 Nowadays, the SAR researchers have successfully applied DL algorithms in STD. Some challenges, however, have occurred in this domain even though applied DL algorithms, which mainly focus on three aspects: (i) ships often have a large aspect ratio and arbitrary directionality in SAR images. Traditional detection algorithms can unconsciously cause redundant detections, which make it difficult to accurately locate the target in complex scenes (such as background interference, clutter, inshore and outshore scenes, e.g., Chinese Gaofen-3 (GF-3) Imagery has 86 scenes \cite{Yuanyuan231}); (ii) ships in ports are often densely arranged, and the effective identification of these ships is complicated; and (iii) ships in SAR images have various scales due to the multi-resolution imaging modes employed in SAR (such as GF-3 Imagery has four resolutions, i.e., 3 m, 5 m, 8 m, and 10 m \cite{Yuanyuan231}) and various ship shapes, which pose a considerable challenge for STD. In this section, we will do a comprehensive survey on DL-based STD, which mainly focuses on solving these challenges aforementioned.

 In \cite{fater_RCNN_SAR}, faster R-CNN architecture \cite{Ren213} was investigated in STD. A new dataset and four strategies (feature fusion, transfer learning, hard negative mining, and other implementation details) were proposed to achieved better accuracy and less test computation costs than the standard faster R-CNN algorithm. A densely connected multi-scale neural network based on faster R-CNN framework was proposed in\cite{Jiao232} to solve the multi-scale and multi-scene STD problems. In\cite{Nie241}, the authors proposed a ship detection and segmentation method based on an improved mask R-CNN model\cite{He214}. This method could accurately detect and segment ships at the pixel level. In\cite{Yuanyuan231}, RetinaNet \cite{Lin221} model was used as the object detector, to automatically determine features representation learning effectively for multi-scale ships detection.

 However, the common target detection models originate from the optical image detection tasks in CV, which maybe degrade their performances when applied to STD more or less, because of special imaging principles of SAR images. Many new algorithms have been proposed to specially address the challenges in STD. These new ideas remain depending on basic targets detection models, such as FPNs\cite{Lin216}, R-CNN\cite{Girshick212}.

 \emph{\textbf{i) Improving Accurately Location of Ship Targets}}

 As for the first problem, i.e., it is difficult to accurately locate the targets in complex scenes. RetinaNet was applied to\cite{Yuanyuan231} to alleviate the limitation of highly depending on the statistical models of sea clutter in STD, which achieved more than a mean average precision (MAP) of 96\%, and could efficiently detect multi-scale ships with high effectiveness in GF-3 SAR images. A new land masking strategy based on the statistical model of sea clutter and neural network model was employed in\cite{An228} to detect ships in GF-3 SAR images. The fully convolutional network (FCN) was applied to separate the sea area from the land. Then, choosing the probability distribution model of CFAR detector based on a tradeoff between the sea clutter modeling accuracy and the computational complexity. In addition, truncated statistic and iterative censoring scheme were used to better implement CFAR detection for boosting the performance of detector. Due to the multi-resolution imaging mode and complex background, multi-level sparse optimization method of SAR image was studied in \cite{Gao225} to handle clutters and sidelobes, so as to extract discriminative features of SAR images. A segmentation method based on a U-Net was developed in\cite{Fan224} to address the problems of false alarms caused by ocean clutter. This algorithm was designed specifically for pixel-wise STD from compact polarimetric SAR images. A novel object detection network was employed in\cite{Kang227,Chen235} to extract contextual features of images. This model also used attention mechanism to rule out false alarms in complex scenarios. A new training strategy was adopted in\cite{Jiao232} to reduce the weights of easy examples in the loss function, so that more attention focused on the hard examples in training process to reduce false alarm. Two parallel sub-channels based multi-feature learning framework was proposed in \cite{two_sub_channels_ship_detection}, including DL-based extracted features and hand-crafted features. Two sub-channels features were concatenated to extract fused deep features to achieve high performance.

 \emph{\textbf{ii) Accurately Detection of Densely Arranged Ships}}

 As for second problem, it is difficult to detect densely arranged ships. Non-maximum suppression (NMS) method was widely used to address this issue. A soft-NMS method was introduced into the detection network model in\cite{Chen235,Wei242} to reduce the number of missed detections of ship targets in the presence of severe overlap for improving the detection performance of the dense ships. In addition, the modified rotation NMS was developed in\cite{Chen238} to solve the problem of the large overlap ratio of the detection box.

 \emph{\textbf{iii) Solving the Problems of Multi-scale Variations}}

 More importantly, it is very vital to design a optimal solution to solve the problems of multi-scale variations in design of STD algorithms. A FPN was used in\cite{Yuanyuan231} to extract multi-scale features for both ship location and classification, and focal loss was also used to address the class imbalance to increase the importance of the hard examples during training process. A densely connected multi-scale neural network based on faster R-CNN was proposed in\cite{Jiao232} to densely connect one feature map to each other feature maps from top to down. In this way, the positive proposals were generated from each fused feature map based on multi-scale SAR images in multi-scene. Similarly, combining with densely connecting convolutional block attention module, a dense attention pyramid network was developed in\cite{Cui237,Li240} to concatenate feature maps from top to bottom of the pyramid network. In this way, sufficient resolution and semantic information features were extracted. In addition, convolutional block attention module refined concatenated feature maps to fuse highlight salient features with global unblurred features of multi-scale ships, and the fused features were as the inputs of detection network to accurately obtain the final detection results.

 To address the diverse scales of ship targets, a loss function incorporated the generalized intersection over union (GIoU) loss to reduce the scale sensitivity of the network\cite{Chen235}. In\cite{Zhang236}, a new bi-directional feature fusion module was incorporated in a lightweight feature optimizing network to enhance the salient features representation of both low and high features representation layers. Aiming to fast achieve positioning rotation detection, the authors proposed a multiscale adaptive recalibration network in \cite{Chen238} to detect multiscale and arbitrarily oriented ships in complex scenarios. The recalibration of the extracted multiscale features improved the sensitivity of the network to the target angle through global information. In particular, a pyramid anchor and a loss function were designed to match the rotated target to accelerate the rotation detection.

 To eliminate the missing detection of small-sized ships targets in SAR imagery, a contextual region convolutional hierarchical neural network with multilayer fusion strategy was designed in\cite{Kang227},  which consisted of a high resolution RPN and an object detection network to extract contextual features. This framework fused the deep contextual semantic features and shallow high-resolution features to improve the detection performance for small-sized ships. A novel split convolution block was used in \cite{Gao225} to enhance the feature representation of small targets, which divided the SAR images into smaller sub-images as the inputs of the network. Also, a spatial attention block was embedded in FPN to reduce the loss of spatial information during the dimensionality reduction process.

 Based on TL method, a pre-trained YOLOv2 model \cite{Redmon218} was applied to STD in\cite{Hamza233}. The experiments on three different datasets showed the effectiveness of the pre-trained YOLOv2. TL strategy was also used to train the detection model in \cite{Yuanyuan229} due to the limited number of datasets. Instead of a single feature map, a scale-transferrable pyramid network was employed in\cite{Liu239} for multi-scale detection. A latent connection based FPN was constructed to inject more semantic information into feature maps with high resolution, and densely connected each feature maps from top to down by using scale-transfer layer. Therefore, the dense scale-transfer connection could expand the resolution of feature maps and explicitly explore valuable information contained in channels. A scale transfer module was also used in\cite{Gui234} to connect with several feature maps to extract multiscale features for STD. In addition, RoIAlign was adapted to calibrate the accuracy of the bounding boxes, and the context features were employed to assist the detection of complex targets in detection subnetwork.

 Nowadays, the existing methods of SAR STD mainly depend on low-resolution representations obtained by classification networks or recover high-resolution representations from low-resolution representations in SAR images. These methods, however, are difficult to obtain accurate prediction results in spatial accuracy of region-level. Based on a high-resolution STD network, a novel framework was proposed in\cite{Wei242} for high-resolution SAR imagery ships detection. This architecture adopted a novel high-resolution FPN connecting with several high-to-low resolution subnetworks in parallel, to make full advantage of the high-resolution feature maps and low-resolution convolutions to maintain high resolution STD. In addition, soft-NMS was also used to improve the detection performance of the dense ships and the Microsoft COCO evaluation metrics was introduced for performance evaluation.

 Most of STD algorithms are focus on detection accuracy. Detection speed, however, is usually neglected. The speed of SAR STD is extraordinarily important, especially in real-time maritime rescue and emergency military decision-making. To improve the detection speed, a pyramid anchor and a loss function were designed in\cite{Chen238} to match the rotated targets to speed up the arbitrary ships rotation detection. A novel grid CNN was developed in\cite{Zhang243} for high-speed STD, which mainly consisted of a backbone CNN and a detection CNN. Inspired by the idea of YOLO algorithm, this method improved the detection speed by meshing the input images and using the depthwise separable convolutions. The experiments results on SSDD dataset and two SAR images from RadarSat-1 and Gaofen-3 showed that the detection speed of this model was faster than the other existing methods, such as faster R-CNN, SSD, and YOLO under the same computing resource, and the detection accuracy was kept within an acceptable range.  To infer a large volume of SAR images with high detection accuracy and relatively high speed, SSD was adopted in\cite{Yuanyuan229} to address STD in complex backgrounds. TL strategy was also adopted to train the detection model.

 In sections 3) and 4), we make a comprehensive survey of SAR-ATR and STD based on SAR imagery. In addition, SAR imagery segmentation is also researched. Targets segmentation tries to separate the target from the background thus eliminating the interference of background noises or clutters. However, it may also discard a part of the target characteristics and target shadows during the segmentation process, which also contains discriminative information for target recognition. Then the tradeoff between interference elimination and discriminability loss will degrade target recognition to some extent\cite{Ding246}. Therefore, the comprehensive evaluation for the effectiveness of segmentation on target recognition is very important. A novel architecture for SAR segmentation based on convolutional wavelet neural network (CWNN) and MRF \cite{Wu245} were proposed in\cite{Duan244}, which could suppress the noise and keep the structures of the learned features complement. In addition, a ship detection and segmentation method based on an improved mask R-CNN model was developed in\cite{Nie241}, which could accurately detect and segment ships at the pixel level. To allow lower layers features to be more effectively utilized at the top layer, a channel-wise and spatial attention mechanisms based bottom-up structure was added to FPN structure of mask R-CNN, so as to shorten the paths between lower layers and the topmost layer. The experiments results showed that the MAPs of detection and segmentation increased from 70.6\% and 62.0\% to 76.1\% and 65.8\%, respectively.

    \vspace{-1.0mm}
\subsection{ISAR Images Processing}\label{S4.3}

\textbf{1) ISAR Imaging}.

 To address the problem of low-resolution (LR) ISAR imaging, the authors employed deep ResNet as an end-to-end framework to directly learn the mapping between the input LR images and the output high-resolution (HR) images with respect to the point spread function (PSF) in \cite{ResNet_low_resolution_high_ISAR}. An amount of multiplicative noise or clutter may be present in real-world ISAR measurement scenarios. The current linear imaging methods are not generally well suitable to alleviate the effects of noise, such as MUSIC, compressive sensing (CS). Since these algorithms rely on phase information significantly which can be heavily distorted or randomized under the imaging process. The authors introduced CNNs model to deal with this issue in \cite{CNNs_ISAR_imaging_speckle_robustness}. In order to exploit a real-time ISAR imaging algorithm, the authors proposed an efficient sparse aperture ISAR autofocusing algorithm in \cite{sparse_Bayesian_learning_divided_subproblems_ISAR_imaging}, which adopted divided simpler subproblems by alternating direction method of multipliers and auxiliary variable to alleviate the complex computation of ISAR imaging used sparse Bayesian learning (SBL) method. This method achieved 20-30 times faster than the SBL-based approach.

 To address the problem of basis mismatch of CS based HR ISAR imaging of fast rotating targets,  a pattern-coupled sparse Bayesian learning method for multiple measurement vectors, i.e., the PC-MSBL algorithm, was proposed in \cite{High_resolution_rotating_object_ISAR_imaging}. A multi-channel pattern-coupled hierarchical Gaussian prior was introduced to model the pattern dependencies among the neighboring range cells and correct the migration through range cells problem. The expectation-maximization (EM) algorithm was used to infer the maximum a posterior estimate of the hyperparameters. To tackle the issue of destroyed coherence between the undersampled pulses caused by sparse aperture radar echoes, the authors proposed a novel Bayesian ISAR autofocusing and scaling algorithm for sparse aperture in \cite{variational_Bayesian_inference_sparse_aperture_ISAR_imaging}.

\textbf{2) ISAR Targets Detection and Recognition}.

 In order to tackle the challenges of ISAR objects detection, a fast and efficient weakly semi-supervised method, called deep ISAR object detection (DIOD), was proposed in \cite{fast_semi_supervised_ISAR_object_detection_1}, which was based on advanced region proposal networks (ARPNs) and weakly semi-supervised deep joint sparse learning. This framework used i) ARPN to generate high-level region proposals and localize potential ISAR objects robustly and accurately in minimal time, ii) a convenient and efficient weakly semi-supervised training method was proposed to solve the problem of small annotated training data, and iii)  a novel sharable-individual mechanism and a relational-regularized joint sparse learning strategy were introduced to further improve the accuracy and speed of the whole system. Similarly, the authors proposed a novel DIOD method, which was based on fully convolutional region candidate networks and DCNNs in \cite{fast_semi_supervised_ISAR_object_detection_2}.

 A TL-based novel method of multiple heterogeneous pre-trained DCNN (P-DCNN) ensemble with stacking algorithm was firstly proposed in \cite{TL_stacking_algorithm_ISAR_space_targets_recognition}, which could realize automatic recognition of space targets in ISAR images with high accuracy under the condition of the small samples. The stacking algorithm was used to realize the ensemble of multiple heterogeneous P-DCNNs, which effectively overcame weak robustness and difficulty in classification accuracy existing in a single weights fine-tuned P-DCNN. A semantic knowledge based deep relation graph learning was proposed in \cite{deep_relative_graph_learning_ISAR_object_recognition} for real-world ISAR object recognition and relation discovery. Dilated deformable CNN was introduced to greatly improve sampling and transformation ability of CNN, and increase the output resolutions of feature maps significantly. Deep graph attribute-association learning method was proposed to obtain semantic knowledge to exploit inter-modal relationships among features, attributes, and classes. Multi-scale relational-regularized convolutional sparse learning was employed to further improve the accuracy and speed of the whole system. In addition, CNNs and CAEs were also used to classify ISAR objects in \cite{CNNs_CAEs_ISAR_objects_SAR}.

 Three ML algorithms were introduced in \cite{three_ML_ISAR_objects_classifcation} for ISAR targets classification, i.e., DT, Bayes, and SVM. A SAE learning algorithm was employed in \cite{SAE_ISAR_objects_classifcation} to solve the classification issue of non-cooperative airplane targets with ISAR images.

    \vspace{-1.0mm}
\subsection{HRRP-based Automatic Target Recognition}\label{S4.4}

 With the advantages of easily acquisition, processing and abundant target feature information, unidimensional high resolution range profile (HRRP) is a specially concern research direction of ATR. HRRP is the projection of target echo scatter vectors in the direction of radar sight line, at the condition of big transmitted signal bandwidth and big target shape. The HRRP-ATR research domain mainly concerns solving three aspect problems: noise robustness, discriminative and informative features extraction, and optimal classifier design. In practice, the first two problems are usually tackled simultaneously.

 There are three stages for HRRP-ATR: image preprocessing \cite{Wu253}, feature extraction and classifier design respectively. Image preprocessing mainly includes denoising \cite{Li250,Li252,Jiang259,Du260} and alleviates sensitivity problems: gesture, translation, and amplitude \cite{Guo251,Li252,Liao257,Xu265}. Feature extraction process extracts low dimensional inherent features from preprocessed HRRP, which are easily identifiable for HRRP of the target, including PCA, expert-based feature engineering. A fine classifier is used to achieve ATR tasks, such as SVM, DNNs.

 Three stages are not rigorously operated sequentially. Some algorithms can achieve multi-operation simultaneously, e.g., PCA has denoise and dimensionality reduction functions. Take DNNs as an example, the DNNs are end-to-end learning architectures, operating the feature extraction and classification simultaneously \cite{Liao257,Xu265,Mian266}.

\textbf{1) Feature Extraction}.

 Probabilistic principal component analysis (PPCA) model was proposed in\cite{Li247,Yuan248} for noise robust feature extraction, which provided prior information for robust features from statistic modeling perspective. In\cite{Xu249}, the authors adopted Bernoulli-Beta prior to learn the needed atoms to determine relatedness between frames of training data. A feature extraction dictionary was used to extract the local and global features of target's HRRP \cite{Li250,Li261} for multi-feature joint learning method based on sparse representation and low-rank representation. Support vector data description was developed in\cite{Guo251} to extract non-linear boundary of dataset as classification features. In addition, orthogonal maximum margin projection subspace (OMMPS) was employed in\cite{Li252} for HRRP's feature extraction to reduce redundancy. To improve recognition performance, multiple kernel projection subspace fusion method was introduced in\cite{Li252,Li254} for feature extraction of HRRP, this method can guarantee the integrity of target information and robustness.

 As for dealing with the challenge of noncooperative target recognition with imbalanced training datasets, t-distributed stochastic neighbor embedding (t-SNE) and synthetic sampling for data preprocessing were utilized in\cite{Pan255} to provide a well segmented and balanced HRRP dataset. Scatter matching algorithm was proposed in\cite{Jiang259,Du260} for dominant scatters features extraction of HRRP with noise robustness. Multi-scale fusion sparsity preserving projections approach was also proposed in\cite{Dai262} to construct multi-scale fusion features in each scale and their sparse reconstructive relationship, which contained more discriminative information. To exploit potential features of HRRP, scale space theory based feature extraction method was employed in\cite{Liu263}, which extended from single scale to multiple scales.

 \textbf{2) Classifier Design}.

 The classifiers of HRRP-based classification mainly include SVMs \cite{Dai262,Liu263}, DBNs \cite{Pan255,Peng256}, SAEs \cite{Zhao258,Feng269}, and RNNs \cite{Xu265}. The learning strategies of classifier include multitask learning \cite{Xu249}, multi-feature learning \cite{Li250}, and multi-scale learning \cite{Dai262,Liu263,FPN_CNN}.

 PPCA based dictionary learning method was proposed in\cite{Yuan248} for HRRP recognition. OMMPS is used to maximize the margin of inter-class by increasing the scatter distance of inter-class and reducing the scatter distance of intra-class simultaneously \cite{Li252}, to improve the recognition performance. T-SNE based discriminant DBN was proposed in\cite{Pan255} as an efficient recognition framework with imbalanced HRRP data, which not only made full use of dataset inherent structure among HRRP samples for segmentation, but also utilized high-level features for recognition. Moreover, the model shared latent information of HRRP data globally, which could enhance the ability of modeling the aspect sectors with few HRRP data.

 In order to reduce preprocessing works, discriminative infinite RBM (Dis-iRBM) was proposed in\cite{Peng256} as an end-to-end adaptive feature learning model to recognize HRRP data. Concatenated DNN was used in \cite{Liao257} for HRRP recognition. Multi-evidence fusion strategy was also adopted for recognition of multiple samples to improve performance.

 Although the deep structure has high accuracy, it is hard to achieve the performance of good generalization and fast learning. In \cite{Zhao258}, the authors combined SAE with regularized ELM to recognize HRRP data with a fast learning speed and better generalization performance. SVM was employed to verify the classification performance of features extracted by MSFSPP and related feature extraction methods in \cite{Dai262}. SVM and three nearest neighbor classifiers demonstrated that the application of scale-space theory in multi-scale feature extraction could effectively enhance the classification performance\cite{Liu263}. A TL-based feature pyramid fusion lightweight CNN model was proposed in \cite{FPN_CNN} to conduct multi-scale representation of HRRP target recognition with small samples at low SNR scenario. Reconstructive and discriminative dictionary learning based on sparse representation classification criteria was developed in\cite{Zhou264}, which incorporated the reconstructive and discriminative powers of atoms during the update of atoms. This algorithm was more robust to the variation of target aspect and noise effect.

 To extract fine discriminative and informative features of HRRP, target-aware recurrent attentional network (TARAN) was used in\cite{Xu265} to make use of temporal dependence and find the informative areas in HRRP. This network utilized RNN to explore the sequential relationship between the range cells within a HRRP sample, and employed the attention mechanism to weight up each time step in the hidden state, so as to discover the target area. To extract high dimensional features and generally contain more target inherent characteristics, discriminant sparse deep AE framework was proposed in\cite{Mian266} to classify HRRPs with small data samples. This framework was inspired by multitask learning and trained by the radar HRRP samples to share inherent structure patterns among the targets. In\cite{Feng269}, the authors built stacked corrective AE to recognize HRRP, which employed the average profile of each HRRP frame as the correction term.

 Considering the noise robust recognition of noncooperative targets, Gaussian kernel and Morlet wavelet kernel were combined in\cite{Xiong267} to form a multiscale kernel sparse coding-based classifier to recognize radar HRRP, which had comparable performance with well-studied template based methods, such as SVM, sparse coding-based classifiers (SCCs) and kernel SCCs. To classify the FFT-magnitude features of complex HRRP, least square support vector data description classifier was developed in\cite{Guo268} to classify HRRP with low computational complexity and overcame the shortcoming of poor capacity of variable targets in support vector data description.

 \vspace{-1.0mm}
\subsection{Micro-doppler Signature Recognition}\label{S4.4}

 Micro-doppler (MD) technique aims to extract the micro-motions of subjects, that may be unique to a particular subject class or activity, to distinguish probable false alarms from real detections or to increase the valuable information extracted from the sensor. Using the available MD returns from sensor for recognition can significantly reduce the false alarm rate, thereby improving the utility of the sensor system \cite{Tahmoush286}. Radar MD signatures, derived from these motions, illustrate the potential ability of the joint time-frequency analysis for exploiting kinetic and dynamic properties of objects \cite{Chen287}, such as drones \cite{Patel288,Ritchie289,Rahman290}, unmanned aerial vehicle (UAV) \cite{Patel288}, human motion \cite{Thayaparan292}, deceptive jamming \cite{Shi291}.

 The MD-based classification and recognition of object's postures and activities has widely absorbed research concerns in the past few years, such as human detection and activity classification \cite{Kim270}, human gesture \cite{Zhang276} and gait recognition \cite{Seyfioglu274,Fioranelli275}, UAV detection \cite{Patel288}. This section will review the achievements of ML-based radar MD signature processing in target classification and recognition.

 Similar to HRRP-ATR, feature extraction and classifier design are mainly stages for MD signature based recognition tasks. The features of targets' activities are extracted from the radar MD spectrogram, such as single vector decomposition (SVD) vectors of raw data \cite{Fioranelli273,Fioranelli275}. The optimal classifier design is based on ML models, such as SVM \cite{Kim270}, ANN \cite{Kim271}, CNN \cite{Kim272,Zhang276,Sherif278,Kim285}, CAE \cite{Sherif278,Seyfioglu280,Parashar282}, RNN \cite{Klarenbeek284}.

 A novel robust MD signal representation method based on both magnitude and phase information of the first Fourier transform was proposed in\cite{Ren277} for UAV detection, i.e., 2D regularized complex-log-Fourier transform and an object-oriented dimensionality reduction technique-subspace reliability analysis. The latent space representation was extracted and interpreted in\cite{Sherif278} from 2D CAEs and t-SNE, respectively. In addition, CAE architecture was employed in\cite{Parashar282} for MD feature extraction. Three features extraction algorithms were proposed in\cite{Miller283}, spectrogram frequency profile (SFP) algorithm, cadence velocity diagram frequency profile (CVDFP) algorithm, and SFP-CVDFP-PCA algorithm, respectively.

 As for classifier design, the SVM and ANN classifier were developed in\cite{Kim270,Kim271} for seven-class human activities classification based on six features extracted from radar doppler spectrogram, respectively. These features included running, walking, walking while holding a stick, crawling, boxing while moving forward, boxing while standing in place, and sitting still. Compared to SVM classifier, a 3-layer AE structure was proposed in\cite{Seyfioglu274}, which achieved a accuracy rate of 89\%, 17\% improvement compared to the benchmark SVM with 127 pre-defined features. SVM was also used in\cite{Miller283} based on multi-feature integration.

 Deep CNN structure was proposed in\cite{Kim272} for human detection and activities classification, jointly learned the necessary features and classification boundaries. MD-based human hand gestures recognition achieved accuracy rate of 98\% using CNN in \cite{Zhang276}. A 50-layer ResNet was trained to identify the walking subject based on the 2D signature \cite{Sherif278}. TL-based DNN model was also used to classify MD signatures of gait motion in \cite{Seyfioglu279}.

 To seek the efficient MD analysis algorithm to distinguish the gaits of subjects, even the MD signatures of those gaits are not visually distinguishable, a 3-layer deep CAE was proposed in \cite{Seyfioglu280}, which utilized unsupervised pre-training to initialize the weights in the subsequent convolutional layers, and yielded a correct classification rate of 94.2\%, 17.3\% improvement over the SVM. These MD signatures ere were measured from the 12 different human indoors activities using a 4 GHz continuous wave radar. CAE is more efficient than other deep models, such as CNN, AE, and traditional classifiers, such as SVM, RF and Xgboost. To compare the efficiency of ANN initialization technologies in classification of MD signals, an unsupervised TL-based pretraining method was applied to CAE \cite{Seyfio281}. VGGNet and GoogleNet were employed to classify human activities with small training samples. In order to address the measurements of a variable observation time and transition between classes over time, a sequence-to-sequence classification method, i.e., RNN with LSTM architecture, was developed in\cite{Klarenbeek284}. In addition, to make full use of time and frequency domain features of MD signatures, merging time and frequency-cadence-velocity diagram was proposed in\cite{Kim285} for drone classification with GoogleNet.

 \vspace{-1.0mm}
\subsection{Range-doppler Image Processing}\label{S4.5}

 Range-doppler (RD) images is also used for classification and recognition of target's motions. A RD image contains information of range units and doppler features. In linear frequency modulation continuous wave (LFMCW) radar, the RD imaging process is as the following: firstly removing the slope of echo signal, then obtaining the radical range information by FFT of fast time domain signal, after that, acquiring the energy distribution of doppler domain by FFT of slow time direction in the same range unit.

 Two different classification architectures based on SAE were developed in \cite{Jokanovi293} for human fall detection and classification, which used RD images and MD images as the inputs of the cascade and parallel connection models, respectively. Firstly, RD images and MD images were as the inputs of initial SAEs to extract identifiable features, respectively. Then, the extracted features were fused as the inputs of a final SAE to finish classification task. The results of experiment showed that the detection probabilities were 89.4\% and 84.1\% for cascade and parallel detection architecture on same dataset, respectively.

 Combining with convolutional and memory functions, an end-to-end learning architecture based on CNN and LSTM was developed in\cite{Wang294} for 11 kinds of dynamic gestures recognition. The RD images of gestures at a time point were as the inputs of CNN and then the RD images sequences at different time points were as inputs of LSTM to finish recognition. This novel recognition model achieved average accuracy rate of 87\% on a 11 kinds of dynamic gestures data and generalized well across 10 users.

 A novel detection method was developed in\cite{Li295} for remotely identifying a potential active shooter with a concealed rifle/shotgun based on radar MD and RD signatures analysis. Special features were extracted and applied for detecting people with suspicious behaviors. ANN model was also adopted for the classification of activities, and achieved an accuracy rate of 99.21\% in distinguishing human subjects carrying a concealed rifle from other similar activities.

\section{Anti-jamming and Interference Mitigation}\label{S5}

 This section will review the radar anti-jamming and interference mitigation technologies in ML-related RSP domain, including jamming or interference classification and recognition and anti-jamming and interference mitigation strategies. The examples of 2D time-frequency images of traditional jamming signals (including radio frequency (RF) noise, frequency-modulation (FM) noise, amplitude-modulation (AM) noise, constant range gate pull off (RGPO), velocity gate pull off (VGPO), convolutional modulation (CM), intermittent sampling (IS)) and the time and frequency domain images of novel jamming signals (including smeared spectrum (SMSP), chopping and interleaving (CI), smart noise jamming (SNJ), range deception jamming signal - amplitude modulation noise (RD-AM), and range deception jamming - frequency modulation noise (RD-FM)) are shown in Fig. 19 and Fig. 20 respectively.

 \begin{figure}[h] 
\vspace{-2.0mm}
\begin{center}
\includegraphics[width=1.0\columnwidth, keepaspectratio]{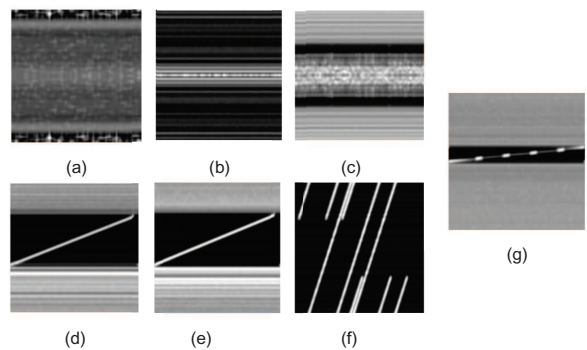}
\end{center}
\captionsetup{font = {footnotesize}, name = {Fig.}, singlelinecheck = off, justification = raggedright, labelsep = period}\
\vspace{-6.0mm}
\caption{The jamming signals(2D time-frequency images), (a)RF, (b)FM, (c)AM, (d)RGPO, (e)VGPO, (f)CM, (g)IS.}
\label{FIG3}
\vspace{-3.0mm}
\end{figure}

 \vspace{-1.0mm}
\subsection{Jamming or Interference Classification and Recognition }\label{S5.2}
 Jamming or interference recognition is very important in radar target detection, tracking, recognition, and anti-jamming or interference suppression tasks. In \cite{interference_classification}, the authors employed SVM classifier to classify six different types of radar-to-radar interference waveforms, including  time-frequency domain signal and range-doppler profiles of different types of interference. As the black-and-white problem of jammer classification in Global Navigation Satellite System (GNSS), SVM and CNN were used to classify six types jammed signals, which achieved an accuracy rate of 94.90\% and 91.36\%, respectively in \cite{jammer_classification}. A method of dense false targets jamming recognition was proposed in\cite{dense_false_targets_jamming_recognition}, which was based on a Gabor time-frequency atomic decomposition and SVM classifier. SVMs were also used to recognize the satellite interference\cite{satellite_interference_recognition} and radio ground-to-air interference signals\cite{radio_interference_recognition}. A robust neural network classifier was employed in \cite{jammer_classification_NN} for classification of frequency-modulated wideband radar jammer signals. In\cite{compound_jamming_recognition}, the neural networks were also developed to recognize compound jamming signals based on features extracted in time domain, frequency domain and fractal dimensions. These signals included additive, multiplicative and convolution signals of typical blanket jamming and deception jamming.

 DL algorithms were also exploited to apply in jamming signals classification and recognition. In\cite{jamming_classification_CNN}, the authors proposed an automatic jamming signal classification method based on CNN model, including audio jamming, narrowband jamming, pulse jamming, sweep jamming and spread spectrum jamming. CNN model, based on time-frequency image as the inputs, was developed in\cite{jamming_recognition_CNN} to classify 9 typical jammings with an accuracy rate of 98.667\% under the jammer-to-noise ratio (JNR) of 0dB-8dB. A LeNet-5 model based on spectrum waterfall was proposed to recognize the jamming patterns in\cite{jamming_recognition_DL}. Similarly, a fine tuning LeNet, with 1D sequences (size of 1*896) as inputs, also employed for 7 kinds of jammings identification in\cite{Chongqing82}, which achieved an accuracy rate of 98\%. In\cite{jamming_identification_DL}, a DL architecture was proposed to identify the jamming factors of electronic information system. The recognition method of four active jamming signal, based on CNN and STFT images as inputs, was proposed in\cite{active_jamming_recognition_CNN}, which achieved an accuracy rate of 99.86\%, including blanket jamming, multiple false target jamming, narrow pulse jamming, and pure signal. The shadow features based on CNN algorithm were proposed for SAR deception jamming recognition in\cite{SAR_deception_jamming_recognition_CNN}. As a multi-user automatic modulation classification task, compound jamming signals recognition based on multi-label CNN model was proposed in \cite{jamming_recognition_MLCNN}. In addition, a jamming prediction method based on DNN and LSTM algorithm was proposed in\cite{jamming_prediction_DNN_LSTM}. The jamming features extracted from PWDs list by DNN and were as the inputs of LSTM for jamming prediction. The AE network consisted of several layers of RNNs was proposed to detect interference signals based on time-frequency images in\cite{interference_recognition_AE_RNN}.

 \begin{figure}[h] 
\vspace{-2.0mm}
\begin{center}
\includegraphics[width=0.8\columnwidth, keepaspectratio]{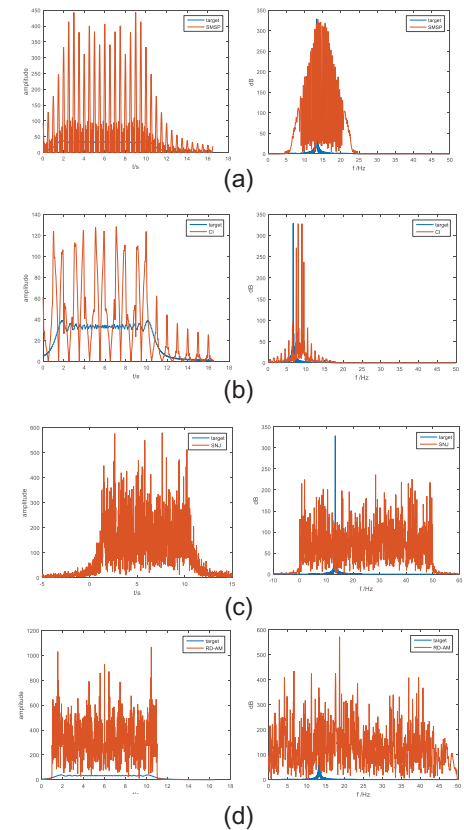}
\end{center}
\captionsetup{font = {footnotesize}, name = {Fig.}, singlelinecheck = off, justification = raggedright, labelsep = period}\
\vspace{-6.0mm}
\caption{The novel jamming signals, time domain (left) and frequency (right), (a)SMSP, (b)CI, (c)SNJ, (d)RD-AM.}
\label{FIG3}
\vspace{-3.0mm}
\end{figure}

  \vspace{-1.0mm}
\subsection{Anti-jamming and Interference Mitigation Strategies}\label{S5.3}

 As a strategy-making process, RL algorithms are usually adopted to make anti-jamming and interference strategies for designing intelligent algorithms. A DQN-based Q-learning algorithm was employed in\cite{DQN_anti_jamming_frequency_hopping} to learn the jammer's strategies to design optimal frequency hopping strategies as the anti-jamming strategy of cognitive radar. Spatial anti-jamming scheme for Internet of satellites  based on deep RL and Stackelberg game was proposed in \cite{spatial_anti_jamming_game_RL}, which regarded routing anti-jamming problem as a hierarchical anti-jamming Stackelberg game. The available routing subsets for fast anti-jamming decision-making were determined by deep RL algorithm to meet high dynamics caused by the unknown interrupts and the unknown congestion.

 As for interference mitigation, a decentralized spectrum allocation strategy was developed in\cite{RL_LSTM_anti_mutual_interference_automotive_radars}, which was based on RL and LSTM model to avoid mutual interference among automotive radars. LSTM was used to aggregate the radar's observations for obtaining more information contributed to RL algorithm. Similarly, a GRU-based RNN algorithm was used for interference mitigation of automotive radar in\cite{GRU_anti_mutual_interference_automotive_radars}.

\section{Other ML-based RSP-related Research}\label{S6}

 In addition to the applications detailed in sections III-V, there are some other that are worth  , such as radar waveform optimization design by RL \cite{Wang296,end_to_end_learning_waveform_detection}, radar spectrum allocation \cite{RL_LSTM_anti_mutual_interference_automotive_radars,Naparstek299,Faganello300}, CEW \cite{You301}, cognitive radar detection \cite{Metcalf302}, antenna array selection via DL \cite{Elbir303}, and moving target indication (MTI) using CNN \cite{Liu304}.

 Compared to DL, RL performs well when used for cognitive decision-making. Therefore, RL is suitable for strategy-making based RSP and radar system design, such as waveform design, anti-jamming, CEW. An intelligent waveform optimization design based on RL method was developed in\cite{Wang296} for multi-target detection of MIMO radar. The sum of target detection probability in range and azimuth cells was as the reward of the learning agent in each learning iteration, to estimate the target detection range and azimuth information. The optimized weighted vector matrix of the transmitted waveform was as the action space of the learning agent. This novel method can improve the performance in detection probability, compared to all-direction waveform design methods. In addition, an end-to-end learning method for joint design the waveform detector was proposed in \cite{end_to_end_learning_waveform_detection}. Both transmitter and receiver were implemented as feedforward neural networks, while the detector and the transmitted waveform were trained alternately. This algorithm  achieved better robustness in clutter and colored noise scenario than traditional methods.

 In\cite{You301}, the authors applied deep RL in CEW for target searching, which built a 3D simulation CEW environment to address the spatial sparsity, continuous action, and partially observable environment existing in CEW. A method of ML-based adaptive optimal target detection threshold estimation in non-Gaussian clutter environment was proposed in \cite{Metcalf302}, which was effective even when the clutter distribution is unknown. A DL method was used for phased array antenna selection to better estimate direction-of-arrival (DOA) in \cite{Elbir303}, which constructed a CNN as a multi-class classification framework. The network determined a new antenna subarray for each radar echo data in a cognitive operation pattern. A CNN-MTI structure was developed in\cite{Liu304} to overcome the constrains of STAP-MTI, which performed feature extraction and classification directly from airborne radar echo. CNN-MTI has proven more robust compared to traditional STAP-CNN and POLY methods.

\section{Discussion and Future Promising Trends}\label{S7}

 Most of the current researches effort has concentrated on the applications of ML to classification and recognition problems. Nevertheless, ML can be further exploited for different RSP applications, such as target detection and tracking. Moreover, unified architectures for detection, tracking and recognition may be conceived that exploit ML as a common framework. This chapter will firstly put forward the future research directions, i.e., possible promising research topics. Then, the detail research contents related to above research topics will be profoundly discussed.

 \vspace{-1.0mm}
\subsection{Research Topics}\label{S6.1}

 \textbf{1) End-to-End Unified Intelligent Detection Architecture}.

 In \cite{NN_Neyman_Pearson}, the authors certified that the outputs of neural network with MSE or cross entropy as loss function satisfied the Neyman-Pearson detection criterion. Therefore, it is promising to exploit an intelligent end-to-end architecture by taking full use of the general non-linear fitting ability of DNNs for radar target detection. The functions of this scheme include pulse compression, coherent accumulation, and CFAR in an unified end-to-end learning manner. The challengeable research problems include the intelligent CFAR, environment identification (such as noise and clutter background automatic classification\cite{sea_clutter_recognition}) techniques. For example, because of the extensive distributed mapping ability of RNN with attention mechanism, problems such as target sidelobe masking, multi-target interference, and target model mismatch may be solved using RNN-related architecture.

 \textbf{2) Target Detection and Tracking-Unified Intelligent Processing}.

 Building an effective closed loop network of unified target detection and tracking can improve the performance of stable target tracking with clutter in the background. It is important to study on the performance evaluation metrics and parameter optimization techniques of target detection and tracking. For example, it is possible to optimally adjust the detection threshold via prior knowledge-based online learning techniques, which is based on the feedback from target tracking information (such as motion trends, covariance estimation) to target detection units. This operation maybe contribute to track the target flight trajectory and improve the detection probability of subsequent point trajectory for confirmed targets.

\textbf{3) End-to-End Framework of Unified Target Detection, Tracking and Recognition}.

 Based on previous two research, it is promising to study an end-to-end architecture to achieve unified intelligent processing for clutter suppression, radar target detection, tracking, and recognition by ANNs-based multi-task learning. Because of powerful non-linear fitting ability, ANNs have high performance in classification and recognition tasks. According to the targets (valuable targets, clutter or noise background) recognition information, radar can program the optimal tracking route based on the ANNs-based prediction of target flight trajectory. In addition, target detection is a special type of target recognition, therefore, it can assist target detection task. However, it is extremely changeable about how to effectively build integrated signal processing mechanism and detection framework, which can promote each other, uniformly make decision-making.

 \vspace{-1.0mm}
\subsection{Promising Research Contents}\label{S6.1}

 \textbf{1) The Solutions of the Limitation of Dataset}.

 Classification and recognition of radar targets suffers of the typical problem of a small amount of labeled samples. To improve the performance with limited data samples, it is necessary to augment the limited data or design effective learning algorithms with limited data. In addition, in order to reduce the cost of obtaining real data, simulation dataset, closely to simulate real complex electromagnetic environment, is also needed to train DL model by transfer learning pattern.

 \emph{\textbf{Data\ augmentation}} The existing data augmentation methods mainly focus on the manipulation of original data samples, e.g., manual extraction of sub-images, add noise, filtering, and flipping \cite{Wang103}. In addition, a method of generating new data samples with GANs was also used in \cite{Cui101,Ma102}. Practical operational conditions, however, are usually neglected when applying these methods to some extent, which make the new data retain the same characteristics. Environmental conditions are a significant component in the radar echo signal, such as different scattering centers with different illuminating directions, which produce different amplitude and phases in the echo signal. Therefore, data augmentation techniques with consideration of radar practical operational conditions (such as SOCs and EOCs) are required. For example, the combination of electromagnetic scattering computation with dynamic or static measurements may be used to improve the accuracy and robustness of target recognition algorithms. Moreover, exploiting the evaluation metrics of generated data equality to efficiently assist the data generations. In this way, the learning model can learn more discriminative features of unknown targets and improve performances in terms of accuracy and generalization.

 \emph{\textbf{Few/zero\ shot\ learning}} This research direction mainly exploits how to effectively extract discriminative features out of small training samples, to improve accuracy and generalization performances. At present, some achievements in this direction have been obtained, such as the design efficient learning model. For example, a feature fusion framework was presented in\cite{Yu105} based on the Gabor features and information of raw SAR images, to fuse the feature vectors extracted from different layers of the proposed neural network. A TL method was employed in\cite{Huang106} to transfer knowledge learned from sufficient unlabeled SAR scene images into labeled SAR target data. A-ConvNets was proposed in\cite{Wang109} to significantly reduce the number of free parameters and the degree of overfitting with small datasets. Group squeeze excitation sparsely connected CNN was developed in\cite{Huang195} to perform dynamic channel-wise feature recalibration with less model parameters. To balance the feature extraction and anti-overfitting, a CAE-HL-CNN model was proposed in\cite{CAE-HL-CNN} to perform effective learning for the classification of MSTAR targets with small training samples.

 However, this research direction is just at the beginning, and needs to have prolonged insight into it. Based on few/zero-shot learning methods from the DL domain, some works \cite{few-shot-learning-survey,few-shot-learning-optimization,zero-shot-learning-1,zero-shot-learning-2} has been done to design effective learning algorithms to address learning issues with small data samples. Few-shot learning can rapidly generalize to new tasks of limited supervised experience by turning to prior knowledge, which mimics human's ability to acquire knowledge from few examples through generalization and analogy \cite{few-shot-learning-survey}. Zero-shot learning aims to precisely recognize unseen categories through a shared visual-semantic function, which is built on the seen categories and expected to well adapt to unseen categories \cite{zero-shot-learning-2}.

\textbf{2) The Design of Lightweight Algorithms}.

 Since the requirements of real-time signal processing and high sampling rate in RSP domain are quite demanding, a large volume of parameters of DL model is still a severe challenge for real-time optimal training, which results in high storage and computation complexity. Lightweight DL models have been proposed, such as MobileNets (v1-v3)\cite{MobileNet_v1,MobileNet_v2,MobileNet_v3}), ShuffleNets (v1-v2)\cite{ShuffleNet_v1,ShuffleNet_v2}), to be embedded in mobile phones or other portable device. Nevertheless, these models are do not fully meet the requirements of RSP, such as low memory resource and strict latency requirements. Therefore, the design of lightweight models or efficient DNN models \cite{efficient_learning_DNN} is necessary for DL model to be efficiently applied to RSP domain. Research on novel lightweight architecture design and deep model compression and accelerating methods, specialized for RSP, is mandatory for enabling this technology.

 \emph{\textbf{Neural\ architecture\ search (NAS)}} Currently employed architectures in DL have mostly been developed manually by human experts, which is a time-consuming and error-prone process. \emph{NAS} provides a promising solution to alleviate this issue \cite{NAS_survey}, being an automatical architecture design system. NAS includes three steps: \emph{search\ space}, \emph{search\ strategy}, and \emph{performance\ estimation}. The purpose of NAS is typically to find the best architectures from a \emph{search\ space} that can highly achieve predictive performance on unknown data. The application of NAS to identify optimal architectures for RSP is an interesting future trend.

 \emph{\textbf{Deep\ model\ compression\ and\ accelerating methods}}  DNNs have achieved great success in many CV tasks. Computation costs and storage intensity, however, are the main challenges of existing DNN models that hinder their deployment in devices with low memory resources or in applications with strict latency requirements. Deep model compression and accelerating methods have been developed to address these challenges in recent years, including parameter pruning and sharing, low-rank factorization, transferred/compact convolutional filters, and knowledge distillation \cite{model_compression_survey,model_compression_hardware_acceleration}. These methods are still in the early stages, as most of techniques only aim at CNN models and classification tasks. It is critical to extensively develop these compression and accelerating techniques in RSP domain.

\textbf{3) Explainable Machine Learning Algorithms}.

 ML has achieved great success in many domains. Black-box property of the DNN model \cite{black_box}, however, demonstrates a severe challenge in practical applications of ML algorithms, such as medical image precessing domain and bank investment decision making. For example, a special doctor needs to clearly know the model how to make decisions in a explainable manner. When we do some ML-based works, especially DNNs, some questions naturally emerge in our mind. For example, how does the network model work? What does the inner structure of the model do about the inputs? Why does the model have the ability to classify, detect, liked human brain does? These questions are triggered by the black box property of non-interpretable ML models. Therefore, to widely apply ML in practice, explainable artificial intelligence (XAI) is a key factor\cite{XAI,medical_XAI}. Especially, XAI techniques are extremely important in RSP, such as the classification and recognition of valuable targets and recognition of jamming types in situation awareness domain. A military commander needs to clearly understand the process of decision-making of ML models to believe the model, so as to deploy highly effective decision strategies, e.g., anti-jamming countermeasures, weapon deployment strategies in electronic warfare.

 The present published literatures about XAI can be roughly categorized into four classes:

 i) post-hoc explanations, such as local/global proxy models assisting explanation\cite{LIME,global_additive_explanation,R-CNN_Unfolding_Latent_Structures}, visualization of latent presentation\cite{visualization_1,visualization_2,visualization_3}, analysis of attributes for prediction\cite{attributes_analysis_1,attributes_analysis_2};

 ii) transparent/interpretable model design, such as embedding transparent/interpretable model in DNNs\cite{interpretable_model_1,interpretable_model_2,interpretable_model_3}, regularization-based design\cite{interpretable_model_3,regularized_explanation_1,regularized_explanation_2}, disentangled representation learning\cite{disentangled_1,disentangled_2}, attention mechanism based design\cite{attention_mechanism_1,attention_mechanism_2};

 iii) interdisciplinary knowledge embedded, such as information theory\cite{information_bottleneck_1,information_bottleneck_2}, physics theory\cite{physics_1,physics_2}, human brain cognitive science\cite{human_cognitive_1,human_cognitive_2,human_cognitive_3};

 iv) combining symbolism and connectionism\cite{symbolism_connectionism_1,Neural_symbolic}.

 In this section, we will take a brief discussion for combining symbolism and connections as an extremely promising method for XAI. The present ML algorithms are just used in an information sensing (i.e., pattern recognition) domain to large extent, which do not have the abilities of casual reasoning\cite{casual_inference} and interpretability. Therefore, it has many challenges in practical applications, such as the requirement of large training data, overfitting, robustness, adversarial attacks\cite{adversarial_learning_DNN_classification}. The deep model has an excellent ability to learn features with large dataset. These features, however, are usually high dimensional and redundant.

 Good representation of data should be that these features, extracted from the learning model, are low-dimensional, abstract, and discrete\cite{representation_learning}, i.e., concept features \cite{human_cognitive_3}, which are similar to the characteristics of the symbolism learning representation method. Symbolism learning \cite{symbolic_AI} has the ability, through logical reasoning, to produce semantic features, which can be logically understood by human beings. Therefore, it is possible to combine symbolism learning (with the ability of logical reasoning) with connectionism learning (i.e., deep learning with ability of powerful features extraction) to achieve human-level concept learning\cite{human_cognitive_3}. Yoshua Bengio has recently proposed consciousness prior as a suitable tool to bridge the gap between the symbolism and connectionism\cite{consciousness_prior}, which can combine attention mechanism to extract consciousness features from semantic features of RNNs with consciousness prior.

 \textbf{4) Cognitive Waveform Design}.

 As the main situation awareness sensing system in EW, a radar system is primarily responsible for surveillance and tracking of the EW environment, including target detection and recognition, jamming/interference countermeasures, and counter-countermeasures, which are consistent with the missions of EW, i.e., electronic support measure (ESM), electronic countermeasure (ECM), and electronic counter-countermeasure (ECCM) systems\cite{EW}. With the development of the cognitive electronic warfare (CEW) in recent years\cite{You301}, many challenges have emerged that affect radar systems. As a possible solution, cognitive radar (CR) has been proposed in\cite{cognitive_radar_1}, which is an interdiscipline research domain of neuroscience, cognitive science, brain science and RSP\cite{cognitive_radar_2}. Three basic ingredients of CR are i) intelligent signal processing; ii) feedback from receiver to transmitter; and iii) preservation of information content of radar returns\cite{cognitive_radar_1}. The basic concepts of CR mainly focus on knowledge-based adaptive radar\cite{cognitive_radar_3}. With the rapid development of ML, especially DL and RL, CR should have novel promising research contents based on advanced ML algorithms in the future.

 Radar waveform design is one of the significant tasks in the design of radar system. Traditional radar usually transmits only one or few types of waveforms to optimise target detection. As a key task of CR, cognitive waveform optimization design has attracted a lot of attention \cite{optimal_waveform_design}. CR makes full use of the knowledge of the external environment and targets, to design optimal waveforms to optimise the tasks of target detection, anti-jamming/interfenrence, at the conditions of radar constraints, objective optimization principles, and advanced optimization theory. The optimization problem of waveform design, however, is a non-convex, high dimension, and multi-constraint optimization problem, whose global optimal solution is usually difficult to find at low computational costs. ML-based optimization methods may indicate alternative directions to address this challenge. Moreover, the optimization process is an iterative search procedure to find the optimal solution, which can be regarded as a problem of sequence decision-making. RNN and RL are good at sequential data processing and optimal strategy-making, respectively. Therefore, it is possible to combine optimization theory with ML to improve the performance in radar waveform optimization design. Some initial works about this theme have emerged, such as branch-and-bound algorithm of mixed-integer linear programming with ML technologies in\cite{learning_branching,learning_branch,RL_integer_programming}, ML  for combinatorial optimization\cite{combinatorial_optimization,combinatorial_optimization_graphs}, RL for solving the vehicle routing problem\cite{RL_vehicle_routing}, and pointer networks with one-layer RNN\cite{pointer_networks}.

 \textbf{5) Intelligent Anti-jamming}.

 The efficient anti-jamming techniques are concerned with toward challenges in an increasingly complex electromagnetic environment. It is difficult for traditional anti-jamming techniques to face current requirements of modern radar systems equipments. The vision of intelligent anti-jamming methods is increasingly intensive with the rapid development of artificial intelligence. In recent years, a new research wave has advanced in this field based on ML algorithms, such as RL-based anti-jamming or interference\cite{DQN_anti_jamming_frequency_hopping,spatial_anti_jamming_game_RL}. This new direction needs to be deeply exploited to address the existing challenges, including jamming recognition, anti-jamming strategy, and the definition of performance metrics.

 \emph{\textbf{Jamming\ recognition}} This aspect has been deeply discussed in the first three parts, which is similar to radar target recognition tasks. Multi-task, multi-view, multi-scale learning techniques, and attention mechanism learning method should be considered.

 \emph{\textbf{Anti-jamming\ strategy}} The selection of efficient anti-jamming measurements is a decision-making process. Deep RL seems to be a promising research lead when training an agent to automatically select adaptive anti-jamming measures with the assistance of knowledge of external environment and targets.

 \emph{\textbf{Performance Evaluation\ metrics}} Although some RL-based achievements in terms of intelligent anti-jamming have been reached, there is little research done in terms of performance evaluation metrics. This direction is vital to evaluate the performances of anti-jamming techniques, which also can assist to select optimal anti-jamming measures.

\section{Conclusion}\label{S8}

 There is a strong evidence of the extensive development of ML-based RSP algorithms that have found application in several radar-related fields. Some areas seem to be more targeted than others due to the direct application of ML-based techniques and because of the strong interest of many researchers, which is likely driven by strong interests from stakeholders. Particularly radar image processing and relative classification is one area where ML-based algorithms may prove a valid solution to current challenges. In this paper, we have provided a structured and amply commented literature survey, followed by indications about future leads, which may be used by many researchers and practitioners to inspire them and help them progressing with their work in this field.

\begin{appendices}
\section{List of acronyms}

ACF autocorrelation function

AWGN additive white Gaussian noise

ASCs attributed scattering centers

Adaboost adaptive boosting

ANN artificial neural network

AEs autoencoders

ATR automatic target recognition

ARPNs advanced region proposal network

BCRN bidirectional convolution-recurrent network

BM3D blocking matching 3D

CNNs convolutional neural networks

CEW cognitive electronic warfare

CDAE convolutional denoising autoencoder

CV computer vision

COCO common objects in context

CWNN convolutional wavelet neural network

CTFD Cohen's time-frequency distribution

CVDFP cadence velocity diagram frequency profile

CR cognitive radar

CV-CNN complex-value CNN

CFAR constant false alarm rates

CPON class probability output network

CWTFD Choi-Williams time-frequency distribution

ConvLSTM convolutional LSTM

CS  compressive sensing

DT decision tree

Dis-iRBM discriminative infinite RBM

DeAE denoising autoencoder

DRL deep reinforcement learning

DBNs deep belief networks

DLFM dual linear frequency modulation

DNNs deep neural networks

DCGANs Deep convolutional GANs

DQN deep Q network

DCC-CNNs despeckling and classification coupled CNNs

DCFM-CNN dual channel feature mapping CNN

DARPA Defense Advanced Research Projects Agency

DSC  depthwise separable convolution

EW electronic warfare

EL Ensemble Learning

ESM electronic support measure

ECM electronic countermeasure

ECCM electronic counter-countermeasure

ECOs extended operating conditions

EQFM even quadratic frequency modulation

ENN Elman neural network

FCN fully convolutional network

FCBF fast correlation-based filter

FPNs feature pyramid networks

FRN-MSF feature recalibration network with multi-scale spatial features

GNN graphical neural network

GIoU generalized intersection over union

GF-3 Gaofen-3

GRU gated recurrent unit

GBDT gradient boosting decision tree

GANs generative adversarial networks

GPUs graphical processing units

GCN global convolutional network

HAPs high-altitude platforms

HMMs Hidden Markov Models

HR high-resolution

HRRP high resolution range profile

HL hinge loss

IDCNN image despeckling convolutional neural network

InSAR interferometric SAR

IDCNN image despeckling convolutional neural network

ILSVRC ImageNet large-scale visual recognition challenge

ICS iterative censoring scheme

ISAR inverse synthetic aperture radar

IoU  intersection of union

JSDC joint supervised dictionary and classifier

JNR jammer-to-noise ratio

K-NN K-nearest neighbor

KSR kernel sparse representation

LSTM long short-term memory

LIME local interpretable model-agnostic explanations

LSGANs least squares generative adversarial networks

LFM linear frequency modulation

LFO-Net lightweight feature optimizing network

LPI low probability intercept

LFMCW linear frequency modulation continuous wave

LOS  line of sight

ML machine learning

MS-CNN multi-stream CNN

MDP Markov decision process

MSTAR moving and stationary target acquisition and recognition

MP mono-pulse

MLP multi-layer perceptron

MTI moving target indication

MRF Markov random field

MLFM multiple linear frequency modulation

MIMO multi-input and multi-output

MVL multi-view learning

MAP mean average precision

MTL multi-task learning

MSCDAE  modified stacked convolutional denoising auto-encoder

MPDPL  multilayer projective dictionary pair learning

MRDED multi-resolution dense encoder and decoder

NLP natural language processing

NN neural network

NAS neural architecture search

NMS non-maximum suppression

OMMPS orthogonal maximum margin projection subspace

OWN optimized Wishart networ

PRI pulse repetition interval

PRI Pulse repetition interval

PSR probability of successful recognition

PTCNN  probability transition convolutional neural network

PWDs pulse description words

PPCA probabilistic principal component analysis

PCA principal component analysis

PSDNN  patch-sorted deep neural network

PGBN  poisson gamma belief network

PSF  point spread function

RSP radar signal processing

RFMLS radio frequency machine learning system

RRSCR radar radiation sources classification and recognition

RVM relevant vector machine

RL reinforcement learning

RF random forest

R-CNN regional convolutional neural network

RFI radio frequency identification

RMA range migration algorithm

RNNs recurrent neural networks

RS remote sensing

RESISC remote sensing image scene classification

RBF radial basis function

REC radar emitter classification

RBM restricted Boltzmann machine

RVFL random vector functional link

RAD range doppler algorithm

SSP speech signal processing

SFM sinusoidal frequency modulation

SAR synthetic aperture radar

SEI specific emitter identification

SCR signal-to-clutter ratio

STAP-MTI spatial time adaptive processing and motion target indication

SSD shot multiBox detector

STD ship targets detection

SNR signal noise ratio

SE-Net sequeeze-and-excitation network

SVMs support vector machines

SAR-DRN SAR dilated residual network

SCCs sparse coding-based classifiers

SOCs standard operating conditions

SMO sequence minimization optimization

SAE sparse autoencoder

SVD single vector decomposition

SFP spectrogram frequency profile

STFT short time fourier transformation

SPP spatial pyramid pooling

SRDNN superpixel restrained DNN

SBL sparse Bayesian learning

TL transfer learning

TARAN Target-aware recurrent attentional network

TPLBP three patch local binary pattern

t-SNE t-distributed stochastic neighbor embedding

TFD time-frequency distribution

TPOT tree-based pipeline optimization tool

TOA time of arrival

TLM texture level map

TPUs tensor processing units

U-CNN unidimensional convolutional neural network

UAV unmanned aerial vehicle

VAE variational autoencoder

WGAN Wasserstein GAN

WGAN-GP Wasserstein GAN with a gradient penalty

WKCNN weighted kernel CNN

WKM weighted kernel module

XGBoost extreme gradient boosting decision tree

XAI explainable artificial intelligence

YOLO You Only Look Once

\end{appendices}


\begin{thebibliography}{306}

\bibitem{Stimson_Airborne_radar} 
 George W. Stimson, Introduction to airborne radar, second edition, SciTech Publishing, Inc.ISBN: 1-891121-01-4, 1998.

\bibitem{Bassem_radar_system} 
 Bassem R. Mahafza, Radar systems analysis and design using MATLAB, Chapman and Hall/CRC Publishing, 2000.

\bibitem{Richards_RSP} 
 Mark A.Richards, Fundamentals of radar signal processing, second edition, McGraw-Hill Publishing, ISBN 13: 9780071798327, 2014.

\bibitem{Vakin_EW} 
 Sergei A. Vakin, Lev N. Shustov, Robert H. Dunwell, Fundamentals of electronic warfare, Artech House Publishing, Inc.ISBN: 1-58053-052-4, 2001.

\bibitem{signal_information_processing} 
 D. Yu, L. Deng, ``Deep learning and its applications to signal and information processing [Exploratory DSP],''  \emph{IEEE Signal Processing Magazine}, vol. 28, no. 1, pp. 145-154, Jan. 2011. Doi: 10.1109/MSP.2010.939038.

\bibitem{Bishop_PRML} 
 Christopher M. Bishop, Pattern recognition and machine learning, Springer India Publishing, ISBN: 9788132209065, 2013.

\bibitem{Goodfellow_DL} 
 I.Goodfellow, Y. Bengio, A. Courville, Deep learning, MIT Press, ISBN: 9780262035613, 2016.

\bibitem{LeCun_DL} 
 LeCun, Y., Bengio, Y., Hinton, G., ``Deep learning,'' Nature, 521 (7553), pp. 436${-}$444, 2015.

\bibitem{RFMLS_1} 
 ``Radio frequency machine learning systems,'' [online]. Available:
 http://www.darpa.mol/program/radio-frequency-machine-learning-systems. Aug. 2018.

\bibitem{RFMLS_2} 
 ``Cognitive electronic warfare: Radio frequency spectrum meets machine learning,'' [online]. Available:
 http://interactive.aviationtoday.com/cognitive-electronic-warfare
 -radio-frequency-spectrum-meets-machine-learning, Aug. 2018.

\bibitem{RFMLS_3} 
 ``DARPA. Contact to apply machine learning to the radio frequency spectrum,'' [online]. Available:
 http://www.baesystems.com/en-us/article/darpa-contact-to-apply-machine-learning
 -to-the-radio-frequency-spectrum, Nov. 2018.

\bibitem{RFMLS_4} 
 ``The radio frequency spectrum + machine learning = a new wave in radio technology,'' [online]. Available: http://www.darpa.mil/news-events/2017-08-11a, Aug. 2017.

\bibitem{BLADE_1} 
 ``DARPA. Behavior learning for adaptive electronic warfare (BLADE). (DARPA-BAA-10-79) [P/OL],'' [online]. Available: http://www.darpa.mil/i2o/solicit /solicit-closed.asp. Jul. 2010.

\bibitem{BLADE_2} 
 ``DARPA. Notice of intent to award sole contract: Behavioral learning for adaptive electronic warfare (BLADE) phase 3 (DARPA-SN-14-24) [P/OL]. [online] Available: http://www.fbo.gov. Feb. 2014.

\bibitem{ARC} 
 ``DARPA. Adaptive radar countermeasures (ARC). (DARPA-BAA-12-54) [P/OL]. [online] Available: http://www.darpa.mil. Aug. 2012.

\bibitem{Arulkumaran_DR} 
 Arulkumaran K, Deisenroth M P, Brundage M, et al., ``Deep reinforcement learning: A brief survey,'' \emph{IEEE\ Signal\ Processing\ Magazine}, vol. 34, no. 6, pp. 26-38, 2017.

\bibitem{Hatcher_DL} 
 W. G. Hatcher and W. Yu, ``A survey of deep learning: platforms, applications and emerging research trends," \emph{IEEE Access}, vol.6, pp. 24411-24432, 2018.

\bibitem{Tan_tansfer} 
 Tan C, Sun F, Kong T, et al., ``A survey on deep transfer learning,'' in \emph{Proc. International Conference on Artificial Neural Networks}, 2018, PP. 270-279.

\bibitem{Liu_DNN} 
 L. Weibo, W. Zidong, L. Xiaohui, et al., ``A survey of deep neural network architectures and their applications,'' \emph{Neurocomputing}, vol. 234, pp. 11-26, 2017.

\bibitem{GANs} 
 Z. Pan, W. Yu, X. Yi, et al., ``Recent progress on generative adversarial networks (GANs): A survey,'' \emph{IEEE Access}, pp. 36322-36333, 2019.

\bibitem{Litjens_med.} 
 G. J. Litjens, T. Kooi, B.E. Bejnordi, et al., ``A survey on deep learning in medical image analysis,'' \emph{Medical Image Analysis}, pp. 60-88, 2017.

\bibitem{DL_ultrasound_imaging} 
 R. J. G. van Sloun, R. Cohen, Y. C. Eldar, ``Deep learning in ultrasound imaging,'' \emph{Proceedings of the IEEE}, vol. 108, no. 1, pp. 11-29, Jan. 2020. Doi: 10.1109/JPROC.2019.2932116.

\bibitem{Kamilaris_agriculture} 
 A. Kamilaris, F.X. Prenafetaboldu, ``Deep learning in agriculture: A survey,'' \emph{Computers and Electronics in Agriculture}, pp. 70-90, 2018.

\bibitem{Zhang_sentiment} 
 L. Zhang, S. Wang, B. Liu, et al., ``Deep learning for sentiment analysis: A survey,'' \emph{Wiley Interdisciplinary Reviews-Data Mining and Knowledge Discovery}, vol. 8, no. 4, 2018.

\bibitem{capsules_network} 
  Sabour S, Frosst N, Hinton G E, et al, `` Dynamic routing between capsules,'' \emph{ arXiv: Computer Vision and Pattern Recognition}, 2017.

\bibitem{Harry_robotics} 
 A. Harry. S. Pierson, Michael, Gashler, ``Deep learning in robotics: A review of recent research,'' \emph{ArXiv:} 1707.07217, 2017.

\bibitem{DL_RS_1} 
 X. X. Zhu, D. Tuia, L. Mou, et al., ``Deep learning in remote sensing: A comprehensive review and list of resources,'' \emph{IEEE Geoscience and Remote Sensing Magazine}, vol. 5, no. 4, pp. 8-36, 2017.

\bibitem{DL_RS_2} 
 L. Zhang, L.Zhang, B. Du, et al., ``Deep learning for remote sensing data: A technical tutorial on the state of the art,'' \emph{IEEE Geoscience and Remote Sensing Magazine}, vol. 4, no. 2, pp. 22-40, 2016. .

\bibitem{DL_RS_3} 
 J. E. Ball, D. T. Anderson, C. S. Chan, et al. ``Comprehensive survey of deep learning in remote sensing: Theories, tools, and challenges for the community,'' \emph{Journal of Applied Remote Sensing}, vol. 11. no. 4, 2017.

\bibitem{ML_RS} 
 A. E. Maxwell, T. A. Warner, F. Fang, et al., ``Implementation of machine-learning classification in remote sensing: An applied review,'' \emph{International Journal of Remote Sensing}, vol. 39, no. 9, pp. 2784-2817, 2018.

\bibitem{DL_RS_4} 
 L. Ma, Y. Liu, X. Zhang, et al., ``Deep learning in remote sensing applications: A meta-analysis and review,'' \emph{Isprs Journal of Photogrammetry and Remote Sensing}, pp. 166-177, 2019.

\bibitem{scene_classify_RS} 
 G. Cheng, J. Han, X. Lu, et al., ``Remote sensing image scene classification: Benchmark and state of the art,'' in \emph{ Proc. Computer Vision and Pattern Recognition, (CVPR)}, 2017, pp. 1865-1883.

\bibitem{AEs_SAR_RS} 
 G. Dong, G. Liao, H. Liu, et al., ``A review of the autoencoder and its variants: A comparative perspective from target recognition in synthetic-aperture radar images,'' \emph{IEEE Geoscience and Remote Sensing Magazine}, vol. 6, no. 3, pp. 44-68, 2018.

\bibitem{Darymli146} 
 K. El-Darymli, E. W. Gill, P. Mcguire, et al., ``Automatic target recognition in synthetic aperture radar imagery: A state-of-the-art review,'' \emph{IEEE Access}, vol. 4, pp. 6014-6058, 2016.

\bibitem{vessel_detection_optical} 
 U. Kanjir, H. Greidanus, K. Ostir, ``Vessel detection and classification from spaceborne optical images: A literature survey,''  \emph{Remote Sensing of Environment}, vol. 207, pp. 1-26, 2018. https://doi.org/10.1016/j.rse.2017.12.033.

\bibitem{multi-view} 
 Y. Li, M. Yang, Z. Zhang, et al., ``A survey of multi-view representation learning,'' \emph{IEEE Transactions on Knowledge and Data Engineering}, vol. 31, no. 10, pp. 1863-1883, 2019.

\bibitem{multi-task} 
 S. Ruder, A. Ltd, Dublin, ``An overview of multi-task learning in deep neural network,'' \emph{ArXiv}: 170605089, 2017.

\bibitem{CNN_latest} 
 Z. Li, W. Yang, S. Peng, et al., ``A survey of convolutional neural networks: Analysis, applications, and prospects,'' \emph{ArXiv}: 2004.02806v1, 2020.

\bibitem{efficient_learning_DNN} 
 V. Sze, Y. Chen, T. Yang, J. S. Emer, ``Efficient processing of deep neural networks: A tutorial and survey,'' \emph{Proceedings of the IEEE}, vol. 105, no. 12, pp. 2295-2329, Dec. 2017, doi: 10.1109/JPROC.2017.2761740.

\bibitem{adversarial_learning_DNN_classification} 
 D. J. Miller, Z. Xiang, G. Kesidis, ``Adversarial learning targeting deep neural network classification: A comprehensive review of defenses against attacks,'' \emph{Proceedings of the IEEE}, vol. 108, no. 3, pp. 402-433, March 2020. Doi: 10.1109/JPROC.2020.2970615.

\bibitem{model_compression_hardware_acceleration} 
 B. L. Deng, G. Li, S. Han, L. Shi, Y. Xie, ``Model compression and hardware acceleration for neural networks: A comprehensive survey,'' \emph{Proceedings of the IEEE}, vol. 108, no. 4, pp. 485-532, April 2020. Doi: 10.1109/JPROC.2020.2976475.

\bibitem{obj_detection_1} 
 Z. Zhao, P. Zheng, S. Xu, et al., ``Object detection with deep learning: A review,'' \emph{IEEE Transactions on Neural Networks and Learning Systems}, vol. 30, no. 11, pp. 3212-3232, Nov. 2019.

\bibitem{obj_detection_2} 
 L. Liu, W. Ouyang, X. Wang, et al., Deep learning for generic object detection: A survey,'' \emph{ International Journal of Computer Vision}, vol. 128, no. 2, pp. 261-318, 2020.

\bibitem{SVM} 
 Burges C J, ``A tutorial on support vector machines for pattern recognition,'' \emph{Data Mining and Knowledge Discovery}, vol. 2, no. 2, pp. 121-167, 1998.

\bibitem{kernal_SVM} 
 T. Zhang, ``An introduction to support vector machines and other kernel-based learning methods,'' \emph{ Ai Magazine}, vol. 22, no. 2, pp. 103-104, 2001.

\bibitem{logical_regression} 
 J. Fleiss, B. Levin, M. Paik, ``Logistic Regression,'' 10.1002/0471445428.ch11, 2004.

\bibitem{SMO} 
 T. Long, T. Yingjie, Y. Chunyan, ``Nonparallel support vector regression model and its SMO-type solver,'' \emph{Neural Networks}, vol. 105, pp. 431-446, 2018.

\bibitem{DT} 
 J. R. Quinlan, ``Induction of decision trees,'' \emph{Machine Learning}, vol. 1, no. 1, pp. 81-106,  1986.

\bibitem{C4.5} 
 J. R. Quinlan, ``C4.5: Programs for machine learning,'' Morgan Kauffman, 1993.

\bibitem{CART} 
 W. Loh, ``Classification and regression trees,'' \emph{Wiley Interdisciplinary Reviews-Data Mining and Knowledge Discovery}, vol. 1, no. 1, pp. 14-23, 2011. https://doi.org/10.1002/widm.8.

\bibitem{ensemble_learning} 
 T. G. Dietterich, ``Ensemble methods in machine learning,'' \emph{Multiple Classifier Systems}, pp. 1-15, 2000.

\bibitem{RFs} 
 B. Leo, ``Random Forests,''  \emph{Machine Language}, 2001. https://doi.org/10.1023/A:1010933404324

\bibitem{adaboost_SAR_ATR} 
 Y. Sun, Z. Liu, S. Todorovic et al., ``Adaptive boosting for SAR automatic target recognition," \emph{IEEE Transactions on Aerospace and Electronic Systems}, vol. 43, no. 1, pp. 112-125, Jan. 2007.

\bibitem{Xgboost} 
 C. Tianqi, G. Carlos, ``XGBoost: A scalable tree boosting system,'' \emph{ArXiv}:1603.02754v1, 2016.

\bibitem{gradient_boosting} 
 J. Friedman. ``Greedy function approximation: A gradient boosting machine,'' \emph{Annals of Statistics}, pp. 1189-1232, 2001.

\bibitem{ANN} 
 F. M. Foody, M. K. Arora, ``An evaluation of some factors afecting the accuracy of classification by an artificial neural network,'' \emph{International Journal of Remote Sensing} vol. 18, pp. 799-810, 1997. Doi:10.1080/014311697218764.

\bibitem{KNN} 
 E. Leif Peterson, ``K-nearest neighbor,'' \emph{Scholarpedia}, vol. 4, no. 2, 1883. doi:10.4249/scholarpedia.

\bibitem{Hinton_2006} 
 G. E. Hinton, S. Osindero Y. Teh, ``A fast learning algorithm for deep belief nets,''  \emph{Neural Computation}, vol. 18, no. 7, pp. 1527-1554, Jul. 2006.

\bibitem{Imagenet_2012} 
 A. Krizhevsky, I. Sutskever, G. Hinton, 2012. ``Imagenet classification with deep convolutional neural networks,'' \emph{Advances in Neural Information Processing Systems}, pp. 1097-1105, 2012.

\bibitem{speech_recognition} 
 L. Deng, J. Li, J. Huang, et al., ``Recent advances in deep learning for speech research at Microsoft,'' in \emph{Proc. International conference on acoustics, speech, and signal processing}, 2013, pp.  8604-8608.

\bibitem{objection} 
 C. Szegedy, A. Toshev, D. Erhan, et al., ``Deep neural networks for object detection,'' in \emph{Proc. Neural Information Processing Systems}, 2013, pp. 2553-2561.

\bibitem{Scene_Recognition} 
 B, Zhou, A. Lapedriza, A. Khosla, et al., ``Places: A 10 million image database for scene recognition,'' \emph{IEEE Transactions on Pattern Analysis and Machine Intelligence}, vol. 40, no. 6, pp. 1452-1464,  2018.

\bibitem{Semantic_Segmentation} 
 A. Garciagarcia, S. Ortsescolano, S. Oprea, et al.,  ``A review on deep learning techniques applied to semantic segmentation,'' in \emph{Proc. Computer Vision and Pattern Recognition (CVPR)}, 2017.

\bibitem{Acoustic_Modeling} 
 A. Mohamed, G. E. Dahl, G. E. Hinton, et al., ``Acoustic modeling using deep belief networks,'' \emph{IEEE Transactions on Audio, Speech, and Language Processing}, vol. 20, no. 1, pp. 14-22, 2012.

\bibitem{music_recommendation} 
 A. V., S. Dieleman, B. Schrauwen, et al., ``Deep content-based music recommendation,'' in \emph{Proc.  Neural Information Processing Systems}, 2013, pp. 2643-2651.

\bibitem{Video_analysis_1} 
 Q. Abbas, M. E. Ibrahim, M. A. Jaffar, et al., ``Video scene analysis: An overview and challenges on deep learning algorithms,'' \emph{Multimedia Tools and Applications}, vol. 77, no. 16, pp. 20415-20453, 2018.

\bibitem{Video_analysis_2} 
 L. Wang, D. Sng, ``Deep learning algorithms with applications to video analytics for a smart city: A survey,'' in \emph{Proc. Computer Vision and Pattern Recognition}, 2015.

\bibitem{NLP_1} 
 Y. Xie, L. Le, Y. Zhou, et al., ``Deep learning for natural language processing,'' \emph{Handbook of Statistics}, 2018.

\bibitem{NLP_2} 
 M. M. Lopez, J. K. Kalita, ``Deep Learning applied to NLP,'' \emph{Computation and Language}, 2017.

\bibitem{NLP_3} 
 T. Young, D. Hazarika, S. Poria, et al., ``Recent trends in deep learning based natural language processing,'' \emph{Computation and Language}, 2017.

\bibitem{RBM_1} 
 H. Lee, R. Grosse, R. Ranganath. et al., ``Convolutional deep belief networks for scalable unsupervised learning of hierarchical representations,'' in \emph{Proc. International Conference on Machine Learning (ICML)}, 2009, 609-616.

\bibitem{RBM_2} 
 N. Srivastava, R. Salakhutdinov, ``Multimodal learning with deep boltzmann machines,'' in \emph{Proc. Neural Information Processing Systems}, 2012, pp. 2222-2230.

\bibitem{DBN_1} 
 E. Geoffrey, Hinton, ``Deep belief networks,''. Scholarpedia, vol. 4, no. 5, pp. 5947, 2009.

\bibitem{denoise_AE} 
 M. Ranzato, C. Poultney, S. Chopra, et al., ``Efficient learning of sparse representations with an energy-based model,'' in {Proc. Advances in Neural Information Processing Systems}, Vancouver, Canada, 2007, pp. 1137-1144.

\bibitem{VAE_1} 
  D. Kingma, S. Mohamed, D. Rezende, et al., ``Semisupervised learning with deep generative models,'' in \emph{Proc. Advances in Neural Information Processing Systems}, Montreal, Canada, 2014, pp. 3581-3589.

\bibitem{VAE_2} 
 D. Rezende, S. Mohamed, and D. Wierstra, ``Stochastic backpropagation and approximate inference in deep generative models,'' in \emph{Proc. ACM 31st Int. Conf. Machine Learning}, Jun. 2014, pp. 1278-1286.

\bibitem{contractive_AE} 
 S. Rifai, P. Vincent, X. Muller, X. Glorot, and Y. Bengio, ``Contractive auto-encoders: Explicit invariance during feature extraction,'' in \emph{Proc. ACM 28th Int. Conf. Machine Learning}, Bellevue, Washington, 2011, pp. 833-840.

\bibitem{conv_AE} 
 J. Masci, U. Meier, D. Cirean, and J. Schmidhuber, ``Stacked convolutional auto-encoders for hierarchical feature extraction,'' in \emph{Proc. Int. Conf. Artificial Neural Networks}, 2011, pp. 52-59.

\bibitem{LeNet} 
 B. L. Lecun Y , Bengio Y , et al., ``Gradient-based learning applied to document recognition,'' \emph{IEEE Proceedings}, vol. 86, no. 11, pp. 2278-2324, 1998.

\bibitem{AlexNet} 
 A. Krizhevsky, I. Sutskever, and G. Hinton, ``ImageNet classification with deep convolutional neural networks,'' \emph{Advances in neural information processing systems}, vol. 25, no. 2, 2012.

\bibitem{VGGNet} 
 K. Simonyan, and A. Zisserman, ``Very deep convolutional networks for large-scale image recognition,'' \emph{Computer Science}, 2014.

\bibitem{GoogleNet_v1} 
 Szegedy C, Liu W, Jia Y, et al, ``Going deeper with convolutions,'' in \emph{Proc. Computer Vision and Pattern Recognition}, 2015, pp. 1-9.

\bibitem{GoogleNet_v2} 
 S. Ioffe, and C. Szegedy, ``Batch normalization: Accelerating deep network training by reducing internal covariate shift,''  \emph{ArXiv}: 1502.03167, 2015.

\bibitem{GoogleNet_v3} 
 C. Szegedy, V. Vanhoucke, Ioffe S, et al., ``Rethinking the inception architecture for computer vision,'' in \emph{Proc. Computer Vision and Pattern Recognition}, 2016, pp. 2818-2826.

\bibitem{GoogleNet_v4} 
 C. Szegedy, S. Ioffe, V. Vanhoucke, et al., ``Inception-v4, inception-resNet and the impact of residual connections on learning,'' in \emph{ Proc. National Conference on Artificial Intelligence}, 2016, pp. 4278-4284.

\bibitem{ResNet} 
 K. He, X. Zhang, S. Ren, and J. Sun, ``Deep residual learning for image recognition,'' \emph{ArXiv}: 1512.03385, 2015.

\bibitem{MobileNet_v1} 
 A. G. Howard, M. Zhu, B. Chen, et al, ``Mobilenets: Efficient convolutional neural networks for mobile vision applications,'' \emph{ArXiv}: 1704.04861, 2017.

\bibitem{MobileNet_v2} 
 M. Sandler, A. Howard, M. Zhu, et al, ``Mobilenetv2: Inverted residuals and linear bottlenecks,'' in \emph{Proc. Computer Vision and Pattern Recognition}, 2018, pp. 4510-4520.

\bibitem{MobileNet_v3} 
 M. S. Andrew Howard, Grace Chu, Liang-Chieh Chen, et al., ``Searching for MobileNetV3,'' \emph{ ArXiv}: 1905.02244, 2019.

\bibitem{ShuffleNet_v1} 
 Zhang X, Zhou X, Lin M, et al. ``ShuffleNet: An extremely efficient convolutional neural network for mobile devices,'' in \emph{Proc. Computer Vision and Pattern Recognition}, 2018, pp. 6848-6856.

\bibitem{ShuffleNet_v2} 
 N. Ma, X. Zhang, H. Zheng, et al. ``ShuffleNet V2: Practical guidelines for efficient CNN architecture design,'' in \emph{Proc. European Conference on Computer Vision}, 2018, pp. 122-138.

\bibitem{EfficientNet} 
 T. Mingxing, V. L. Quoc,`` EfficientNet: Rethinking model scaling for convolutional neural networks,'' \emph{ArXiv}: 1905.11946, 2019.

\bibitem{GhostNet} 
 K. Han, Y. Wang, Q. Tian, et al., ``GhostNet: More features from cheap operations,'' \emph{ArXiv}: 1911.11907, 2019.

 \bibitem{LSTM} 
 S. Hochreiter, J. Schmidhuber, ``Long short-term memory,'' \emph{Neural Computation}, vol. 9, no. 8, pp. 1735-1780, 1997.

\bibitem{GRU} 
 M. Ravanelli, P. Brakel, M. Omologo, et al., ``Light gated recurrent units for speech recognition,'' \emph{Audio and Speech Processing}, vol. 2, no. 2, pp. 92-102, 2018.

\bibitem{GAN} 
 I. Goodfellow, J. Pougetabadie, M. Mirza, et al., ``Generative adversarial nets,'' in \emph{Proc. Neural Information Processing Systems}, 2014, pp. 2672-2680.

\bibitem{GAN_overview_1} 
 A. Creswell, T. White, V. Dumoulin, et al., ``Generative adversarial networks: An overview,'' \emph{IEEE Signal Processing Magazine}, vol. 35, no. 1, pp. 53-65, 2018.

\bibitem{GAN_overview_2} 
 Z, Wang, Q, She, T. E. Ward, et al., ``Generative adversarial networks: A survey and taxonomy,'' \emph{ ArXiv}: Learning, 2019.

\bibitem{DC_GAN} 
 A. Radford, L. Metz, S. Chintala, et al., ``Unsupervised representation learning with deep convolutional generative adversarial networks,'' \emph{ArXiv}: Learning, 2015.

\bibitem{WGAN} 
 M. Arjovsky, S. Chintala, and L. Bottou, ``Wasserstein gan,'' \emph{ArXiv}:1701.07875, 2017.

\bibitem{Improve_WGAN} 
 I. Gulrajani, F. Ahmed, M. Arjovsky, et al., ``Improved training of wasserstein GANS,'' in \emph{Proc. Advances Neural Information Processing Systems Conf.}, 2017.

\bibitem{conditional_GAN} 
 M. Mirza, S. Osindero, ``Conditional generative adversarial nets,'' \emph{ArXiv}: Learning, 2014.

\bibitem{cycle_GAN} 
 J. Y. Zhu, T. Park, P. Isola, et al., ``Unpaired image-to-image translation using cycle-consistent adversarial networks,'' in \emph{Proc. International Conference on Computer Vision (ICCV)}, 2017.

\bibitem{conditional_cycleGAN} 
 L. Yongyi, T. Yu-Wing, T. Chi-Keung, ``Attribute-guided face generation using conditional cycleGAN, \emph{ArXiv}: 1705.09966, 2018.

\bibitem{InfoGAN} 
 X. Chen, Y. Duan, R. Houthooft, et al.,  ``InfoGAN: Interpretable representation learning by information maximizing generative adversarial nets,'' \emph{ArXiv}: Learning, 2016.


\bibitem{EW_Spezio16} 
 A. E. Spezio, ``Electronic warfare system,'' \emph{IEEE Trans. Microw. Theory Tech.}, vol. 50, no. 3, pp. 633-644, 2002.

\bibitem{histogram_Zak2} 
 J. Zak, P. Zavorka, ``The identification of radar signals by histogram description leads to cluster analysis processing,'' in \emph{Proc. 15th Int. Radar Symp. (IRS)}, Dresden, Germany, 2014, pp. 1-4.

\bibitem{knowledge_J3} 
 J. Matuszewski and L. Paradowski, ``The knowledge based approach for emitter identification,'' in \emph{Proc. 12th International Conference on Microwaves and Radar}, Krakow, Poland, 1998, pp. 810-814.

\bibitem{K_means_Yang4} 
 Z. Yang, Z. Wu, Z. Yin, et al, ``Hybrid radar emitter recognition based on rough k-means classifier and relevance vector machine,'' \emph{Sensors}, vol. 13, no. 1, pp. 848-864, 2013.

\bibitem{emission_Dudczyk} 
 J. Dudczyk, J. Matuszewski, M. Wnuk, ``Applying the radiated emission to the emitter identification,'' in \emph{Proc. 15th Int. Conf. on Microwaves, Radar and Wireless Communications}, Warsaw, Poland, 2004,  pp. 431-434.

\bibitem{RF_DNA} 
 L. T. Sun, Z. T. Huang, X. Wang, et al, ``Overview of radio frequency fingerprint extraction in specific emitter identification,'' \emph{Journal of Radars}, in press. Doi: 10.12000/JR19115. (in Chinese).

\bibitem{intra_pulse_sparse_representation} 
 J. S. Xie, ``Robust intra-pulse modulation recognition via sparse representation,''  \emph{IEEE International Radar Conference}, 2016, pp. 1-4.

\bibitem{waveform_recognition_auto_image} 
 Z. Huang, Z. Ma, G. Huang, ``Radar waveform recognition based on multiple autocorrelation Images,,'' \emph{IEEE Access}, vol. 7, pp. 98653-98668, 2019. Doi: 10.1109/ACCESS.2019.2930250.

\bibitem{waveform_recognition_CWD} 
 J. Unden, V. Koivunen, ``Automatic radar waveform recognition,'' \emph{IEEE Journal of Selected Topics in Signal Processing}, vol. 1, no. 1, pp. 124-136, 2007.

\bibitem{LPI_recognition_CWD} 
 H. Wang, M. Diao, L. Gao, ``Low probability of intercept radar waveform recognition based on dictionary learning,'' in \emph{Proc. 2018 10th International Conference on Wireless Communications and Signal Processing (WCSP)}, Hangzhou, 2018, pp. 1-6,. Doi: 10.1109/WCSP.2018.8555906.

\bibitem{radar_signal_recognition_CWD} 
 J. Gao, L. Shen, F. Ye, S. Wang, et al., ``Multi-feature radar signal modulation recognition based on improved PSO algorithm,,''  \emph{The Journal of Engineering}, vol. 2019, no. 19, pp. 5588-5592, 2019. Doi: 10.1049/joe.2019.0439.

\bibitem{radar_emitter_recognition_CWD} 
 W. Gongming, C. Shiwen, H. Xueruobai, et al., ``Radar emitter sorting and recognition based on time-frequency image union feature,'' in \emph{Proc. 2019 IEEE 4th International Conference on Signal and Image Processing (ICSIP)}, Wuxi, China, 2019, pp. 165-170. Doi: 10.1109/SIPROCESS.2019.8868524.

\bibitem{LPI_recognition_DT} 
 T. Alrubeaan, K. Albagami, A. Ragheb, et al., ``An investigation of LPI radar waveforms classification in RoF channels,'' \emph{IEEE Access}, vol. 7, pp. 124844-124853, 2019. Doi: 10.1109/ACCESS.2019.2938317.

\bibitem{radar_signal_recognition_DT} 
 V. Iglesias, J. Grajal, P. Royer, et al., ``Real-time low-complexity automatic modulation classifier for pulsed radar signals,''  \emph{IEEE Transactions on Aerospace and Electronic Systems}, vol. 51, no. 1, pp. 108-126, January 2015. Doi: 10.1109/TAES.2014.130183.

\bibitem{radar_signal_recognition_adaboost} 
 J. Guo, P. Ge, W. Jin, et al., ``Radar signal recognition based on FCBF and adaBoost algorithm,'' in \emph{Proc. 2018 37th Chinese Control Conference (CCC)}, Wuhan, 2018, pp. 4185-4190. Doi: 10.23919/ChiCC.2018.8483351.

\bibitem{radar_emitter_recognition_clustering} 
 G. Revillon, A. Mohammad-Djafari, C. Enderli, ``Radar emitters classification and clustering with a scale mixture of normal distributions,'' in \emph{Proc. 2018 IEEE Radar Conference (RadarConf18)}, Oklahoma City, OK, 2018, pp. 1371-1376. Doi: 10.1109/RADAR.2018.8378764.

\bibitem{radar_emitter_recognition_online_clustering} 
 Jun Liu, J. P. Y. Lee, Lingjie Li, et al., ``Online clustering algorithms for radar emitter classification,'' \emph{IEEE Transactions on Pattern Analysis and Machine Intelligence}, vol. 27, no. 8, pp. 1185-1196, Aug. 2005. Doi: 10.1109/TPAMI.2005.166.

\bibitem{non_cooperative_radar_emitter_recognition_online_clustering} 
 J. Sui, Z. Liu, L. Liu, et al., ``Online non-cooperative radar emitter classification from evolving and imbalanced pulse streams,'' \emph{IEEE Sensors Journal}. Doi: 10.1109/JSEN.2020.2981976.

\bibitem{radar_emitter_identification_K-means} 
 Z. Yang, Z. Wu, Z. Yin, et al.,  ``Hybrid radar emitter recognition based on rough k-means classifier and relevance vector machine,'' \emph{Sensors}, vol. 13, no. 1, pp. 848-864, 2013.

\bibitem{radar_emitter_recognition_AF} 
 Y. Zhai, S. Fang, Z. Cheng and X. Ji, ``Radar emitter recognition based on ambiguity function contour plot,"  in \emph{Proc. 2019 IEEE 7th International Conference on Computer Science and Network Technology (ICCSNT)}, Dalian, China, 2019, pp. 232-235. Doi: 10.1109/ICCSNT47585.2019.8962468.

\bibitem{RF_fingerprint_KNN} 
 S. Ur Rehman, K. Sowerby, C. Coghill, ``RF fingerprint extraction from the energy envelope of an instantaneous transient signal,'' in \emph{ 2012 Australian Communications Theory Workshop (AusCTW)}, Wellington, 2012, pp. 90-95. Doi: 10.1109/AusCTW.2012.6164912.

\bibitem{radar_signal_recognition_KNN} 
 Y. Huang, W. Jin, B. Li, et al., ``Automatic modulation recognition of radar signals based on Manhattan distance-based features,'' \emph{IEEE Access}, vol. 7, pp. 41193-41204, 2019. Doi: 10.1109/ACCESS.2019.2907159.

\bibitem{radar_emitter_recognition_KNN} 
 S. Guo, R. E. White, M. Low, ``A comparison study of radar emitter identification based on signal transients,'' in \emph{Proc. 2018 IEEE Radar Conference (RadarConf18)}, Oklahoma City, OK, 2018, pp. 0286-0291. Doi: 10.1109/RADAR.2018.8378572.

\bibitem{radar_emitter_identification_KNN} 
 A. Aubry, A. Bazzoni, V. Carotenuto, et al., ``Cumulants-based Radar Specific Emitter Identification,'' in \emph{Proc. 2011 IEEE International Workshop on Information Forensics and Security}, Iguacu Falls, 2011, pp. 1-6. Doi: 10.1109/WIFS.2011.6123155.

\bibitem{radar_emitter_identification_CWD_CNN} 
 Z. Liu, Y. Shi, Y. Zeng, et al., ``Radar emitter signal detection with convolutional neural network,'' in \emph{Proc. 2019 IEEE 11th International Conference on Advanced Infocomm Technology (ICAIT)}, Jinan, China, 2019, pp. 48-51. Doi: 10.1109/ICAIT.2019.8935926.

\bibitem{intra_pulse_identification_CTFD} 
 Z. Qu, C. Hou, C. Hou, et al.,  ``Radar signal intra-pulse modulation recognition based on convolutional neural network and deep Q-learning network,'' \emph{IEEE Access}, vol,8, pp.49125 - 4913, 2020.

\bibitem{FST_bivariate_image} 
 G. Kong, V. Koivunen, ``Radar waveform recognition using Fourier-based synchrosqueezing transform and CNN,''  in \emph{Proc. 2019 IEEE 8th International Workshop on Computational Advances in Multi-Sensor Adaptive Processing (CAMSAP)}, Le gosier, Guadeloupe, 2019, pp. 664-668. Doi: 10.1109/CAMSAP45676.2019.9022525.

\bibitem{radar_pulse_detection_STFT} 
 E. Yar, M. B. Kocamis, A. Orduyilmaz, ``A complete framework of radar pulse detection and modulation classification for cognitive EW,'' in \emph{Proc. 2019 27th European Signal Processing Conference (EUSIPCO)}, A Coruna, Spain, 2019, pp. 1-5. Doi: 10.23919/EUSIPCO.2019.8903045.

\bibitem{intra_pulse_identification_STFT} 
 F.C. Akyon, Y. K. Alp, G. Gok, et al., ``Deep learning in electronic warfare systems: Automatic intra-pulse modulation recognition,'' in \emph{Proc. 2018 26th Signal Processing and Communications Applications Conference (SIU)}, Izmir, 2018, pp. 1-4. Doi: 10.1109/SIU.2018.8404294.

\bibitem{radar_emitter_identification_energy_cumulant_STFT} 
 X. Wang, G. Huang, Z. Zhou, et al., ``Radar emitter recognition based on the energy cumulant of short time fourier transform and reinforced deep belief network,'' \emph{Sensors}, vol. 18, no. 9, 2018.

\bibitem{radar_emitter_recognition_DBN_LR} 
 H. Li, W. Jing, Y. Bai, ``Radar emitter recognition based on deep learning architecture,''  in \emph{Proc. 2016 CIE International Conference on Radar (RADAR)}, Guangzhou, 2016, pp. 1-5. Doi: 10.1109/RADAR.2016.8059512.

\bibitem{waveform_identification_resnet} 
 X. Qin, X. Zha, J. Huang, et al, ``Radar waveform recognition based on deep residual network,'' in \emph{Proc. 2019 IEEE 8th Joint International Information Technology and Artificial Intelligence Conference (ITAIC)}, Chongqing, China, 2019, pp. 892-896, doi: 10.1109/ITAIC.2019.8785588.

\bibitem{radar_emitter_recognition_IQ} 
 H. P. Khanh Nguyen, V. Long Do, Q. T. Dong, ``A parallel neural network-based scheme for radar emitter recognition,'' in \emph{Proc. 2020 14th International Conference on Ubiquitous Information Management and Communication (IMCOM)}, Taichung, Taiwan, 2020, pp. 1-8. Doi: 10.1109/IMCOM48794.2020.9001727.

\bibitem{automatic_modulation_classification_CNN} 
 F. Wang, C. Yang, S. Huang, et al., ``Automatic modulation classification based on joint feature map and convolutional neural network,,''  \emph{IET Radar, Sonar and Navigation}, vol. 13, no. 6, pp. 998-1003, Jun. 2019. Doi: 10.1049/iet-rsn.2018.5549.

\bibitem{automatic_modulation_classification_AF} 
 A. Dai, H. Zhang, H. Sun, ``Automatic modulation classification using stacked sparse auto-encoders,'' in \emph{Proc. 2016 IEEE 13th International Conference on Signal Processing (ICSP)}, Chengdu, 2016, pp. 248-252. Doi: 10.1109/ICSP.2016.7877834.

\bibitem{radar_waveform_classification_AF} 
 P. Itkin and N. Levanon, ``Ambiguity function based radar waveform classification and unsupervised adaptation using deep CNN models,'' in \emph{Proc. 2019 IEEE International Conference on Microwaves, Antennas, Communications and Electronic Systems (COMCAS)}, Tel-Aviv, Israel, 2019, pp. 1-6. Doi: 10.1109/COMCAS44984.2019.8958242.

\bibitem{radar_emitter_classification_RNN} 
 X. Li, Z. Liu, Z. Huang, et al., ``Radar emitter classification with attention-based multi-RNNs,''  \emph{IEEE Communications Letters}, doi: 10.1109/LCOMM.2020.2995842.

\bibitem{PRI_modulation_recognition_RNN} 
 X. Li, Z. Liu, Z. Huang, ``Attention-based radar PRI modulation recognition with recurrent neural networks,'' \emph{IEEE Access}, vol. 8, pp. 57426-57436, 2020. Doi: 10.1109/ACCESS.2020.2982654.

\bibitem{waveform_identification_GRU} 
 S. A. Shapero, A. B. Dill, B. O. Odelowo, ``Identifying agile waveforms with neural networks,'' in \emph{[Proc. 2018 21st International Conference on Information Fusion (FUSION)}, Cambridge, 2018, pp. 745-752. Doi: 10.23919/ICIF.2018.8455370.

\bibitem{intra_pulse_identification_CNN_LSTM} 
 S. Wei, Q. Qu, H. Su, et al., ``Intra-pulse modulation radar signal recognition based on CLDN network,''  \emph{IET Radar, Sonar and Navigation}, vol. 14, no. 6, pp. 803-810, Jun. 2020. Doi: 10.1049/iet-rsn.2019.0436.

\bibitem{kernel_Shi} 
 Y. Shi, H. Ji, ``Kernel canonical correlation analysis for specific radar emitter identification,'' \emph{Electron. Lett.}, vol. 50, no. 18, pp. 1318-1320, 2014.

\bibitem{Huang} 
 G. Huang, Y. Yuan, X. Wang, et al., ``Specific emitter identification based on nonlinear dynamical characteristics,'' \emph{Can. J. Electr. Comput. Eng.}, vol. 39, no. 1, pp. 34-41, 2016.

\bibitem{Dudczyk} 
 J. Dudczyk, A. Kawalec, ``Identification of emitter sources in the aspect of their fractal features,'' \emph{Bull. Pol. Acad. Sci., Tech. Sci.}, vol. 61, no. 3, pp. 623-628, 2013.

\bibitem{Dudczyk1} 
 J. Dudczyk, ``Radar emission sources identification based on hierarchical agglomerative clustering for large data sets,'' \emph{J. Sens.}, pp. 1-9, 2016.

\bibitem{Conning} 
 M. Conning, F. Potgieter, ``Analysis of measured radar data for Specific Emitter Identification,'' in \emph{Proc. IEEE radar conference}, pp. 35-38, 2010.

\bibitem{Roe11} 
 J. Roe, ``A review of applications of artificial intelligence techniques to naval ESM signal processing,'' in \emph{Proc. IEE Colloquium on Application of Artificial Intelligence Techniques to Signal Processing}, London, UK,  1989, pp. 5/1-5/5.

\bibitem{Zhang} 
 M. Zhang, M. Diao, L. Gao, and L. Liu, ``Neural networks for radar waveform recognition,'' \emph{ Symmetry}, vol. 9, no. 5, pp. 75, 2017.

\bibitem{Qu} 
 Z. Qu, W. Wang, C. Hou and C. Hou, ``Radar signal intra-Pulse modulation recognition based on convolutional denoising autoencoder and deep convolutional neural network,'' \emph{IEEE Access}, vol. 7, pp. 112339-112347, 2019.


\bibitem{Matuszewski} 
 J. Matuszewski, A. Kawalec, ``Knowledge-based signal processing for radar identification,'' in \emph{Proc. International Conference on Modern Problems of Radio Engineering, Telecommunications and Computer Science}, 2008, pp. 302-305.

\bibitem{emitter_database} 
 J. Dudczyk, J. Matuszewski, M. Wnuk, ``Applying the relational modelling and knowledge-based technologies to the emitter database design,'' in \emph{Proc. 14th IEEE International Conference on Microwaves, Radar and Wireless Communications}, MIKON, 2002, pp. 172-175.

\bibitem{Wang16} 
 X. Wang, Y. Xiong, B. Tang, et al., ``An approach to the modulation recognition of MIMO radar signals,'' \emph{Eurasip Journal on Wireless Communications and Networking}, 2013. Doi: 10.1186/1687-1499-2013-66.

\bibitem{Nguyen17} 
 L. H. Nguyen, T. D. Tran. ``RFI-radar signal separation via simultaneous low-rank and sparse recovery,'' in \emph{Proc. IEEE radar conference}, 2016, pp. 1-5.

\bibitem{Zhu18} 
 W. Zhu, M. Li, C. Zeng, et al.,  ``Research on Online Learning of Radar Emitter Recognition Based on Hull Vector,'' in \emph{Proc. IEEE International Conference on Data Science in Cyberspace}, 2017, pp.  328-332.

\bibitem{Conning19} 
 M. Conning, F. Potgieter. ``Analysis of measured radar data for Specific Emitter Identification,'' in \emph{Proc. IEEE radar conference}, 2010, pp. 35-38.

\bibitem{Helicopter20} 
 S. T Nguyen, S. Kodituwakku, R. Melino et al., ``Signal separation of helicopter radar returns using wavelet-based sparse signal optimization,'' 2016.

\bibitem{Li21} 
 Li J, Ying Y. ``Radar signal recognition algorithm based on entropy theory,'' in \emph{Proc.  International Conference on Systems}, 2014, pp. 718-723.

\bibitem{Kamgar22} 
 K.P.Behrooz, K.P.Behzad, J.Sciortino, ``Automatic Data Sorting Using Neural Network Techniques,''  DTIC Document, 1996.

\bibitem{Ataa23} 
 A. W. Ataa, S. N. Abdullah, ``Deinterleaving of radar signals and PRF identification algorithms,'' \emph{IET Radar Sonar and Navigation}, vol. 1, no. 5, pp. 340-347, 2007.

\bibitem{Liu24} 
 Z. Liu, P. S. Yu. ``Classification, denoising, and deinterleaving of pulse streams with recurrent neural networks,'' \emph{IEEE Transactions on Aerospace and Electronic Systems}, vol. 55, no. 4, pp. 1624-1639, 2019.

\bibitem{PRI_deinterleaving} 
 Z. Ge, X. Sun, W. Ren, et al, ``Improved algorithm of radar pulse repetition interval deinterleaving based on pulse correlation,'' \emph{IEEE Access}, vol. 7, pp. 30126-30134, 2019. Doi: 10.1109/ACCESS.2019.2901013.

\bibitem{TOA_deinterleaving} 
 O. Torun, M. B. Kocamis, H. Abaci, et al, ``Deinterleaving of radar signals with stagger PRI and dwell-switch PRI types,'' \emph{ 2017 25th Signal Processing and Communications Applications Conference (SIU)}, Antalya, 2017, pp. 1-4, doi: 10.1109/SIU.2017.7960383.

\bibitem{deinterleaving_cluster_SVM} 
 H. Mu, J. Gu and Y. Zhao, ``A deinterleaving method for mixed pulse signals in complex electromagnetic environment,'' \emph{ 2019 International Conference on Control, Automation and Information Sciences (ICCAIS)}, Chengdu, China, 2019, pp. 1-4. Doi: 10.1109/ICCAIS46528.2019.9074573.

\bibitem{MLP_deinterleaving} 
 A. Erdogan, K. George, ``Deinterleaving radar pulse train using neural networks,'' \emph{2019 IEEE International Conference on Computational Science and Engineering (CSE) and IEEE International Conference on Embedded and Ubiquitous Computing (EUC)}, New York, NY, USA, 2019, pp. 141-147. Doi: 10.1109/CSE/EUC.2019.00036.

\bibitem{AEs_denoising} 
 X. Li, Z. Liu and Z. Huang, "Denoising of Radar Pulse Streams With Autoencoders," in IEEE Communications Letters, vol. 24, no. 4, pp. 797-801, April 2020, doi: 10.1109/LCOMM.2020.2967365.

\bibitem{Jordanov25} 
 I. Jordanov, N. Petrov, A. Petrozziello , et al., ``Supervised radar signal classification,'' in \emph{Proc. International Joint Conference on Neural Network}, 2016, pp. 1464-1471.

\bibitem{Anjaneyulu26} 
 L. Anjaneyulu, N. S. Murthy, N. V. Sarma, et al., ``Identification of LPI radar signals by higher order spectra and neural network techniques,'' in \emph{ Proc. International Conference on Electronic Design}, 2008, pp. 1-6.

\bibitem{Azimisadjadi27} 
 M. R. Azimisadjadi, D. Yao, Q. Huang, et al., ``Underwater target classification using wavelet packets and neural networks,'' \emph{IEEE Transactions on Neural Networks}, vol. 11, no. 3, pp. 784-794, 2000.

\bibitem{Yin28} 
 Z. Yin, W. Yang, Z. Yang, et al., ``A study on radar emitter recognition based on spds neural network,'' \emph{Information Technology Journal}, vol. 10, no. 4, pp. 883-888, 2011.

\bibitem{Zhang29} 
 Z. Zhang, X. Guan, Y. He, et al., ``Study on radar emitter recognition signal based on rough sets and RBF neural network,'' in \emph{Proc. International Conference on Machine Learning and Cybernetics}, 2009, pp. 1225-1230.

\bibitem{Shieh30} 
 C. Shieh, C. Lin, ``A vector neural network for emitter identification,'' \emph{IEEE Transactions on Antennas and Propagation}, vol. 50, no. 8, pp. 1120-1127, 2002.

\bibitem{Granger31} 
 E. Granger, M. A. Rubin, S. Grossberg, et al., ``A what-and-where fusion neural network for recognition and tracking of multiple radar emitters,'' \emph{Neural Networks}, vol. 14, no. 3, pp. 325-344, 2001.

\bibitem{waveform32} 
 D. R Pape, J. A Anderson, J. A Carter, et al., ``Advanced signal waveform classifier,'' In \emph{Proc.  of the 8th Optical Technology for Microwave Applications}, 1997, pp. 162-169.

\bibitem{Liu33} 
 H. Liu, Z. Liu, W. Jiang, et al., ``Approach based on combination of vector neural networks for emitter identification,'' \emph{IET Signal Processing}, vol. 4, no. 2, pp. 137-148, 2010.

\bibitem{Matuszewski34} 
 J. Matuszewski. ``Applying the decision trees to radar targets recognition,'' in \emph{Proc.  International radar symposium}, 2010, pp. 1-4.

\bibitem{Chen35} 
 W. Chen, K. Fu, J. Zuo, et al., ``Radar emitter classification for large data set based on weighted-xgboost,'' \emph{IET Radar Sonar and Navigation}, vol. 11, no. 8, pp. 1203-1207, 2017.

\bibitem{Rehman36} 
 S. U. Rehman , K. W. Sowerby, C. Coghill, et al., ``RF fingerprint extraction from the energy envelope of an instantaneous transient signal,''  in \emph{Proc. Australian communications theory workshop}, 2002, pp. 90-95.

\bibitem{Petrov37} 
 N. Petrov, I. Jordanov, J. Roe, et al.,  ``Radar emitter signals recognition and classification with feedforward networks,''  \emph{Procedia Computer Science}, pp. 1192-1200, 2013.

\bibitem{Sun38} 
 J. Sun, G. Xu, W. Ren, et al., ``Radar emitter classification based on unidimensional convolutional neural network,''  \emph{IET Radar Sonar and Navigation}, vol. 12, no. 8, pp. 862-867, 2018.

\bibitem{Matuszewski39} 
 J. Matuszewski. ``The Specific Radar Signature in Electronic Recognition System,''  \emph{Przeglad Elektrotechniczny}, 2013.

\bibitem{Zeng40} 
 D. Zeng, X. Zeng, G. Lu,et al.,``Automatic modulation classification of radar signals using the generalised time-frequency representation of zhao, atlas and marks,'' \emph{IET radar, sonar and navigation}, vol. 5, no. 4, pp. 507-516, 2011.

\bibitem{Zeng41} 
 D. Zeng, X. Zeng, H. Cheng, et al., ``Automatic modulation classification of radar signals using the rihaczek distribution and hough transform,'' \emph{IET Radar, Sonar and Navigation}, vol. 6, no. 5, pp. 322-331, 2012.

\bibitem{Mingqiu42} 
 R. Mingqiu, C. Jinyan, and Z. Yuanqing, ``Classification of radar signals using time-frequency transforms and fuzzy clustering,'' in \emph{Proc. IEEE International Conference on In Microwave and Millimeter Wave Technology (ICMMT)}, 2010, pp. 2067-2070.

\bibitem{Konopko43} 
 K. Konopko, Y. P. Grishin, and D. Janczak, ``Radar signal recognition based on time-frequency representations and multidimensional probability density function estimator,'' in \emph{Proc. IEEE in Signal Processing Symposium (SPSympo)}, 2015, pp. 1-6.

\bibitem{Rigling44} 
 B. D. Rigling, C. Roush, ``Acf-based classification of phase modulated waveforms,'' in \emph{Proc. IEEE in Radar Conference}, 2010, pp. 287-291.

\bibitem{Wang45} 
 C. Wang, H. Gao, and X. Zhang, ``Radar signal classificcation based on auto-correlation function and directed graphical model,'' in \emph{Proc. IEEE International Conference on Signal Processing, Communications and Computing (ICSPCC)}, 2016, pp. 1-4.

\bibitem{Yu46} 
 Z. Yu, C. Chen, and W. Jin, ``Radar signal automatic classification based on pca,'' in \emph{Proc. WRI Global Congress on Intelligent Systems}, 2009, pp. 216-220.

\bibitem{zhang47} 
 G. Zhang, W. Jin, L. Hu, et al., ``Radar emitter signal recognition based on support vector machines,'' in \emph{Proc. International Conference on Control, Automation, Robotics and Vision}, 2004, pp. 826-831.

\bibitem{Yuan48} 
 Y. Yuan, Z. Huang, Z. Sha, et al., ``Spenific emitter identification based on transient energy tracjectory,'' \emph{Progress in Electromagnetics Research C}, vol. 44, no. 44, pp. 67-82, 2013.

\bibitem{Xu49} 
 J. Xu, M. He, J. Han and C. Chen, ``A comprehensive estimation method for kernel function of radar signal classifier,'' \emph{In Chinese Journal of Electronics}, vol. 24, no. 1, pp. 218-222, Jan. 2015.

\bibitem{Zhu50} 
 W. Zhu, M. Li, C. Zeng, et al., ``Research on online learning of radar emitter recognition based on hull vector,'' in \emph{Proc. IEEE International Conference on Data Science in Cyberspace}, 2017, pp. 328-332.

\bibitem{Liu51} 
 S. Liu, X. Yan, P. Li, et al., ``Radar emitter recognition based on SIFT position and scale features,''  \emph{IEEE Transactions on Circuits and Systems Ii-express Briefs}, vol. 65, no. 12, pp. 2062-2066, 2018.

\bibitem{Shieh52} 
 C. Shieh, C. Lin. ``A vector neural network for emitter identification,'' \emph{IEEE Transactions on Antennas and Propagation}, vol. 50, no. 8, pp. 1120-1127, 2002.

\bibitem{Liu53} 
 H. Liu, Z. Liu, W. Jiang, et al., ``Approach based on combination of vector neural networks for emitter identification,'', \emph{IET Signal Processing}, vol. 4, no. 2, pp. 137-148, 2010.

\bibitem{Yin54} 
 Z. Yin, W. Yang, Z. Yang, et al., ``A study on radar emitter recognition based on spds neural network,'' \emph{Information Technology Journal}, vol. 10, no. 4, pp. 883-888, 2011.

\bibitem{Zhang55} 
 Z. Zhang, X. Guan, Y. He, et al., ``Study on radar emitter recognition signal based on rough sets and RBF neural network,'' in \emph{Proc. International conference on machine learning and cybernetics}, 2009, pp. 1225-1230.

\bibitem{Granger56} 
 E. Granger, M. A. Rubin, S. Grossberg, et al., ``A what-and-where fusion neural network for recognition and tracking of multiple radar emitters,'' \emph{Neural Networks}, vol. 14, no. 3, pp. 325-344, 2001.

\bibitem{Wong57} 
 M. D. Wong, A. K. Nandi, ``Automatic digital modulation recognition using artificial neural network and genetic algorithm,'' \emph{Signal Process}. vol. 84, pp. 351-365, 2004.

\bibitem{Matuszewski58} 
 J. Matuszewski,  ``Applying the decision trees to radar targets recognition,'' in \emph{Proc. International radar symposium}, 2010, pp. 1-4.

\bibitem{Kim60} 
 K. Kim, C. M. Spooner, I. Akbar, et al., ``Specific emitter identification for cognitive radio with application to IEEE 802.11,'' in \emph{Proc. Global Communications Conference}, 2008, pp. 1-5.

\bibitem{Yang61} 
 Z. Yang, Z. Wu, Z. Yin, et al., ``Hybrid radar emitter recognition based on rough k-Means classifier and relevance vector machine,'' \emph{Sensors}, vol. 13, no. 1, pp. 848-864, 2013.

\bibitem{Yang62} 
 Z. Yang, W. Qiu, H. Sun, et al., ``Robust radar emitter recognition based on the three-dimensional distribution feature and transfer learning,'' \emph{Sensors}, vol. 16, no. 3, pp. 289-289, 2016.

\bibitem{Pavlyshenko64} 
 B. M. Pavlyshenko, ``Linear, machine learning and probabilistic approaches for time series analysis,''  in \emph{Proc. IEEE First Int. Conf. on Data Stream Mining and Processing (DSMP)}, Lviv, Ukraine, 2016, pp. 377-381.

\bibitem{Kim65} 
 L. S. Kim, H. B. Bae, R. M. Kil, et al., ``Classification of the trained and untrained emitter types based on class probability output networks,'' \emph{Neurocomputing}, pp. 67-75, 2017.

\bibitem{Guo66} 
 J. Guo, P. Ge, W. Jin, et al., ``Radar signal recognition based on FCBF and adaBoost algorithm,'' in \emph{Proc. Chinese control conference}, 2018.

\bibitem{Sun67} 
 J. Sun, G. Xu, W. Ren, et al., ``Radar emitter classification based on unidimensional convolutional neural network,'' \emph{IET Radar Sonar and Navigation}, vol. 12, no. 8, pp. 862-867, 2018.


\bibitem{Selim68} 
 A. Selim, F. Paisana, J. A. Arokkiam, et al., ``Spectrum monitoring for radar bands using deep convolutional neural networks,'' in \emph{Proc. IEEE Global Communications Conference}, Singapore, 2017, pp. 1-6.

\bibitem{Chongqing69} 
 ``Radar signal recognition based on squeeze-and-excitation networks,'' in \emph{Proc. 2019 IEEE Interference on Signal, Data and Signal Processing (ISDSP)}, Chongqing, 2019.

\bibitem{PRI_modulation_recognition} 
 S. J. Wei, Q. Z. Qu, Y. Wu, et al., ``PRI modulation recognition based on squeeze-and-excitation networks,'' \emph{IEEE Communications Letters}, vol. 24, no. 5, pp. 1047-1051, May 2020. Doi: 10.1109/LCOMM.2020.2970397.

\bibitem{ACSE_PRI_autocorrelation} 
 Q. Qu, S. Wei, Y. Wu, M. Wang, ``ACSE networks and autocorrelation features for PRI modulation recognition,'' \emph{IEEE Communications Letters}, 2020. Doi: 10.1109/LCOMM.2020.2992266.

\bibitem{multi_branch_ACSE_radar_signal_recognition} 
 S. Wei, Q. Qu, M. Wang, et al., ``Automatic modulation recognition for radar signals via multi-branch ACSE networks,'' \emph{IEEE Access}, vol. 8, pp. 94923-94935, 2020. Doi: 10.1109/ACCESS.2020.2995203.

\bibitem{Liu70} %
 Z. Liu, P. S. Yu, ``Classification, Denoising, and deinterleaving of pulse treams with recurrent neural networks,'' \emph{IEEE Transactions on Aerospace and Electronic Systems}, vol. 55, no. 4, pp. 1624-1639 2019.

\bibitem{Ming71} 
 M. D. Ming Zhang, L. Guo, ``Convolutional neural networks for automatic cognitive radio waveform recognition,'' \emph{IEEE Access}, vol. 5, pp. 11074-11082, 2017.

\bibitem{Zhang72} 
 M. Zhang, M. Diao, L. Gao, L. Liu, ``Neural networks for radar waveform recognition,''  \emph{Symmetry}, vol. 9, no. 5, pp. 75, 2017.

\bibitem{Time_frequency73} 
 Radar Emitter Signal Detection with Time-frequency image processing and Convolutional Neural Network

\bibitem{Intra_Pulse74} 
 F. C. Akyon, Y. K. Alp, G. Gok, et al., ``Classification of intra-pulse modulation of radar signals by feature fusion based convolutional neural networks,'' in \emph{Proc. European Signal Processing conference}, 2018, PP. 2290-2294.

\bibitem{Qu75} 
 Z. Qu, X. Mao, and Z. Deng, ``Radar signal intra-pulse modulation recognition based on convolutional neural network,'' \emph{IEEE Access}, vol. 6, pp. 43874-43884, 2018.

\bibitem{Wang76} 
 Z. Qu, W. Wang, C. Hou and C. Hou, ``Radar signal intra-pulse modulation recognition based on convolutional denoising autoencoder and deep convolutional neural network,'' \emph{IEEE Access}, vol. 7, pp. 112339-112347, 2019.

\bibitem{waveform_recognition_CDAE} 
 Z. Liu, X. Mao, Z. Deng, et al., ``Radar signal waveform recognition based on convolutional denoising autoencoder,'' in \emph{Proc. International Conference on Communications}, 2018, pp. 752-761.

\bibitem{waveform_recognition_DCDAE} 
 Z. Zhou, G. Huang, H. Chen, et al., ``Automatic radar waveform recognition based on deep convolutional denoising auto-encoders,'' \emph{Circuits Systems and Signal Processing}, vol. 37, no. 9, pp. 4034-4048, 2018.

\bibitem{LPI_waveform_recognition_CNN_TPOT} 
 J. Wan, X. Yu, Q. Guo, et al., ``LPI radar waveform recognition based on CNN and TPOT,'' \emph{Symmetry}, vol. 11, no. 5, 2019.

\bibitem{waveform_recognition_CNN_LSTM} 
 Q. Wang, P. Du, J. Yang, et al., ``Transferred deep learning based waveform recognition for cognitive passive radar,'' \emph{Signal Processing}, pp. 259-267, 2019.

\bibitem{waveform_recognition_CNN_SVM} 
 Q. Guo, X. Yu, G. Ruan, et al., ``LPI radar waveform recognition based on deep convolutional neural network transfer learning,''  \emph{Symmetry}, vol. 11, no. 4, 2019.

\bibitem{radar_signal_modulation_recognition_CNN_transfer_learning} 
 D. Li, R. Yang, X. Li and S. Zhu, ``Radar signal modulation recognition based on deep joint learning,'' \emph{IEEE Access}, vol. 8, pp. 48515-48528, 2020. Doi: 10.1109/ACCESS.2020.2978875.

\bibitem{Wang77} 
 X. Wang, G. Huang, Z. Zhou, et al.,  ``Radar emitter recognition based on the short time fourier transform and convolutional neural networks,'' in \emph{Proc. International Congress on Image and Signal Processing}, 2017, pp. 1-5.

\bibitem{Kong78} 
 M. Kong, J. Zhang, W. Liu, G. Zhang, ``Radar emitter identification based on deep convolutional neural network,'' in \emph{Proc. 2018 International Conference on Control, Automation and Information Sciences (ICCAIS)}, Hangzhou, 2018, pp. 309-314.

\bibitem{Ye79} 
 W.Q. Ye and C. Peng, ``Recognition algorithm of emitter signals based on pca + cnn,'' in \emph{ Proc. 2018 IEEE 3rd Advanced Information Technology, IEEE Electronic and Automation Control Conference (IAEAC)}, 2018, pp. 2410-2414.

\bibitem{Zhang80} 
 W. Zhang, P. Ge, W. Jin, et al.,  ``Radar signal recognition based on TPOT and LIME,'' in \emph{Proc. Chinese Control Conference}, 2018.

\bibitem{Zhou81} 
 Z. Zhou, J. Zhang, T. Zhang, et al., ``A novel algorithm of radar emitter identification based convolutional neural network and random vector functional-link,'' in \emph{Proc. International Conference on Natural Computation}, 2019, PP. 833-842.

\bibitem{Jeong83} 
 C. M. Jeong, Y. G. Jung, S. J. Lee, et al., ``Deep belief networks based radar signal classification system,''  \emph{Journal of Ambient Intelligence and Humanized Computing}, pp. 1-13, 2018.

\bibitem{Liu84} 
 S. Liu, X. Yan, P. Li, et al.,  ``Radar emitter recognition based on SIFT position and scale features,'' \emph{IEEE Transactions on Circuits and Systems Ii-express Briefs}, vol. 65, no. 12, pp. 2062-2066, 2018.

\bibitem{Cao85} 
 R. Cao, J. Cao, J. Mei, et al.,  ``Radar emitter identification with bispectrum and hierarchical extreme learning machine,'' \emph{Multimedia Tools and Applications}, vol. 78, no. 20, pp. 28953-28970, 2019.

\bibitem{Li86} 
 X. Li, Z. Huang, F. Wang, X. Wang, T. Liu, ``Toward convolutional neural networks on pulse repetition interval modulation recognition,'' \emph{IEEE Communications Letters}, vol. 22, no. 11, pp. 2286-2289, Nov. 2018.

\bibitem{Keydel87} 
 E. R. Keydel, S. W. Lee, J. T. Moore, ``MSTAR extended operating conditions: A tutorial,'' in \emph{Proc. Aerospace/Defense Sensing and Controls}, 1996, pp. 228-242.

\bibitem{Wang88} 
 H. Wang, S. Chen, F. Xu and Y. Jin, ``Application of deep-learning algorithms to MSTAR data,'' in \emph{Proc. 2015 IEEE International Geoscience and Remote Sensing Symposium (IGARSS)}, Milan, 2015, pp. 3743-3745.

\bibitem{Zhou90} 
 Y. Zhou, H. P. Wang, F. Xu, et al., ``Polarimetric SAR image classification using deep convolutional neural networks,'' \emph{IEEE Geoscience and Remote Sensing Letters}, vol. 13, no. 12, pp. 1935-1939,  2016. Doi: 10.1109/LGRS.2016.2618840.

\bibitem{Feng91} 
 F. Xu, H. P. Wang, Y. Q. Jin,  ``Deep learning as applied in SAR target recognition and Terrain classification,''. \emph{Journal of Radars}, vol. 6, no. 2, pp. 136-148, 2017. Doi: 10.12000/JR16130.(in Chinese)

\bibitem{Zhang92} 
 Z. M. Zhang, H. P. Wang, F. Xu, et al.,  ``Complex-valued convolutional neural network and its application in polarimetric SAR image classification,'' \emph{IEEE Transactions on Geoscience and Remote Sensing}, vol. 55, no. 12, pp. 7177-7188, 2017. Doi: 10.1109/TGRS.2017.2743222.

\bibitem{Sterling93} 
 T. Sterling,  D. S. Katz,  L. Bergman,  ``High  Performance  computing  systems  for  autonomous spaceborne  missions,'' \emph{International  Journal  of  High  Performance  Computing  Applications}, vol. 15, no. 3, pp. 282-296, 2001.

\bibitem{html94} 
 http://earth.eo.esa.int/polsarpro/default.html

\bibitem{MSTAR95} 
 U. A. Force, ``MSTAR overview,'' [online]. Available: http://tinyurl.com/pc8nh3s, 2013.

\bibitem{Xing96} 
 X. Xing, K. Ji, H. Zou, W. Chen, J. Sun, ``Ship classification in TerraSAR-X images with feature space based sparse representation,'' \emph{ IEEE Geosci. Remote Sens. Lett.}, vol. 10, no. 6, pp. 1562-1566, Nov. 2013, [online] Available: http://dx.doi.org/10.1109/LGRS.2013.2262073.

\bibitem{Li97} 
 J. Li, C. Qu, J. Shao, ``Ship detection in SAR images based on an improved faster R-CNN,'' in \emph{Proc. 2017 SAR in Big Data Era: Models, Methods and Applications (BIGSARDATA)}, Beijing, 2017, pp. 1-6.

\bibitem{Wang98} 
 Y. Y. Wang, C. Wang, H. Zhang, et al., ``A SAR dataset of ship detection for deep learning under complex backgrounds,'' \emph{Remote Sensing}, vol. 11, no. 7, 2019. Doi: 10.3390/rs11070765.

\bibitem{Sun99} 
 S. Xian, Z. R. Wang, Y. R. Sun, et al.,  ``AIR-SARShip-1.0: High resolution SAR ship detection dataset,'' \emph{Journal of Radars}, doi: 10.12000/JR19097. (in Chinese)

\bibitem{HRSID_ship_detection_dataset} 
 S. J. Wei, X. F. Zeng, Q. Z. Qu, et al., ``HRSID: A high-resolution SAR images dataset for ship detection and instance segmentation,'' \emph{IEEE Access}, doi: 10.1109/ACCESS.2020.300586.

\bibitem{Cui101} 
 Z. Y. Cui, M. R. Zhang, Z. J. Cao, et al., ``Image data augmentation for SAR sensor via generative adversarial nets,'' \emph{IEEE Access}, vol. 7, pp. 42255-42268, 2019.

\bibitem{Ma102} 
 Y. Ma, L. Yan, W. Y. Zhang, S. Yan, ``SAR target recognition based on transfer learning and data augmentation with LSGANs,'' in \emph{Proc. Chinese Automation Congress (CAC) 2019}, 2019, pp.  2334-2337.

\bibitem{Wang103} 
 Z. Wang, L. Du, J. Mao, B. Liu, et al., ``SAR target detection based on SSD with data augmentation and transfer learning,'' \emph{IEEE Geoscience and Remote Sensing Letters}, vol. 16, no. 1, pp. 150-154, Jan. 2019.

\bibitem{Lv104} 
 J. Y. Lv, Y. Liu, ``Data augmentation based on attributed scattering centers to train robust CNN for SAR ATR,'' \emph{IEEE Access}, vol. 7, pp. 25459-25473, 2019.

\bibitem{Yu105} 
 Q. Z. Yu, H. B. Hu, X. P. et al., ``High-performance SAR automatic target recognition under limited data condition based on a deep feature fusion network,''  \emph{IEEE Access}, vol. 7, pp. 165646-165658, 2019.

\bibitem{Huang106} 
 Z. L. Huang, Z. X. Pan, B. Lei, ``Transfer learning with deep convolutional neural network for SAR target classification with limited labeled data,''  \emph{Remote Sensing}, vol. 9, pp. 907, 2017.

\bibitem{Yan107} 
 Y. Yan, ``Convolutional neural networks based on augmented training samples for synthetic aperture radar target recognition,''  \emph{Journal of Electronic Imaging}, vol. 27, pp. 1, 2018.

\bibitem{Ding108} 
 J. Ding, B. Chen, H. Liu et al., ``Convolutional neural network with data augmentation for SAR target recognition,''  \emph{IEEE Geoscience and Remote Sensing Letters}, vol. 13, no. 3, pp. 364-368, March 2016.

\bibitem{Wang109} 
 H. Wang, S. Chen, F. Xu et al., ``Application of deep-learning algorithms to MSTAR data,'' in \emph{Proc. 2015 IEEE International Geoscience and Remote Sensing Symposium (IGARSS)}, Milan, 2015, pp. 3743-3745.

\bibitem{Everingham110} 
 M. Everingham, S. M. Ali Eslami, L. V. Gool, ``The PASCAL visual object classes challenge: A retrospective,'' \emph{International Journal of Computer Vision}, vol. 111, pp. 98-136, 2015.

\bibitem{Huang111} 
 L. Q. Huang, B. Liu, B. Y. Li, et al., ``OpenSARShip: A dataset dedicated to sentinel-1 ship interpretation,'' \emph{IEEE Journal of Selected Topics in Applied Earth Observations and Remote Sensing}, vol. 11, no. 1, pp. 195-208, 2018.

\bibitem{Oliver112} 
 C. Oliver, S. Quegan, ``Understanding synthetic aperture radar images,'' Norwood, MA, USA: Artech House, 1998.

\bibitem{Baraldi113} 
 A. Baraldi, F. Parmiggiani, ``A refined gamma MAP SAR speckle filter with improved geometrical adaptivity,''  \emph{IEEE Trans. Geosci. Remote Sens.}, vol. 33, no. 5, pp. 1245-1257, Sep. 1995.

\bibitem{Frost114} 
 V. S. Frost, J. A. Stiles, K. S. Shanmugan, et al., ``A model for radar images and its application to adaptive digital filtering of multiplicative noise,''  \emph{IEEE Trans. Pattern Anal. Mach. Intell.}, vol. PAMI-4, no. 2, pp. 157-166, Mar. 1982.

\bibitem{Dabov115} 
 K. Dabov, A. Foi, V. Katkovnik, et al., ``Image denoising with block-matching and 3D filtering,''  in \emph{Proc. SPIE}, vol. 6064, 2006.

\bibitem{Parrilli116} 
 S. Parrilli, M. Poderico, C. V. Angelino, ``A nonlocal SAR image denoising algorithm based on LLMMSE wavelet shrinkage,''  \emph{IEEE Trans. Geosci. Remote Sens.}, vol. 50, no. 2, pp. 606-616, Feb. 2012.

\bibitem{Achim117} 
 A. Achim, P. Tsakalides, A. Bezerianos, ``SAR image denoising via Bayesian wavelet shrinkage based on heavy-tailed modeling,''  \emph{IEEE Trans. Geosci. Remote Sens.}, vol. 41, no. 8, pp. 1773-1784, Aug. 2003.

\bibitem{Argenti118} 
 F. Argenti, L. Alparone, ``Speckle removal from SAR images in the undecimated wavelet domain,''  \emph{IEEE Trans. Geosci. Remote Sens.}, vol. 40, no. 11, pp. 2363-2374, 2002.

\bibitem{Xie119} 
 H. Xie, L. E. Pierce, F. T. Ulaby, ``SAR speckle reduction using wavelet denoising and Markov random field modeling,''  \emph{IEEE Trans. Geosci. Remote Sens.}, vol. 40, no. 10, pp. 2196-2212, 2002.

\bibitem{Patel120} 
 V. M. Patel, G. R. Easley, R. Chellappa, et al., ``Separated component-based restoration of speckled SAR images,''  \emph{IEEE Trans. Geosci. Remote Sens.}, vol. 52, no. 2, pp. 1019-1029, Feb. 2014.

\bibitem{Argenti121} 
 F. Argenti, A. Lapini, T. Bianchi, et al., ``A tutorial on speckle reduction in synthetic aperture radar images,'' \emph{IEEE Geosci. Remote Sens. Mag.}, vol. 1, no. 3, pp. 6-35, Sep. 2013.

\bibitem{despeckling_CNN_ensemble}
 D. Mishra, S. Tyagi, S. Chaudhury, et al, ``Despeckling CNN with ensembles of classical outputs,'' in \emph{Proc. 2018 24th International Conference on Pattern Recognition (ICPR)}, 2018, pp. 3802-3807.

\bibitem{Chierchia122} 
 G. Chierchia, D. Cozzolino, G. Poggi, et al., ``SAR image despeckling through convolutional neural networks,'' in \emph{ Proc. 2017 IEEE International Geoscience and Remote Sensing Symposium}, Fort Worth, TX, USA, 2017, pp. 5438-5441.

\bibitem{Zhao123} 
 J. P. Zhao, W. W. Guo, B. Liu, et al., ``Convolutional neural network-based SAR image classification with noisy labels,''  \emph{Journal of Radars}, vol. 6, no. 5, pp. 514-523, 2017. Doi: 10.12000/JR16140.(in Chinese)

\bibitem{Wang124} 
 P. Y. Wang, H. Zhang, V. M. Patel, ``SAR image despeckling using a convolutional neural network,''  \emph{IEEE Signal Processing Letters}, vol. 24, no. 12, pp. 1763-1767, 2017. Doi: 10.1109/LSP.2017.2758203.

\bibitem{Wang125} 
 J. Wang, T. Zheng, P. Lei, et al., ``Ground target classification in noisy SAR images using convolutional neural networks,'' \emph{IEEE Journal of Selected Topics in Applied Earth Observations and Remote Sensing}, vol. 11, no. 11, pp. 4180-4192, Nov. 2018.

\bibitem{Hemani126} 
 H. Parikh, S. Patel, V. Patel, ``Analysis of denoising techniques for speckle noise removal in synthetic aperture radar images,'' in \emph{Proc. 2018 International Conference on Advances in Computing Communications and Informatics (ICACCI)}, 2018, pp. 671-677.

\bibitem{Martino127} 
 G. Di Martino, M. Poderico, G. Poggi, et al., ``Benchmarking framework for SAR despeckling,''  \emph{IEEE Trans. Geosci. Remote Sens.}, vol. 52, no. 3, pp. 1596-1615, Mar. 2014.

\bibitem{Mukherjee128} 
 S. Mukherjee, A. Zimmer, N. K. Kottayil, et al., ``CNN-Based InSAR denoising and coherence metric,''  \emph{2018 IEEE sensors}, 2018, pp. 1-4.

\bibitem{Nicolas129} 
 N. Brodu, ``Low-rankness transfer for denoising Sentinel-1 SAR images,'' in \emph{ Proc. 2018 9th International Symposium on Signal Image Video and Communications (ISIVC)}, 2018, pp. 106-111.

\bibitem{Sergio130} 
 S. Vitale, G. Ferraioli, V. Pascazio, ``A new ratio image based CNN algorithm for SAR despeckling,'' in \emph{Proc. IEEE International Geoscience and Remote Sensing Symposium IGARSS}, 2019, pp. 9494-9497.

\bibitem{Giampaolo131} 
 G. Ferraioli, V. Pascazio, S. Vitale, ``A novel cost function for despeckling using convolutional neural networks,''  in \emph{Proc. Remote Sensing Event (JURSE)}, Joint Urban, 2019, pp. 1-4.

\bibitem{Ramin132} 
 R. Farhadiani, S. Homayouni, A. Safari, ``Hybrid SAR speckle reduction using complex wavelet shrinkage and non-local PCA-based Ffiltering,''  \emph{IEEE Journal of Selected Topics in Applied Earth Observations and Remote Sensing}, vol. 12, no. 5, pp. 1489-1496, 2019.

\bibitem{Zhou133} 
 Z. H. Zhou, E. Y. Lam, C. Lee, ``Nonlocal means filtering based speckle removal utilizing the maximum a posteriori estimation and the total variation image prior,'' \emph{IEEE Access}, vol. 7, pp. 99231-99243, 2019.

\bibitem{Yang134} 
 X. L. Yang, L. Denis, F. Tupin, et. al, ``SAR image despeckling using pre-trained convolutional neural network models,'' in \emph{Proc. Remote Sensing Event (JURSE)}, Joint Urban, 2019, pp. 1-4.

\bibitem{Ravani135} 
 K. Ravani, S. Saboo, J. S. Bhatt, ``A practical approach for SAR image despeckling using deep learning,'' \emph{2019 IEEE International Geoscience and Remote Sensing Symposium (IGARSS)}, 2019, pp. 2957-2960.

\bibitem{Ronneberger136} 
 O. Ronneberger et al., ``U-net: Convolutional networks for biomedical image segmentation,'' in \emph{Proc. International Conference on Medical Image Computing and Computer-Assisted Intervention}, 2015, pp. 234-241.

\bibitem{Cozzolino137} 
 D. Cozzolino, L. Verdoliva, G. Scarpa, et al., ``Nonlocal SAR image despeckling by convolutional neural networks,'' in \emph{ 2019 IEEE International Geoscience and Remote Sensing Symposium (IGARSS)}, 2019, pp. 5117-5120.

\bibitem{Vitale138} 
 S. Vitale, D. Cozzolino, G. Scarpa, et al., ``Guided patchwise nonlocal SAR despeckling,'' \emph{IEEE Transactions on Geoscience and Remote Sensing}, vol. 57, no. 9, pp. 6484-6498, 2019.

\bibitem{Tupin139} 
 F. Tupin, L. Denis, C. A. Deledalle, et al., ``Ten years of patch-based approaches for sar imaging: A review", in \emph{Proc. 2019 IEEE International Geoscience and Remote Sensing Symposium (IGARSS)}, 2019, pp. 5105-5108.

\bibitem{Gu140} 
 F. Gu, H. Zhang, C. Wang, ``A two-component deep learning network for SAR Iimage denoising,'' \emph{IEEE Access}, vol. 8, pp. 17792-17803, 2020.

\bibitem{Zhang141} 
 Q. Zhang, Q. Yuan, J. Li, et al., ``Learning a dilated residual network for SAR image despeckling,'' \emph{Remote Sens.}, vol. 10, no. 2, pp. 196, Jan. 2018.

\bibitem{Liu142} 
 S. Liu, T. Liu, L. Gao, et al., ``Convolutional neural network and guided filtering for SAR image denoising,'' \emph{Remote Sens.}, vol. 11, no. 6, pp. 702, Mar. 2019.

\bibitem{Foucher143} 
 S. Foucher, M. Beaulieu, M. Dahmane, et al., ``Deep speckle noise filtering,''  in \emph{Proc. IEEE Int. Geosci. Remote Sens. Symp. (IGARSS)}, 2017, pp. 5311-5314.

\bibitem{Gu144} 
 F. Gu, H. Zhang, C. Wang, et al., ``Residual encoder-decoder network introduced for multisource SAR image despeckling,''  in \emph{Proc. SAR in Big Data Era: Models Methods and Applications (BIGSARDATA)}, 2017, pp. 1-5.

\bibitem{Tang145} 
 X. Tang, L. Zhang, X. Ding, ``SAR image despeckling with a multilayer perceptron neural network, \emph{ Int. J. Digit. Earth}, vol. 12, no. 3, pp. 354-374, Mar. 2019.

\bibitem{Wang147} 
 H. Wang, S. Chen, F. Xu et al., ``Application of deep-learning algorithms to MSTAR data,'' in \emph{Proc. 2015 IEEE International Geoscience and Remote Sensing Symposium (IGARSS)}, Milan, 2015, pp. 3743-3745.

\bibitem{Bai148} 
 Y. B. Bai, C. Gao, S. Singh, et al., ``A framework of rapid regional tsunami damage recognition from post-event TerraSAR-X imagery using deep neural networks,''  \emph{IEEE Geoscience and Remote Sensing Letters}, vol. 15, no. 1, pp. 43-47, 2018. Doi: 10.1109/LGRS.2017.2772349.

\bibitem{Zhou149} 
 Y. Zhou, H. P. Wang, F. Xu, et al., ``Polarimetric SAR image classification using deep convolutional neural networks,''  \emph{IEEE Geoscience and Remote Sensing Letters}, vol. 13, no. 12, pp. 1935-1939,  2016. Doi: 10.1109/LGRS.2016.2618840.

\bibitem{Xu150} 
 F. Xu, H. P. Wang, Y. Q. Jin, ``Deep learning as applied in SAR target recognition and Terrain classification,''  \emph{Journal of Radars}, vol. 6, no. 2, pp. 136-148, 2017. Doi: 10.12000/JR16130.(in Chinese)

\bibitem{Zhang151} 
 Z. W. Zhang, H. P. Wang, F. Xu, et al., ``Complex-valued convolutional neural network and its application in polarimetric SAR image classification,''  \emph{IEEE Transactions on Geoscience and Remote Sensing}, vol. 55, no. 12, pp. 7177-7188, 2017. Doi: 10.1109/TGRS.2017.2743222.

\bibitem{FCN_CV_SAR_classification} 
 A. G. Mullissa, C. Persello, A. Stein, ``PolSARNet: A deep fully convolutional network for polarimetric SAR image classification,''  \emph{IEEE Journal of Selected Topics in Applied Earth Observations and Remote Sensing}, vol. 12, no. 12, pp. 5300-5309, Dec. 2019. Doi: 10.1109/JSTARS.2019.2956650.

\bibitem{CV_CAE_complex_PolSAR_classify} 
 R. Shang, G. Wang, M A Okoth, et al., ``Complex-valued convolutional autoencoder and spatial pixel-squares refinement for polarimetric SAR image classification,'' \emph{Remote Sensing}, vol. 11, no. 5, 2019.


\bibitem{Cui152} 
 Z. Y. Cui, M. G. Zhang, Z. J. Cao, ``Image data augmentation for SAR sensor via generative adversarial nets,''  \emph{IEEE Access}, vol. 7, pp. 42255-42268, 2019.

\bibitem{Ma153} 
 Y. Ma, Y. Liang, W. Y. Zhang, ``SAR target recognition based on transfer learning and data augmentation with LSGANs,''  in \emph{Proc. Chinese Automation Congress (CAC)}, 2019, pp. 2334-2337.

\bibitem{Wang154} 
 Z. Wang, L. Du, J. Mao, et al., ``SAR target detection based on SSD with data augmentation and transfer learning,''  \emph{IEEE Geoscience and Remote Sensing Letters}, vol. 16, no. 1, pp. 150-154, Jan. 2019.

\bibitem{Junya155} 
 J. Y. Lv, Y. Liu, ``Data augmentation based on attributed scattering centers to train robust CNN for SAR ATR", \emph{IEEE Access}, vol. 7, pp. 25459-25473, 2019.

\bibitem{Yu156} 
 Q. Z. Yu, H. B. Hu, X. P. Geng, et al., ``High-performance SAR automatic target recognition under limited data condition based on a deep feature fusion network,'' \emph{IEEE Access}, vol. 7, pp. 165646-165658, 2019.

\bibitem{Shang157} 
 R. H. Shang, J. M. Wang, L. C. Jiao, et al., ``SAR targets classification based on deep memory convolution neural networks and transfer parameters,'' \emph{IEEE Journal of Selected Topics in Applied Earth Observations and Remote Sensing}, vol. 11, no. 8, pp. 2834-2846, 2018.

\bibitem{Ding159} 
 J. Ding, B. Chen, H. Liu, et al., ``Convolutional neural network with data augmentation for SAR target recognition,''  \emph{IEEE Geoscience and Remote Sensing Letters}, vol. 13, no. 3, pp. 364-368, March 2016.

\bibitem{Furukawa160} 
 H. Furukawa, ``Deep learning for target Classification from SAR imagery data augmentation and translation invariance,'' \emph{ArXiv}: 1708.07920, 2017.

\bibitem{Zhao161} 
 Z. Zhao, et al., ``Discriminant deep belief network for high-resolution SAR image classification,'' \emph{Pattern Recognition}, 2016. http://dx.doi.org/10.1016/j.patcog.2016.05.028.

\bibitem{Chen162} 
 S. Chen, H. Wang, ``SAR target recognition based on deep learning,'' in \emph{Proc. International Conference Data Science}, 2014, pp. 541-547.

\bibitem{Zhao163} 
 Q. Zhao, J. C. Principe, `` Support vector machines for SAR automatic target recognition,''  \emph{IEEE Transactions on Aerospace and Electronic Systems}, vol. 37, no. 2, pp. 643-654, 2001.

\bibitem{Georgios164} 
 C. Georgios, ``SVM-based target recognition from synthetic aperture radar images using target region outline descriptors,''  \emph{Nonlinear Anal. Theory Meth. Appl.}, vol. 71, no. 12, pp. e2934-e2939, Dec. 2009. [online] Available: http://dx.doi.org/10.1016/j.na.2009.07.030.

\bibitem{Zhao165} 
 J. Zhao. C. Principe, ``Support vector machines for SAR automatic target recognition,'' \emph{IEEE Trans. Aerosp. Electron. Syst.}, vol. 37, no. 2, pp. 643-654, Apr. 2001. [online] Available: http://dx.doi.org/10.1109/7.937475.

\bibitem{Stanhope166} 
 S. A. Stanhope, J. M. Daida, ``Genetic programming for automatic target classification and recognition in synthetic aperture radar imagery,'' \emph{Evol. Programming VII}, Berlin, Germany, Springer, pp. 735-744, 1998. [online] Available: http://dx.doi.org/10.1007/BFb0040824.

\bibitem{Sun167} 
 Y. Sun, Z. Liu, S. Todorovic, et al., ``Adaptive boosting for SAR automatic target recognition,'' \emph{ IEEE Trans. Aerosp. Electron. Syst.}, vol. 43, no. 1, pp. 112-125, Jan. 2007. [online] Available: http://dx.doi.org/10.1109/TAES.2007.357120.

\bibitem{Perciano168} 
 T. Perciano, F. Tupin, R. Hirata, et al., ``A hierarchical Markov random field for road network extraction and its application with optical and SAR data,'' in \emph{ Proc. IEEE Int. Geosci. Remote Sens. Symp. (IGARSS)}, Jul. 2011, pp. 1159-1162,  [online] Available: http://dx.doi.org/10.1109/IGARSS.2011.6049403.

\bibitem{Weisenseel169} 
 R. Weisenseel, W. Karl, D. Castanon, et al., ``Markov random field segmentation methods for SAR target chips,'' in \emph{Proc. SPIE}, Aug. 1999, vol. 3721, pp. 462-473, [online] Available: http://dx.doi.org/10.1117/12.357662.

\bibitem{Clausi170} 
 D. A. Clausi, ``Improved texture recognition of SAR sea ice imagery by data fusion of MRF features with traditional methods,'' in \emph{Proc. IEEE Int. Geosci. Remote Sens. Symp. (IGARSS)}, Jul. 2001, vol. 3, pp. 1170-1172, [online] Available: http://dx.doi.org/10.1109/IGARSS.2001.976781.

\bibitem{Ruohong171} 
 H. Ruohong, Y. Ruliang, ``SAR target recognition based on MRF and gabor wavelet feature extraction,'' in \emph{Proc. IEEE Int. Geosci. Remote Sens. Symp. (IGARSS)}, Jul. 2008, vol. 2, pp. II-907-II-910, , [online] Available: http://dx.doi.org/10.1109/IGARSS.2008.4779142.

\bibitem{Zhao172} 
 Z. Lin, K. F. Ji, M. Kang, et al., ``Deep convolutional highway unit network for SAR target classification with limited labeled training data,''  \emph{IEEE Geoscience and Remote Sensing Letters}, vol. 14, no. 7, pp. 1091-1095, 2017.

\bibitem{Ievgen173} 
 I. M. Gorovyi, D. S. Sharapov, ``Comparative analysis of convolutional neural networks and support vector machines for automatic target recognition,'' in \emph{Proc. IEEE Microwaves Radar and Remote Sensing Symposium (MRRS)}, 2017, pp. 63-66.

\bibitem{Pei174} 
 J. F. Pei, Y. L. Huang, W. B. Huo, ``SAR automatic target recognition based on multiview deep learning framework,''  \emph{IEEE Transactions on Geoscience and Remote Sensing}, vol. 56, no. 4, pp. 2196-2210, 2018.

\bibitem{Zhang175} 
 F. Zhang, C. Hu, Q. Yin, ``Multi-aspect-aware bidirectional LSTM networks for synthetic aperture radar target recognition,''  \emph{IEEE Access}, vol. 5, pp. 26880-26891, 2017.

\bibitem{Ding176} 
 B. Y. Ding, G. J. Wen, X. H. Huang, et. al, ``Data augmentation by multilevel reconstruction using attributed scattering center for SAR target recognition,'' \emph{IEEE Geoscience and Remote Sensing Letters}, vol. 14, no. 6, pp. 979-983, 2017.

\bibitem{Pei177} 
 J. F. Pei, Y. L. Huang, W. B. Huo, ``Multi-view SAR ATR based on networks ensemble and graph search,'' in \emph{Proc. 2018 IEEE Radar Conference (RadarConf18)}, 2018, pp. 0355-0360.

\bibitem{Huang178} 
 Z. L. Huang, Z. X. Pan, B. Lei, ``Transfer Learning with deep convolutional neural network for SAR target classification with limited labeled data,''  \emph{Remote Sensing}, vol. 9, pp. 907, 2017.

\bibitem{Ding179} 
 B. Y. Ding, G. J. Wen, C. H. Ma, ``An efficient and robust framework for SAR target recognition by hierarchically fusing global and local features,'' \emph{IEEE Transactions on Image Processing}, vol. 27, no. 12, pp. 5983-5995, 2018.

\bibitem{Pei180} 
 J. F. Pei, W. B. Huo, Q. H. Zhang, et al., ``Multi-view bistatic synthetic aperture radar target recognition based on multi-input deep convolutional neural network,'' in \emph{Proc. 2018 IEEE International Geoscience and Remote Sensing Symposium (IGARSS) }, 2018, pp. 2314-2317

\bibitem{Cho181} 
 J. H. Cho, C. G. Park, ``Multiple feature aggregation using convolutional neural networks for SAR image-based automatic target recognition,''  \emph{IEEE Geoscience and Remote Sensing Letters}, vol. 15, no. 12, pp. 1882-1886, 2018.

\bibitem{Qi182} 
 W. Qi, G. J. Wen, ``SAR target classification method based on convolutional neural network,''  in \emph{Proc. 2018 IEEE 9th International Conference on Software Engineering and Service Science (ICSESS)} , 2018, pp. 1151-1155.

\bibitem{Bai183} 
 X. R. Bai, R. H. Xue, L. Wang, et al., ``Sequence SAR image classification based on bidirectional convolution-recurrent network,''  \emph{IEEE Transactions on Geoscience and Remote Sensing}, vol. 57, no. 11, pp. 9223-9235, 2019.

\bibitem{Zhang184} 
 S. Zhang, Z. D. Wen, Z. G. Liu, et al., ``Rotation awareness based self-supervised learning for SAR target recognition,'' in \emph{Proc. 2019 IEEE International Geoscience and Remote Sensing Symposium (IGARSS)} 2019, pp. 1378-1381.

\bibitem{Ren185} 
 H. H. Ren, X. L. Yu, L. Zou, et al., ``Joint supervised dictionary and classifier learning for multi-view SAR image classification,''  \emph{IEEE Access}, vol. 7, pp. 165127-165142, 2019.

\bibitem{Ning186} 
 C. Ning, W. B. Liu, G. Zhang, et al., ``Synthetic aperture radar target recognition using weighted multi-task kernel sparse representation,'' \emph{IEEE Access}, vol. 7, pp. 181202-181212, 2019.

\bibitem{Gu187} 
 Y. Gu, Y. Xu, J. Liu, ``SAR ATR by decision fusion of multiple random convolution features,'' in \emph{Proc. 2019 22th International Conference on Information Fusion}, 2019, pp. 1-8.

\bibitem{Kang188} 
 M. Kang, K. F. Ji, X. G. Leng, et al., ``Synthetic aperture radar target recognition with feature fusion based on a stacked autoencoder,''  \emph{Sensors}, vol. 17, pp. 192, 2017.

\bibitem{Zhang189} 
 X. Z. Zhang, Y. J. Wang, Z. Y. Tan, et al., ``Two-stage multi-task representation learning for synthetic aperture radar (SAR) target images classification,'' \emph{Sensors}, vol. 17, pp. 2506, 2017.

\bibitem{Shao190} 
 J. Q. Shao, C. W. Qu, J. W. Li, et al., ``A lightweight convolutional neural network based on visual attention for SAR image target classification,'' \emph{Sensors}, vol. 18, pp. 3039, 2018.

\bibitem{Chen191} 
 S. Chen, H. Wang, F. Xu, et al., ``Target classification using the deep convolutional networks for SAR images,''  \emph{IEEE Trans. Geosci. Remote Sens.}, vol. 54, pp. 4806-4817, 2016.

\bibitem{Zhao192} 
 P. F. Zhao, K. Liu, H. Zou, et al, ``Multi-stream convolutional neural network for SAR automatic target recognition,''  \emph{Remote Sensing}, vol. 10, pp. 1473, 2018.

\bibitem{Touafria193} 
 M. Touafria, Q. Yang, ``A concurrent and hierarchy target learning architecture for classification in SAR application,''  \emph{Sensors}, vol. 18, pp. 3218, 2018.

\bibitem{Tian194} 
 Z. Z. Tian, L. P. Wang, R. H. Zhan, et al., ``Classification via weighted kernel CNN: application to SAR target recognition,''  \emph{International Journal of Remote Sensing}, pp. 1, 2018.

\bibitem{Huang195} 
 G. Q. Huang, X. G. Liu, J. P. Hui, et al., ``A novel group squeeze excitation sparsely connected convolutional networks for SAR target classification,''  \emph{International Journal of Remote Sensing}, pp. 1, 2019.

\bibitem{Lv196} 
 J. Y. Lv, ``Exploiting multi-level deep features via joint sparse representation with application to SAR target recognition,'' \emph{International Journal of Remote Sensing}, pp. 1, 2019.

\bibitem{Qi197} 
 B. Q. Qi, H. T. Jing, H. Chen, et al., ``Target recognition in synthetic aperture radar image based on PCANet,''  \emph{The Journal of Engineering}, vol. 2019, no. 21, pp. 7309-7312, 2019.

\bibitem{patch_sorted_AE_SVM} 
 Z. Ren, B. Hou, Z. Wen, et al., ``Patch-sorted deep feature learning for high resolution SAR Image Classification,''  \emph{IEEE Journal of Selected Topics in Applied Earth Observations and Remote Sensing}, vol. 11, no. 9, pp. 3113-3126, Sept. 2018. Doi: 10.1109/JSTARS.2018.2851023.

\bibitem{spatial_relation_PolSAR_classification} 
 H. Liu, S. Yang, S. Gou, et al., ``Polarimetric SAR feature extraction with neighborhood preservation-based deep learning,'' \emph{IEEE Journal of Selected Topics in Applied Earth Observations and Remote Sensing}, vol. 10, no. 4, pp. 1456-1466, April 2017. Doi: 10.1109/JSTARS.2016.2618891.

\bibitem{MF_SARNet_SAR_ATR} 
 Y. Zhai et al., ``MF-SarNet: Effective CNN with data augmentation for SAR automatic target recognition,'' \emph{ The Journal of Engineering}, vol. 2019, no. 19, pp. 5813-5818, 10 2019. Doi: 10.1049/joe.2019.0218.

\bibitem{deep_Bayesian_GNN_SAR_recognition} 
 D. Guo, B. Chen, ``SAR image target recognition via deep Bayesian generative network,'' in \emph{Proc. 2017 International Workshop on Remote Sensing with Intelligent Processing (RSIP)}, Shanghai, 2017, pp. 1-4. Doi: 10.1109/RSIP.2017.7958814.

\bibitem{unsupervised_SAEs_clustering_SAR_classification} 
 T. Yan, X. Sun, W. Yang, et al., ``Unsupervised classification of polarimetric SAR images using deep embedding network,'' in\emph{Proc. 2017 SAR in Big Data Era: Models, Methods and Applications (BIGSARDATA)}, Beijing, 2017, pp. 1-6, doi: 10.1109/BIGSARDATA.2017.8124939.

\bibitem{ensemble_learning_CNN_PolSAR} 
 W. Wu, H. Li, L. Zhang, et al., ``High-resolution PolSAR scene classification with pretrained deep convnets and manifold polarimetric parameters,'' \emph{IEEE Transactions on Geoscience and Remote Sensing}, vol. 56, no. 10, pp. 6159-6168, 2018.

\bibitem{oil_spillage_detection_CNN_SVM} 
 D. Song et al., ``A novel marine oil spillage identification scheme based on convolution neural network feature extraction from fully polarimetric SAR imagery,''  \emph{IEEE Access}, vol. 8, pp. 59801-59820, 2020. Doi: 10.1109/ACCESS.2020.2979219.

\bibitem{single_layer_efficently_classify} 
 T. Gadhiya, A. K. Roy, ``Optimized Wishart network for an efficient classification of multifrequency PolSAR data,''  \emph{IEEE Geoscience and Remote Sensing Letters}, vol. 15, no. 11, pp. 1720-1724, Nov. 2018. Doi: 10.1109/LGRS.2018.2861081.

\bibitem{CNN_SVM_SAR_ATR} 
 S. A. Wagner, ``SAR ATR by a combination of convolutional neural network and support vector machines,'' \emph{ IEEE Transactions on Aerospace and Electronic Systems}, vol. 52, no. 6, pp. 2861-2872, December 2016. Doi: 10.1109/TAES.2016.160061.

\bibitem{DCNN_SVM_SAR_ATR} 
 F. Gao, T. Huang, J. Wang, et al., ``Combining deep convolutional neural network and SVM to SAR image target recognition,'' in \emph{Proc. 2017 IEEE International Conference on Internet of Things (iThings) and IEEE Green Computing and Communications (GreenCom) and IEEE Cyber, Physical and Social Computing (CPSCom) and IEEE Smart Data (SmartData)}, Exeter, 2017, pp. 1082-1085. Doi: 10.1109/iThings-GreenCom-CPSCom-SmartData.2017.165.

\bibitem{AEs_SVM_SAR_recognition} 
 S. R. Tian, C. Wang, H. Zhang, ``Hierarchical feature exttratction for object recogition in complex SAR image using modified convolutional auto-encoder,'' in \emph{Proc. 2017 IEEE International Geoscience and Remote Sensing Symposium (IGARSS)}, Fort Worth, TX, 2017, pp. 854-857. Doi: 10.1109/IGARSS.2017.8127087.

\bibitem{DCNN_improved_cost_function_SVM} 
 F. Gao, T. Huang, J. Sun, et al., ``A new algorithm of SAR image target recognition based on improved deep convolutional neural network,'' \emph{Cognitive Computation}, vol. 11, no. 6, pp. 809-824, 2019.

\bibitem{TL_CNN_SVM} 
 M. A. Mufti, E. A. Hadhrami, B. Taha, et al., ``Using transfer learning technique for SAR automatic target recognition,'' in \emph{ Proc. SPIE 11197, SPIE Future Sensing Technologies, 111970A}, 12 November 2019, https://doi.org/10.1117/12.2538012.

\bibitem{deep_kernel_learning_SVM_SAR_recognition} 
 X. Chen, X. Peng, R. Duan, et al., ``Deep kernel learning method for SAR image target recognition,'' \emph{Review of Scientific Instruments}, vol. 88, no. 10, 2017.

\bibitem{multistage_fusion_SAR_IR_recognition} 
 Y. Cho, S. Shin, S. Yim, et al., ``Multistage fusion with dissimilarity regularization for SAR/IR target recognition,''  \emph{IEEE Access}, vol. 7, pp. 728-740, 2019. Doi: 10.1109/ACCESS.2018.2885736.

\bibitem{sparse_manifold_regularized_CNN} 
 H. Liu, F. Shang, S. Yang, et al., ``Sparse manifold-regularized neural networks for polarimetric SAR terrain classification,''  \emph{IEEE Transactions on Neural Networks and Learning Systems}, doi: 10.1109/TNNLS.2019.2935027.

\bibitem{sparse_temporal_ensemble_CNN} 
 R. Xue, X. Bai, F. Zhou, ``Spatial-temporal ensemble convolution for sequence SAR target classification,'' \emph{IEEE Transactions on Geoscience and Remote Sensing}, doi: 10.1109/TGRS.2020.2997288.

\bibitem{DBN_InSAR_classification} 
 W. Liu, S. Gou, W. Chen, et al., ``Classification of interferometric synthetic aperture radar image with deep learning approach,'' in \emph{Proc. 2016 CIE International Conference on Radar (RADAR)}, Guangzhou, 2016, pp. 1-3. Doi: 10.1109/RADAR.2016.8059200.

\bibitem{pose_angle_information_CNN_SAR_recognition} 
 J. Oh, G. Youm, M. Kim, ``SPAM-Net: A CNN-based SAR target recognition network with pose angle marginalization learning,'' \emph{IEEE Transactions on Circuits and Systems for Video Technology}, doi: 10.1109/TCSVT.2020.2987346.

\bibitem{crop_classify_SAR_optical_CNN} 
 L. Gu, F. He, S. Yang, ``Crop classification based on deep learning in northeast China using SAR and optical imagery,'' in \emph{Proc. 2019 SAR in Big Data Era (BIGSARDATA)}, Beijing, China, 2019, pp. 1-4. Doi: 10.1109/BIGSARDATA.2019.8858437.

\bibitem{AEs_multi_modal_fusing_SAR_multispectral} 
 J. Geng, H. Wang, J. Fan et al., ``Classification of fusing SAR and multispectral image via deep bimodal autoencoders,'' in \emph{Proc. 2017 IEEE International Geoscience and Remote Sensing Symposium (IGARSS)}, Fort Worth, TX, 2017, pp. 823-826. Doi: 10.1109/IGARSS.2017.8127079.

\bibitem{spatial_polarimetric_information_fusing_PolSAR_classify} 
 F. Gao, T. Huang, J. Wang, et al., ``Dual-branch deep convolution neural network for polarimetric SAR image classification,'' \emph{Applied Sciences}, vol. 7, no. 5, 2017.

\bibitem{Dense_CNN_SAR_recognition} 
 H. Dong, L. Zhang, B. Zou, ``Densely connected convolutional neural network based polarimetric SAR image classification,'' in \emph{Proc. IGARSS 2019 - 2019 IEEE International Geoscience and Remote Sensing Symposium}, Yokohama, Japan, 2019, pp. 3764-3767. Doi: 10.1109/IGARSS.2019.8900292.

\bibitem{Inception_v3_SAR_classiifcation} 
 C. Wang et al., ``Automated geophysical classification of sentinel-1 wave mode SAR images through deep-learning,'' in \emph{Proc. IGARSS 2018 - 2018 IEEE International Geoscience and Remote Sensing Symposium}, Valencia, 2018, pp. 1776-1779. Doi: 10.1109/IGARSS.2018.8518354.

\bibitem{UNet_SAR_classification} 
 Y. Guo, E. Chen, Z. Li, et al., ``Convolutional highway unit network for large-scale classification with GF-3 dual-PolSar data,'' in \emph{IGARSS 2018 - 2018 IEEE International Geoscience and Remote Sensing Symposium}, Valencia, 2018, pp. 2424-2427. Doi: 10.1109/IGARSS.2018.8518737.

\bibitem{2D_temporal_convolution_SAR_recognition} 
 R. Xue, X. Bai, ``2D-temporal convolution for target recognition of SAR sequence image,'' in \emph{Proc. 2019 6th Asia-Pacific Conference on Synthetic Aperture Radar (APSAR)}, Xiamen, China, 2019, pp. 1-4, doi: 10.1109/APSAR46974.2019.9048438.

\bibitem{convolutional_LSTM_rotation_information_SAR_recognition} 
 L. Wang, X. Xu, H. Dong, et al., ``Exploring convolutional lstm for Polsar image classification,'' in \emph{Proc. IGARSS 2018 - 2018 IEEE International Geoscience and Remote Sensing Symposium}, Valencia, 2018, pp. 8452-8455. Doi: 10.1109/IGARSS.2018.8518517.

\bibitem{complex_contourlet_CNN_PolSAR_classify} 
 L. Li, L. Ma, L. Jiao, et al., ``Complex contourlet-CNN for polarimetric SAR image classification,'' \emph{Pattern Recognition}, 2019.

\bibitem{SAR_Net_PolSAR_classification} 
 Z. Huang, M. Datcu, Z. Pan, et al., ``Deep SAR-Net: Learning objects from signals,'' \emph{Isprs Journal of Photogrammetry and Remote Sensing}, pp. 179-193, 2020.

\bibitem{nonlinear_manifold_learning_FCN_SAR_classify} 
 C. He, M. Tu, D. Xiong, et al., ``Nonlinear manifold learning integrated with fully convolutional networks for PolSAR image classification,'' \emph{Remote Sensing}, vol. 12, no. 4, 2020.

\bibitem{contrastive_regulated_CNN_PolSAR_classify} 
 J. Zhao, M. Datcu, Z. Zhang, et al., ``Contrastive-regulated CNN in the complex domain: A method to learn physical scattering signatures from flexible PolSAR images,'' \emph{IEEE Transactions on Geoscience and Remote Sensing}, vol. 57, no. 12, pp. 10116-10135, 2019.

\bibitem{DCNN_expert_knowledge_PolSAR_classify} 
 X. Wang, Z. Cao, Z. Cui, et al., ``PolSAR image classification based on deep polarimetric feature and contextual information,'' \emph{Journal of Applied Remote Sensing}, vol. 13, no. 03, 2019.

\bibitem{water_shadow_classify_dense_network} 
 P. Zhang, L. Chen, Z. Li, et al., ``Automatic extraction of water and shadow from SAR images based on a multi-resolution dense encoder and decoder network,'' \emph{Sensors}, vol. 19, no. 16, 2019.

\bibitem{multi_spatial_features_SAR_scene_classify} 
 L. Chen, X. Cui, Z. Li, et al., ``A new deep learning algorithm for SAR scene classification based on spatial statistical modeling and features re-calibration,'' \emph{Sensors}, vol. 19, no. 11, 2019.

\bibitem{gB_DBN_mixed_model_SAR_classify} 
 Z. Zhao, L. Guo, M. Jia, et al., ``The generalized gamma-DBN for high-resolution SAR image classification,'' \emph{Remote Sensing}, vol. 10, no. 6, 2018.

\bibitem{MPDPL_SAE_PolSAR_classify} 
 Y. Chen, L. Jiao, Y. Li, et al., ``Multilayer projective dictionary pair learning and sparse autoencoder for PolSAR image classification,'' \emph{IEEE Transactions on Geoscience and Remote Sensing}, vol. 55, no. 12, pp. 6683-6694, 2017.

\bibitem{spatial_pyramid_pool_CNN_SAR_ATR} 
 L. J. Peng, X. H. Liu, M. Liu, et al., ``SAR target recognition and posture estimation using spatial pyramid pooling within CNN,'' in \emph{Proc. 2017 International Conference on Optical Instruments and Technology: Optoelectronic Imaging/Spectroscopy and Signal Processing Technology}, 2018. Doi: 10.1117/12.2285913.

\bibitem{deep_RNN_agricultural_classify} 
 E. Ndikumana, D. H. Minh, N. Baghdadi, et al., ``Deep recurrent neural network for agricultural classification using multitemporal SAR Sentinel-1 for Camargue, France,'' \emph{Remote Sensing}, vol. 10, no. 8, 2018.

\bibitem{GANs_TL_SAR_semi_supervised} 
 W. Zhang, Y. Zhu, Q. Fu, ``Semi-supervised deep transfer learning-Based on adversarial feature learning for label limited SAR target recognition,'' \emph{IEEE Access}, vol. 7, pp. 152412-152420, 2019. Doi: 10.1109/ACCESS.2019.2948404.

\bibitem{DQN_PolSAR_image_classification} 
 K. Huang, W. Nie, N. Luo, ``Fully polarized SAR imagery classification based on deep reinforcement learning method using multiple polarimetric features,''  \emph{IEEE Journal of Selected Topics in Applied Earth Observations and Remote Sensing}, vol. 12, no. 10, pp. 3719-3730, Oct. 2019. Doi: 10.1109/JSTARS.2019.2913445.

\bibitem{adversarial_learning_multi_band_SAR_classify} 
 W. Zhang, Y. Zhu and Q. Fu, ``Adversarial deep domain adaptation for multi-band SAR images classification,'' \emph{IEEE Access}, vol. 7, pp. 78571-78583, 2019. Doi: 10.1109/ACCESS.2019.2922844.

\bibitem{adversarial_learning_semi_supervised_SAR_classify} 
 C. Zheng, X. Jiang, X. Liu, ``Semi-supervised SAR ATR via multi-discriminator generative adversarial network," \emph{IEEE Sensors Journal}, vol. 19, no. 17, pp. 7525-7533, 1 Sept.1, 2019. Doi: 10.1109/JSEN.2019.2915379.

\bibitem{WAEs_generation_SAR_images} 
 K. Wang, G. Zhang, Y. Leng, et al., ``Synthetic aperture radar image generation with deep generative models,'' \emph{IEEE Geoscience and Remote Sensing Letters}, vol. 16, no. 6, pp. 912-916, June 2019. Doi: 10.1109/LGRS.2018.2884898.

\bibitem{zero_shot_learning_SAR_images} 
 Q. Song, F. Xu, ``Zero-shot learning of SAR target feature space with deep generative neural networks,''  \emph{IEEE Geoscience and Remote Sensing Letters}, vol. 14, no. 12, pp. 2245-2249, Dec. 2017. Doi: 10.1109/LGRS.2017.2758900.

\bibitem{unsupervised_domain_adaption_SAR} 
 F. Ye, W. Luo, M. Dong, et al., ``SAR image retrieval based on unsupervised domain adaptation and clustering,'' \emph{IEEE Geoscience and Remote Sensing Letters}, vol. 16, no. 9, pp. 1482-1486, Sept. 2019. Doi: 10.1109/LGRS.2019.2896948.

\bibitem{euclidean_distance_constrain_AE_SAR_recognition} 
 S. Deng, L. Du, C. Li, et al., ``SAR automatic target recognition based on euclidean distance restricted autoencoder,''  \emph{IEEE Journal of Selected Topics in Applied Earth Observations and Remote Sensing}, vol. 10, no. 7, pp. 3323-3333, July 2017. Doi: 10.1109/JSTARS.2017.2670083.

\bibitem{DAE_triplet_restriction_SAR_object_classify} 
 S. Tian, C. Wang, H. Zhang, et al., ``SAR object classification using the DAE with a modified triplet restriction,'' \emph{IET Radar, Sonar and Navigation}, vol. 13, no. 7, pp. 1081-1091, 7 2019. Doi: 10.1049/iet-rsn.2018.5413.

\bibitem{Siamese_network_small_samples_SAR_recognition} 
 J. Tang, F. Zhang, Q. Yin, et al., ``Synthetic aperture radar target recognition of incomplete training datasets via Siamese network,'' \emph{The Journal of Engineering}, vol. 2019, no. 20, pp. 6845-6847, 10 2019. Doi: 10.1049/joe.2019.0566.

\bibitem{graph_CNN_SAR_classify} 
 H. Zhu, N. Lin, H. Leung, et al., ``Target classification from SAR imagery based on the pixel grayscale decline by graph convolutional neural network,'' \emph{IEEE Sensors Letters}, vol. 4, no. 6, pp. 1-4, June 2020. Doi: 10.1109/LSENS.2020.2995060.

\bibitem{adaboost_rotation_forest_SAR_limited_data} 
 F. Zhang, Y. Wang, J. Ni, et al., ``SAR target small sample recognition based on CNN cascaded features and adaBoost rotation forest,'' \emph{IEEE Geoscience and Remote Sensing Letters}, vol. 17, no. 6, pp. 1008-1012, June 2020. Doi: 10.1109/LGRS.2019.2939156.

\bibitem{Gabor_DCNN_SAR_recognition} 
 T. Jiang, Z. Cui, Z. Zhou, et al., ``Data augmentation with Gabor filter in deep convolutional neural networks for sar target recognition,'' in \emph{Proc. IGARSS 2018 - 2018 IEEE International Geoscience and Remote Sensing Symposium}, Valencia, 2018, pp. 689-692. Doi: 10.1109/IGARSS.2018.8518792.

\bibitem{adversarial_AEs_generalization_SAR_classify} 
 Q. Song, F. Xu£¬ Y. Jin, ``SAR image representation learning with adversarial autoencoder networks,'' in \emph{Proc. IGARSS 2019 - 2019 IEEE International Geoscience and Remote Sensing Symposium}, Yokohama, Japan, 2019, pp. 9498-9501, doi: 10.1109/IGARSS.2019.8898922.

\bibitem{active_DL_PolSAR_classify} 
 H. Bi et al., ``An active deep learning approach for minimally-supervised polsar image classification,'' in \emph{Proc. IGARSS 2019 - 2019 IEEE International Geoscience and Remote Sensing Symposium}, Yokohama, Japan, 2019, pp. 3185-3188, doi: 10.1109/IGARSS.2019.8899214.

\bibitem{multi_scale_CAE_SAR_object_recognition} 
 S. Tian, C. Wang, H. Zhang, ``SAR object classification with a multi-scale convolutional auto-encoder,'' in \emph{Proc. 2019 SAR in Big Data Era (BIGSARDATA)}, Beijing, China, 2019, pp. 1-4. Doi: 10.1109/BIGSARDATA.2019.8858491.

\bibitem{dual_CNN_PolSAR_classify} 
 W. Hua, S. Wang, W. Xie, et al., ``Dual-channel convolutional neural network for polarimetric SAR images classification,'' in \emph{Proc. IGARSS 2019 - 2019 IEEE International Geoscience and Remote Sensing Symposium}, Yokohama, Japan, 2019, pp. 3201-3204. Doi: 10.1109/IGARSS.2019.8899103.

\bibitem{CompressUnit_deeper_CNN_SAR_classify} 
 Y. Zhang, X. Sun, H. Sun, et al., ``High resolution SAR image classification with deeper convolutional neural network,'' in \emph{Proc. IGARSS 2018 - 2018 IEEE International Geoscience and Remote Sensing Symposium}, Valencia, 2018, pp. 2374-2377. Doi: 10.1109/IGARSS.2018.8518829.

\bibitem{unsupervised_clustering_SAR_object_classify} 
 C. Wang et al., ``A deep unsupervised learning method for SAR object classification,'' in \emph{Proc. 2019 6th Asia-Pacific Conference on Synthetic Aperture Radar (APSAR)}, Xiamen, China, 2019, pp. 1-5, doi: 10.1109/APSAR46974.2019.9048377.

\bibitem{optical_SAR_knowledge_transfer_SAR_classification} 
 M. Rostami, S. Kolouri, E. Eaton, et al., ``SAR image classification using few-shot cross-domain transfer learning,,'' in \emph{Proc. 2019 IEEE/CVF Conference on Computer Vision and Pattern Recognition Workshops (CVPRW)}, Long Beach, CA, USA, 2019, pp. 907-915. Doi: 10.1109/CVPRW.2019.00120.

\bibitem{conv_biLSTM_few_shot_SAR_classification} 
 L. Wang, X. Bai, F. Zhou, ``Few-shot SAR ATR based on Conv-BiLSTM prototypical networks,'' in \emph{Proc. 2019 6th Asia-Pacific Conference on Synthetic Aperture Radar (APSAR)}, Xiamen, China, 2019, pp. 1-5. Doi: 10.1109/APSAR46974.2019.9048492.

\bibitem{class_probability_vector_PolSAR_classify} 
 C. Liu, W. Liao, H. Li, et al., ``A deep-neural-network-based hybrid method for semi-supervised classification of polarimetric SAR data,'' in \emph{Proc. 2019 6th Asia-Pacific Conference on Synthetic Aperture Radar (APSAR)}, Xiamen, China, 2019, pp. 1-5. Doi: 10.1109/APSAR46974.2019.9048529.

\bibitem{transferred_MS_CNN_SAR_recognition} 
 Y. K. Zhai, W. B. Deng, Y. Xu, et al., ``Robust SAR automatic target recognition based on transferred MS-CNN with L2-Regularization,'' in \emph{Computational Intelligence and Neuroscience}, vol. 2019, no. 3, pp. 1-13,  November 2019.

\bibitem{asymmetric_parallel_CNN_SAR_recognition} 
 Y, Hou, Y, Bai, T. Xu, et al., ``Deep convolutional neural network structural design for synthetic aperture radar image target recognition based on incomplete training data and displacement insensitivity,'' \emph{Journal of Electronic Imaging}, vol. 28, no. 5, 2019.

\bibitem{compact_CNN_PolSAR_classify} 
 M. Ahishali, S. Kiranyaz, T. Ince, et al., ''Dual and single polarized SAR image classification using compact convolutional neural networks,'' \emph{Remote Sensing}, vol. 11, no. 11, 2019.

\bibitem{CAE-HL-CNN} 
 Q. Rui,  F. Xiongjun. D. Jian, et al., ``A semi-greedy neural network CAE-HL-CNN for SAR target recognition with limited training data,'' \emph{ International Journal of Remote Sensing}, 2020. doi:10.1080/01431161.2020.1766149.

\bibitem{nonlocal_AEs_VAR_PolSAR_classify} 
 Wang R, Wang Y., ``Classification of PolSAR image using neural nonlocal stacked sparse autoencoders with virtual adversarial regularization,'' \emph{Remote Sensing}, vol. 11, no. 9, 2019.

\bibitem{semi_supervised_GANs_multi_generators_SAR_recognition} 
 F. Gao, F. Ma F, J. Wang, et al., ``Semi-supervised generative adversarial nets with multiple generators for SAR image recognition,'' \emph{Sensors}, vol. 18, no. 8, 2018.

\bibitem{DCGANs_semi_supervise_SAR_recognition} 
 Gao F, Yang Y, Wang J, et al., ``A deep convolutional generative adversarial networks (DCGANs)-based semi-supervised method for object recognition in synthetic aperture radar (SAR) images,'' \emph{Remote Sensing}, vol. 10, no. 6, 2018.

\bibitem{expert_knowledge_CNN_PolSAR_recognition} 
 S. Chen, C. Tao, X. Wang, S. Xiao, ``Polarimetric SAR targets detection and classification with deep convolutional neural network,''  in \emph{Proc. 2018 Progress in Electromagnetics Research Symposium (PIERS-Toyama)}, Toyama, 2018, pp. 2227-2234. Doi: 10.23919/PIERS.2018.8597856.

\bibitem{luminance_analysis_SAR_recognition} 
 H. Zhu, W. Wang, R. Leung, ``SAR target classification based on radar image luminance analysis by deep learning,'' \emph{IEEE Sensors Letters}, vol. 4, no. 3, pp. 1-4, March 2020. Doi: 10.1109/LSENS.2020.2976836.

\bibitem{scale_transformation_denoising} 
 J. Geng, J. Fan, H. Wang, et al., ``High-resolution SAR image classification via deep convolutional autoencoders,''  \emph{IEEE Geoscience and Remote Sensing Letters}, vol. 12, no. 11, pp. 2351-2355, Nov. 2015. Doi: 10.1109/LGRS.2015.2478256.

\bibitem{restrain_noise} 
 J. Geng, H. Wang, J. Fan, et al., ``Deep supervised and contractive neural network for SAR image classification,''  \emph{IEEE Transactions on Geoscience and Remote Sensing}, vol. 55, no. 4, pp. 2442-2459, April 2017. Doi: 10.1109/TGRS.2016.2645226.

\bibitem{superpixel_classification_PolSAR} 
 J. Geng, X. Ma, J. Fan, et al., ``Semisupervised classification of polarimetric SAR image via superpixel restrained deep neural network,'' \emph{IEEE Geoscience and Remote Sensing Letters}, vol. 15, no. 1, pp. 122-126, Jan. 2018. Doi: 10.1109/LGRS.2017.2777450.

\bibitem{noise_invariant_CNN} 
 Y. Kwak, W. Song, S. Kim, ``Speckle-noise-invariant convolutional neural network for SAR target recognition,'' \emph{IEEE Geoscience and Remote Sensing Letters}, vol. 16, no. 4, pp. 549-553, April 2019. Doi: 10.1109/LGRS.2018.2877599.

\bibitem{TL_robustness} 
 Z. Huang, C. O. Dumitru, Z. Pan, et al., ``Classification of large-scale high-resolution SAR images with deep transfer learning,''  \emph{IEEE Geoscience and Remote Sensing Letters}. Doi: 10.1109/LGRS.2020.2965558.

\bibitem{complex_backgroud_SAR} 
 Y. Zhou et al., ``Complex background SAR target recognition based on convolution neural network,'' in \emph{Proc. 2019 6th Asia-Pacific Conference on Synthetic Aperture Radar (APSAR)}, Xiamen, China, 2019, pp. 1-4. Doi: 10.1109/APSAR46974.2019.9048279.

\bibitem{advanced_technology_noise_phase_errors} 
 N. Inkawhich, E. Davis, U. Majumder, et al., ``Advanced techniques for robust SAR ATR: Mitigating noise and phase errors,'' in \emph{Proc. 2020 IEEE International Radar Conference (RADAR)}, Washington, DC, USA, 2020, pp. 844-849. Doi: 10.1109/RADAR42522.2020.9114784.

\bibitem{noise_clutter_robustness} 
 C. Li, L. Du, S. Deng, et al., ``Point-wise discriminative auto-encoder with application on robust radar automatic target recognition,'' \emph{Signal Processing}, 2020.

\bibitem{local_spatial_information_SSAE_denoising} 
 L. Zhang, W. Ma, D. Zhang, et al., ``Stacked sparse autoencoder in PolSAR data classification using local spatial information,'' \emph{IEEE Geoscience and Remote Sensing Letters}, vol. 13, no. 9, pp. 1359-1363, 2016.
\bibitem{scale_invariant_CNN} 
 Y. Jiang, Z. Chi, ``A scale-invariant framework for image classification with deep learning,'' in \emph{Proc. 2017 IEEE International Conference on Systems, Man, and Cybernetics (SMC)}, Banff, AB, 2017, pp. 1019-1024. Doi: 10.1109/SMC.2017.8122744.

\bibitem{SAR_recognition_adversarial_attacks} 
 T. Huang, Q. X. Zhang, J. B. Liu, et al., ``Adversarial attacks on deep-learning-based SAR image target recognition,'' \emph{Journal of Network and Computer Applications}, vol. 162, no. 15, July, 2020.

\bibitem{hySARNet_robustness} 
 R. J Soldin, D. N Macdonald, M. D Reisman, et al., ``HySARNet: A hybrid machine learning approach to synthetic aperture radar automatic target recognition,'' in \emph{Proc. SPIE 10988, Automatic Target Recognition XXIX}, May 2019, https://doi.org/10.1117/12.2518155.

\bibitem{deep_AEs_unbalanced_PolSAR_data_classify} 
 X. Chen, J. Deng, ``A robust polarmetric SAR terrain classification based on sparse deep autoencoder model combined with wavelet kernel-based classifier,,'' \emph{IEEE Access}, vol. 8, pp. 64810-64819, 2020. Doi: 10.1109/ACCESS.2020.2983478.

\bibitem{joint_distribution_adaptation_Net_TL_SAR_classifiy} 
 J. Geng, X. Deng, X. Ma, ``Transfer learning for SAR image classification via deep joint distribution adaptation networks,'' \emph{IEEE Transactions on Geoscience and Remote Sensing}, doi: 10.1109/TGRS.2020.2964679.

\bibitem{incremental_learning_PolSAR_image_classification} 
 J. Fan, X. Wang, X. Wang, et, al., ``Incremental Wishart broad learning system for fast PolSAR image classification,''  \emph{IEEE Geoscience and Remote Sensing Letters}, vol. 16, no. 12, pp. 1854-1858, Dec. 2019. Doi: 10.1109/LGRS.2019.2913999.

\bibitem{fast_Wishart_network} 
 T. Gadhiya, A. K. Roy, ``Superpixel-driven optimized Wishart network for fast PolSAR image classification using global ${k}$ -means algorithm,''  \emph{IEEE Transactions on Geoscience and Remote Sensing}, vol. 58, no. 1, pp. 97-109, Jan. 2020. Doi: 10.1109/TGRS.2019.2933483.

\bibitem{computation_cost_tricks_DNN} 
 Z. Wang, X. Xu, ``Efficient deep convolutional neural networks using CReLU for ATR with limited SAR images,'' \emph{The Journal of Engineering}, vol. 2019, no. 21, pp. 7615-7618, 11 2019. Doi: 10.1049/joe.2019.0567.

\bibitem{computation_cost_spatial_anchor_graph} 
 H. Liu, S. Yang, S. Gou, et al., ``Fast classification for large polarimetric SAR data based on refined spatial-anchor graph,'' \emph{IEEE Geoscience and Remote Sensing Letters}, vol. 14, no. 9, pp. 1589-1593, Sept. 2017. Doi: 10.1109/LGRS.2017.2724844.

\bibitem{pixel_grayscale_graph_CNN} 
 H. Zhu, N. Lin, H. Leung, et al., ``Target classification from SAR imagery based on the pixel grayscale decline by graph convolutional neural network,'' \emph{IEEE Sensors Letters}, vol. 4, no. 6, pp. 1-4, June 2020. Doi: 10.1109/LSENS.2020.2995060.

\bibitem{shallow_features_computation_cost} 
 D. Li, Y. Gu, S. Gou, et al., ``Full polarization SAR image classification using deep learning with shallow feature,'' in \emph{Proc. 2017 IEEE International Geoscience and Remote Sensing Symposium (IGARSS)}, Fort Worth, TX, 2017, pp. 4566-4569. Doi: 10.1109/IGARSS.2017.8128018.

\bibitem{real_time_embedded_device} 
 C. Zhao, P. Wang, J. Wang, et al., ``A maritime target detector based on CNN and embedded device for GF-3 images,'' in \emph{Proc. 2019 6th Asia-Pacific Conference on Synthetic Aperture Radar (APSAR)}, Xiamen, China, 2019, pp. 1-4. Doi: 10.1109/APSAR46974.2019.9048264.

\bibitem{flexible_structure_CNN} 
 S. Wagner, K. Barth, S. Br¨¹ggenwirth, ``A deep learning SAR ATR system using regularization and prioritized classes,'' in \emph{Proc. 2017 IEEE Radar Conference (RadarConf)}, Seattle, WA, 2017, pp. 0772-0777. Doi: 10.1109/RADAR.2017.7944307.

\bibitem{small_samples_SAR_ATR} 
 Z. Ying, C. Xuan, Y. Zhai, et al., ``TAI-SARNET: Deep transferred atrous-inception CNN for small samples SAR ATR,'' \emph{Sensors}, vol. 20, no. 6, 2020.

\bibitem{high_speed_SAR_classify} 
 C. Ozcan, O. K .Ersoy, I. U. Ogul, et al., ''Fast texture classification of denoised SAR image patches using GLCM on spark, \emph{Turkish Journal of Electrical Engineering and Computer Sciences}, vol. 28, no. 1, pp. 182-195, 2020.

\bibitem{parallel_CNN_alleviate_computation_cost} 
 Y. Hou, Y. Bai, T. Xu, et al., ``Deep convolutional neural network structural design for synthetic aperture radar image target recognition based on incomplete training data and displacement insensitivity,'' \emph{Journal of Electronic Imaging}, vol. 28, no. 5, 2019.

\bibitem{semi_random_real_time} 
 H. Syed, R. Bryla, U. K. Majumder, et al., ``Semi-random deep neural networks for near real-time target classification,'' \emph{Algorithms for Synthetic Aperture Radar Imagery XXVI}, 2019. Doi: 10.1117/12.2520237.

\bibitem{micro_CNN_knowledge_distillation} 
 R. Min, H. Lan, Z. Cao, et al., ``A gradually distilled CNN for SAR target recognition,'' \emph{IEEE Access}, pp. 42190-42200, 2019.

\bibitem{NN_acceleration_three_strategies} 
 H. Chen, F. Zhang, B. Tang, et al., ``Slim and efficient neural network design for resource-constrained SAR target recognition,'' \emph{Remote Sensing}, vol. 10, no. 10, 2018.

\bibitem{Zhang198} 
 J. H. Zhang, H. J. Song, B. B. Zhou, ``SAR target classification based on deep forest model,''  \emph{Remote Sensing}, vol. 12, pp. 128, 2020.

\bibitem{Xue199} 
 Y. Xue, J. F. Pei, Y. L. Huang, et al., ``Target recognition for SAR images based on heterogeneous CNN ensemble,'' \emph{2018 IEEE Radar Conference (RadarConf18)}, 2018, pp. 0507-0512.


\bibitem{Coman200} 
 C. Coman, R. Thaens, ``A deep learning SAR target classification experiment on MSTAR dataset,'' in \emph{Proc. 2018 19th International Radar Symposium (IRS)}, pp. 1-6, 2018.

\bibitem{Odysseas201} 
 O. Kechagias-Stamatis, N. Aouf, ``Fusing deep learning and sparse coding for SAR ATR,'' \emph{IEEE Transactions on Aerospace and Electronic Systems}, vol. 55, no. 2, pp. 785-797, 2019.

\bibitem{Kazemi202} 
 S. Kazemi, B. Yonel, B. Yazici, ``Deep learning for direct automatic target recognition from SAR Data,'' in \emph{Proc. 2019 IEEE Radar Conference (RadarConf)}, pp. 1-6, 2019.

\bibitem{Liu203} 
 J. J. Liu, X. J. Fu, K. Q. Liu, et al., ``Spotlight SAR image recognition based on dual-channel feature map convolutional neural network,'' in \emph{Proc. 2019 IEEE 4th International Conference on Signal and Image Processing (ICSIP)}, pp. 65-69, 2019.

\bibitem{Dong204} 
 Y.B. Dong, H. Zhang, C. Wang, Y. Y. Wang, ``Fine-grained ship classification based on deep residual learning for high-resolution SAR images,'' \emph{Remote Sensing Letters}, vol. 10, pp. 1095, 2019.

\bibitem{Gao205} 
 G. Gao, L. Liu, L. Zhao, et al., ``An adaptive and fast CFAR algorithm based on automatic censoring for target detection in high-resolution SAR images,'' \emph{IEEE Trans. Geosci. Remote Sens.}, vol. 47, pp. 1685-1697, 2008.

\bibitem{Farrouki206} 
 A. Farrouki, M. Barkat, ``Automatic censoring CFAR detector based on ordered data variability for nonhomogeneous environments,''  in \emph{Proc. IEE Proc.-Radar Sonar Navig.}, 2005, pp. 43-51.

\bibitem{Huang207} 
 Huang, X.; Yang, W.; Zhang, H.; Xia, G.S. Automatic ship detection in SAR images using multi-scale heterogeneities and an acontrario decision. Remote Sens. 2015, 7, 7695-7711.

\bibitem{Xing208} 
 X. Xing, K. Ji, H. Zou, ``Ship classification in TerraSAR-X images with feature space based sparse representation,''  \emph{IEEE Geosci. Remote Sens. Lett.}, vol. 10, no. 6, pp. 1562-1566, Nov. 2013, [online] Available: http://dx.doi.org/10.1109/LGRS.2013.2262073.

\bibitem{Ma209} 
 F. Ma, F. Gao, J. Wang, et al., ``A novel biologically inspired target detection method based on saliency analysis in SAR imagery,''  \emph{Neurocomputing}, 2019.

\bibitem{Xie210} 
 T. Xie, W. Zhang, L. Yang,  ``Inshore ship detection based on level set method and visual saliency for SAR images,''  \emph{Sensors (Basel)}, vol. 18, no. 11, pp. 3877, 2018. Doi:10.3390/s18113877.

\bibitem{Girshick211} 
 R. Girshick, J. Donahue, T. Darrell, ``Rich feature hierarchies for accurate object detection and semantic segmentation,''  in \emph{Proc. of the IEEE Conference on Computer Vision and Pattern Recognition}, Columbus, OH, USA, 23-28 June 2014, pp. 580-587.

\bibitem{Girshick212} 
 R. Girshick,  ``Fast r-cnn,''  in \emph{Proc. of the IEEE International Conference on Computer Vision}, Santiago, Chile, 7-13 Dec. 2015, pp. 1440-1448.

\bibitem{Ren213} 
 S. Ren, K. He, R. Girshick, ``Faster R-CNN: Towards real-time object detection with region proposal networks,''  \emph{IEEE Transactions on Pattern Analysis and Machine Intelligence}, vol. 39, no. 6, pp. 1137-1149, 2016.

\bibitem{He214} 
 K. He, G. Gkioxari, P. Doll¨¢r, ``Mask r-cnn. \emph{In Proceedings of the IEEE International Conference on Computer Vision}, Venice, Italy, 22-29 Oct. 2017, pp. 2961-2969.

\bibitem{Cai215} 
 Z. Cai, N. Vasconcelos, ``Cascade r-cnn: Delving into high quality object detection,'' \emph{In Proceedings of the IEEE Conference on Computer Vision and Pattern Recognition}, Salt Lake City, UT, USA, 18-23 June 2018, pp. 6154-6162.

\bibitem{Lin216} 
 T. Y. Lin, P. Doll¨¢r, R. Girshick, ``Feature pyramid networks for object detection,'' \emph{In Proceedings of the IEEE Conference on Computer Vision and Pattern Recognition}, Honolulu, HI, USA, 21-26 July 2017, pp. 2117-2125.

\bibitem{Redmon217} 
 J. Redmon, S. Divvala, R. Girshick, `` You only look once: Unified, real-time object detection,'' \emph{In Proceedings of the IEEE Conference on Computer Vision and Pattern Recognition}, Las Vegas, NV, USA, 27-30 June 2016, pp. 779-788.

\bibitem{Redmon218} 
 J. Redmon, A. Farhadi, ``YOLO9000: Better, faster, stronger,''  \emph{In Proceedings of the IEEE Conference on Computer Vision and Pattern Recognition}, Honolulu, HI, USA, 21-26 July 2017, pp. 7263-7271.

\bibitem{Redmon219} 
 J. Redmon, A. Farhadi, ``Yolov3: An incremental improvement,''  \emph{ArXiv}: 1804.02767, 2018.

\bibitem{YOLOv4} 
 B. Alexey, W. Chien-Yao, L. H. Y. Mark,  ``YOLOv4: Optimal speed and accuracy of object detection,'' \emph{ArXiv}: 2004.10934, 2020.

\bibitem{ploy-YOLOv3} 
 P. Hurtik, V. Molek, J. Hula, et al., ``Poly-YOLO: higher speed, more precise detection and instance segmentation for YOLOv3,'' \emph{arXiv}: 2005.13243, 2020.

\bibitem{Liu220} 
 W. Liu, D. Anguelov, D. Erhan, et al., ``Ssd: Single shot multibox detector,'' \emph{In Proceedings of the 14th European Conference on Computer Vision}, Amsterdam, The Netherlands, 8-16 October 2016, pp. 21-37.

\bibitem{Lin221} 
 T. Y. Lin, P. Goyal, R. Girshick, He, ``Focal loss for dense object detection,'' \emph{In Proceedings of the IEEE International Conference on Computer Vision}, Venice, Italy, 22-29 October 2017, pp. 2980-2988.

\bibitem{fater_RCNN_SAR} 
 J. Li, C. Qu, J. Shao, ``Ship detection in SAR images based on an improved faster R-CNN,''  in \emph{Proc.2017 SAR in Big Data Era: Models, Methods and Applications (BIGSARDATA)}, Beijing, 2017, pp. 1-6.

\bibitem{Fan224} 
 Q. Fan, F. Chen, M. Cheng, et al., ``Ship detection using a fully convolutional network with compact polarimetric SAR images,''  \emph{Remote Sensing}, vol. 11, no. 18, 2019.

\bibitem{Gao225} 
 F. Gao, W. Shi, J. Wang, et al., ``Enhanced feature extraction for ship detection from multi-resolution and multi-scene synthetic aperture radar (SAR) images,''  \emph{Remote Sens.}, vol. 11, no. 22, 2019.

\bibitem{Ma226} 
 F. Ma, F. Gao, J. Wang, et al., ``A novel biologically inspired target detection method based on saliency analysis in SAR imagery,''  \emph{Neurocomputing}, 2019.

\bibitem{Kang227} 
 M. Kang, K. Ji, X. Leng, et al.,  ``Contextual region-based convolutional neural network with multilayer fusion for SAR ship detection,'' \emph{Remote Sensing}, vol. 9, no. 8, 2017.

\bibitem{An228} 
 Q. An, Z. Pan, H. You. ``Ship detection in Gaofen-3 SAR images based on sea clutter distribution analysis and deep convolutional neural network,''  \emph{Sensors (Basel)}, vol. 18, no. 2, pp. 334,  2018. Doi:10.3390/s18020334.

\bibitem{Yuanyuan229} 
 Y. Y. Wang, C. Wang, H. Zhang, `` Combining a single shot multibox detector with transfer learning for ship detection using sentinel-1 SAR images,''  \emph{Remote Sensing Letters}, vol. 9, no. 8, pp. 780-788, 2018. Doi: 10.1080/2150704X.2018.1475770.

\bibitem{Xie230} 
 T. Xie, W. Zhang, L. Yang, ``Inshore ship detection based on level set method and visual saliency for SAR images,'' \emph{Sensors (Basel)}, vol. 18, no. 11, 2018. Doi:10.3390/s18113877.

\bibitem{Yuanyuan231} 
 Y. Y. Wang, C. Wang, H. Zhang, et al., ``Automatic ship detection based on RetinaNet using multi-resolution Gaofen-3 imagery,''  \emph{Remote Sensing}, vol. 11, no. 5, 2019.

\bibitem{Jiao232} 
 J. Jiao, Y. Zhang, H. Sun, ``A densely connected end-to-end neural network for multiscale and multiscene SAR ship detection,''  \emph{IEEE Access}, vol. 6, pp. 20881-20892, 2018.

\bibitem{two_sub_channels_ship_detection} 
 T. Zheng, J. Wang, P. Lei, ``Deep learning based target detection method with multi-features in SAR imagery,''  in \emph{Proc. 2019 6th Asia-Pacific Conference on Synthetic Aperture Radar (APSAR)}, Xiamen, China, 2019, pp. 1-4. Doi: 10.1109/APSAR46974.2019.9048509.

\bibitem{Hamza233} 
 H. M. Khan, Y. Z. Cai, ``Ship detection in SAR Image using YOLOv2,'' in \emph{Proc. 2018 37th Chinese Control Conference (CCC)}, 2018, pp. 9495-9499.

\bibitem{Gui234} 
 Y. C. Gui, X. H. Li, L. Xue, ``A scale transfer convolution network for small ship detection in SAR images,'' in \emph{Proc. 2019 IEEE 8th Joint International on Information Technology and Artificial Intelligence Conference (ITAIC)}, 2019, pp. 1845-1849.

\bibitem{Chen235} 
 C. Chen, C. He, C. H. Hu, ``A deep neural network based on an attention mechanism for SAR ship detection in multiscale and complex scenarios,'' \emph{IEEE Access}, vol. 7, pp. 104848-104863, 2019.

\bibitem{Zhang236} 
 X. H. Zhang, H. P. Wang, C. G. Xu, ``A lightweight feature optimizing network for ship detection in SAR image,'' \emph{IEEE Access}, vol. 7, pp. 141662-141678, 2019.

\bibitem{Cui237} 
 Z. Y. Cui, Q. Li, Z. J. Cao, ``Dense attention pyramid networks for multi-scale ship detection in SAR images,''  \emph{IEEE Transactions on Geoscience and Remote Sensing}, vol. 57, no. 11, pp. 8983-8997, 2019.

\bibitem{Chen238} 
 C. Chen, C. He, C. H. Hu, et al., ``MSARN: A deep neural network based on an adaptive recalibration mechanism for multiscale and arbitrary-oriented SAR ship detection,''  \emph{ IEEE Access}, vol. 7, pp. 159262-159283, 2019.

\bibitem{Liu239} 
 N. Y. Liu, Z. Y. Cui, Z. J. Cao, ``Scale-transferrable pyramid network for multi-scale ship detection in Sar images,''  in \emph{Proc. 2019 IEEE International Geoscience and Remote Sensing Symposium (IGARSS)}, 2019, pp. 1-4.

\bibitem{Li240} 
 Q. Li, R. Min, Z. Y. Cui, ``Multiscale ship detection based on dense attention pyramid network in Sar images,'' in \emph{Proc. 2019 IEEE International Geoscience and Remote Sensing Symposium (IGARSS)}, 2019, pp. 5-8.

\bibitem{Nie241} 
 X. Nie, M. Y. Duan, H. X. Ding, ``Attention mask R-CNN for ship detection and segmentation from remote sensing images,''  \emph{IEEE Access}, vol. 8, pp. 9325-9334, 2020.

\bibitem{Wei242} 
 S. J. Wei, H. Su, J. Ming,et al., ``Precise and robust ship detection for high-resolution SAR imagery based on HR-SDNet,''  \emph{Remote Sensing}, vol. 12, pp. 167, 2020.

\bibitem{Zhang243} 
 T. Zhang, X. Zhang,  ``High-speed ship detection in SAR images based on a grid convolutional neural network,''  \emph{Remote Sensing}, vol. 11, no. 1206, 2019.

\bibitem{Duan244} 
 Y. Duan, F. Liu, L. Jiao, et al., ``SAR image segmentation based on convolutional-wavelet neural network and markov random field,'' \emph{Pattern Recognition}, vol. 64, pp. 255-267, 2017. https://doi.org/10.1016/j.patcog.2016.11.015.

\bibitem{Wu245} 
 T. Wu, X. Ruan, X. Chen, et al., ``A modified method for the estimation of SAR target aspect angle based on MRF segmentation,''  in \emph{Proc. SPIE}, Oct. 2009, vol. 7495, pp. 74. [online] Available: http://dx.doi.org/10.1117/12.830709.

\bibitem{Ding246} 
 B. Y. Ding, G. J. Wen, C. H. Ma, et al.,  ``Evaluation of target segmentation on SAR target recognition,''  in \emph{Proc. 2017 4th International Conference on Information Cybernetics and Computational Social Systems (ICCSS)} , 2017, pp. 663-667.


\bibitem{ResNet_low_resolution_high_ISAR} 
 X. Gao, D. Qin, J. Gao, et al., ''Resolution enhancement for inverse synthetic aperture radar images using a deep residual network,''  \emph{Microwave and Optical Technology Letters}, vol. 62. no. 4, pp. 1588-1593, 2020.

\bibitem{CNNs_ISAR_imaging_speckle_robustness} 
 J. Mitchell, S. Tjuatja, ``ISAR imaging in the presence of quasi-random multiplicative noise using convolutional deep learning,'' in \emph{Proc. International Geoscience and Remote Sensing Symposium}, 2019, pp. 2583-2586.

\bibitem{sparse_Bayesian_learning_divided_subproblems_ISAR_imaging} 
 S. Zhang, Y. Liu, X. Li, ``Fast sparse aperture ISAR autofocusing and imaging via ADMM based sparse Bayesian learning,'' \emph{IEEE Transactions on Image Processing}, vol. 29, pp. 3213-3226, 2020. Doi: 10.1109/TIP.2019.2957939.

\bibitem{High_resolution_rotating_object_ISAR_imaging} 
 X. He, N. Tong, X. Hu, et al., ``High-resolution ISAR imaging of fast rotating targets based on pattern-coupled Bayesian strategy for multiple measurement vectors,'' \emph{Digital Signal Processing}, pp.151-159, 2019.

\bibitem{variational_Bayesian_inference_sparse_aperture_ISAR_imaging} 
 S. Zhang, Y. Liu, X. Li, et al., ``Joint sparse aperture ISAR autofocusing and scaling via modified newton method-based variational Bayesian inference,'' \emph{IEEE Transactions on Geoscience and Remote Sensing}, vol. 57, no. 7, pp. 4857-4869, 2019.

\bibitem{fast_semi_supervised_ISAR_object_detection_1} 
 B. Xue, N. Tong,  ``DIOD: Fast and efficient weakly semi-supervised deep complex ISAR object detection,'' \emph{IEEE Transactions on Systems, Man, and Cybernetics}, vol. 49, no. 11, pp. 3991-4003, 2019.

\bibitem{fast_semi_supervised_ISAR_object_detection_2} 
 Xue B, Tong N, Xu X, et al. DIOD: Fast, Semi-Supervised Deep ISAR Object Detection[J]. IEEE Sensors Journal, 2019, 19(3): 1073-1081.

\bibitem{TL_stacking_algorithm_ISAR_space_targets_recognition} 
 H. Yang, Y. Zhang, W. Ding, ``Multiple heterogeneous P-DCNNs ensemble with stacking algorithm: A novel recognition method of space target ISAR images under the condition of small sample set,''  \emph{IEEE Access}, vol. 8, pp. 75543-75570, 2020. Doi: 10.1109/ACCESS.2020.2989162.

\bibitem{deep_relative_graph_learning_ISAR_object_recognition} 
 B. Xue, N. Tong, ``Real-world ISAR object recognition and relation discovery using deep relation graph learning,''  \emph{IEEE Access}, vol. 7, pp. 43906-43914, 2019. Doi: 10.1109/ACCESS.2019.2896293.

\bibitem{CNNs_CAEs_ISAR_objects_SAR} 
 S. Zaied, A. Toumi, A. Khenchaf, et al., ``Target classification using convolutional deep learning and auto-encoder models,'' in \emph{Proc. International Conference on Advanced Technologies for Signal and Image Processing}, 2018, pp. 1-6.

\bibitem{three_ML_ISAR_objects_classifcation} 
 K. D. Uttecht, C. X. Chen, J. C. Dickinson, et al., ``A comparison of machine learning methods for target recognition using ISAR imagery,'' in \emph{Proc. Proceedings of SPIE}, 2011.

\bibitem{SAE_ISAR_objects_classifcation} 
 X. He, N. Tong, X. Hu, ``Automatic recognition of ISAR images based on deep learning,'' in \emph{Proc. 2016 CIE International Conference on Radar (RADAR)}, Guangzhou, 2016, pp. 1-4. Doi: 10.1109/RADAR.2016.8059230.

\bibitem{Li247} 
 B. Li, H. Li, W. B. Huang, ``Modified PPCA methods for radar HRRP robust recognition,'' \emph{Journal of Northwestern Polytechnical University} , vol. 34, no. 6, 2016.(in Chinese)

\bibitem{Yuan248} 
 J. Yuan, W. Liu, G. Zhang, ``Application of dictionary learning algorithm in HRRP based on statistical modeling,''  \emph{XI Tong Gong Cheng Yu Dian Zi Ji Shu/systems Engineering \& Electronics}, vol. 40, no. 4, pp. 762-767, 2018.

\bibitem{Xu249} 
 D. L. Xu, L. Du, P. H. Wang, et al., ``Radar HRRP target recognition by utilizing multitask sparse learning with a small training data size,'' \emph{Journal of Xidian University}, vol. 43, no. 2, pp. 23-28, 2016.(in Chinese)

\bibitem{Li250} 
 L. Li, Z. Liu, ``Noise-robust multi-feature joint learning HRRP recognition method,'' \emph{Journal of Xidian University}£¬ vol, 45, pp. 57-62, 2018.

\bibitem{Guo251} 
 Y. Guo, H. Xiao, Y. Kan, et al.,  ``Learning using privileged information for HRRP-based radar target recognition,''  \emph{IET Signal Processing} vol. 12, no. 2, pp. 188-197, 2017.

\bibitem{Li252} 
 M. Li, G. Zhou, B. Zhao, et al., ``Sparse representation denoising for radar high resolution range profiling,''  \emph{International Journal of Antennas and Propagation}, pp. 1-8, 2014.

\bibitem{Wu253} 
 J. N. Wu, Y. G. Chen, D. J. Feng, et al., ``Target recognition for polarimetric HRRP based on pre-classification and model matching,'' \emph{Systems Engineering and Electronics}, 2010. http://www.sys-ele.com/EN/10.3969/j.issn.1001-506X.2016.09.01

\bibitem{Li254} 
 L. Li, Z. Liu, T. Li, et al., ``Radar high-resolution range profile feature extraction method based on multiple kernel projection subspace fusion,''  \emph{IET Radar Sonar and Navigation}, vol. 12. no. 4, pp. 417-425, 2018.

\bibitem{Pan255} 
 M. Pan, J. Jiang, Q. Kong, et al., ``Radar HRRP target recognition based on t-SNE segmentation and discriminant deep belief network,''  \emph{IEEE Geoscience and Remote Sensing Letters}, vol. 14, no. 9, pp. 1609-1613, 2017.

\bibitem{Peng256} 
 X. Peng, X. Gao, Y. Zhang, et al.,  ``An adaptive feature learning model for sequential radar high resolution range profile recognition,''  \emph{Sensors}, vol. 17, no. 7, 2017.

\bibitem{Liao257} 
 K. Liao, J. Si, F. Zhu, et al., ``Radar HRRP target recognition based on concatenated deep neural networks,''  \emph{IEEE Access}, pp. 29211-29218, 2018.

\bibitem{Zhao258} 
 F. Zhao, Y. Liu, K. Huo, et al., ``Radar HRRP target recognition based on stacked autoencoder and extreme learning machine,'' \emph{Sensors}, vol. 18, no. 1, 2018.

\bibitem{Jiang259} 
 Y. Jiang, Y. Li, J. Cai, et al., ``Robust automatic target recognition via HRRP sequence based on scatterer matching,'' \emph{Sensors}, vol. 18, no. 2, 2018. Doi: 10.3390/s18020593.

\bibitem{Du260} 
 L. Du, H. He, L. Zhao, et al., ``Noise robust radar HRRP target recognition based on scatterer matching algorithm,'' \emph{ IEEE Sensors Journal}, vol. 16, no. 6, pp. 1743-1753, March15, 2016.

\bibitem{Li261} 
 L. Li, Z. Liu,  ``Noise-robust HRRP target recognition method via sparse-low-rank representation,''  \emph{Electronics Letters}, vol. 53, no. 24, pp. 1602-1604, 2017.

\bibitem{Dai262} 
 W. Dai, G. Zhang, Y. Zhang, et al.,  ``HRRP classification based on multi-scale fusion sparsity preserving projections,''  \emph{Electronics Letters}, vol. 53, no. 11, pp. 748-750, 2017.

\bibitem{Liu263} 
 J. Liu, N. Fang, Y. J. Xie, et al., ``Scale-space theory-based multi-scale features for aircraft classification using HRRP,'' \emph{ Electronics Letters}, vol. 52, no. 6, pp. 475-477, 2016.

\bibitem{FPN_CNN} 
 C. Guo, H. Wang, T. Jian, et al., ``Radar target recognition based on feature pyramid fusion lightweight CNN,'' \emph{IEEE Access}, vol. 7, pp. 51140-51149, 2019. Doi: 10.1109/ACCESS.2019.2909348.

\bibitem{Zhou264} 
 D. Zhou, ''Radar target HRRP recognition based on reconstructive and discriminative dictionary learning,'' \emph{Signal Processing}, pp. 52-64, 2016.

\bibitem{Xu265} 
 B. Xu, B. Chen, J. Wan, et al.,  ``Target-aware recurrent attentional network for radar HRRP target recognition,''  \emph{Signal Processing}, pp. 268-280, 2019.

\bibitem{Mian266} 
 P. Mian, J. Jie, L. Zhu, et al., ``Radar HRRP recognition based on discriminant deep autoencoders with small training data size,''  \emph{lectronics Letters}, vol. 52, no. 20, pp. 1725-1727, 2016.

\bibitem{Xiong267} 
 W. Xiong, G. Zhang, S. Liu, et al.,  ``Multiscale kernel sparse coding-based classifier for HRRP radar target recognition,''  \emph{IET Radar Sonar and Navigation}, vol. 10, no. 9, pp. 1594-1602, 2016.

\bibitem{Guo268} 
 Y. Guo, H. Xiao, Q. Fu, et al.,  ``Least square support vector data description for HRRP-based radar target recognition,'' \emph{Applied Intelligence}, vol. 46, no. 2, pp. 365-372, 2017.

\bibitem{Feng269} 
 B. Feng, B. Chen, H. Liu, et al., ``Radar HRRP target recognition with deep networks,''  \emph{Pattern Recognition}, pp. 379-393, 2017.

\bibitem{Kim270} 
 Y. Kim, H. Ling, `` Human activity classification based on micro-doppler signatures using a support vector machine,''  \emph{IEEE Transactions on Geoscience and Remote Sensing}, vol. 47, no. 5, pp. 1328-1337, 2009.

\bibitem{Kim271} 
 Y. Kim, H. Ling, `` Human activity classification based on micro-Doppler signatures using an artificial neural network,''  in \emph{Proc. IEEE Antennas and Propagation Society International Symposium}, 2008, pp. 1-4.

\bibitem{Kim272} 
 Y. Kim, T. Moon, ``Human detection and activity classification based on micro-doppler signatures using deep convolutional neural networks,'' \emph{IEEE Geoscience and Remote Sensing Letters}, vol. 13, no. 1, pp. 8-12, 2016.

\bibitem{Fioranelli273} 
 F. Fioranelli, M. Ritchie, H. D. Griffiths, et al., ``Classification of unarmed/armed personnel using the NetRAD multistatic radar for micro-doppler and singular value decomposition features,''  \emph{IEEE Geoscience and Remote Sensing Letters}, vol. 12, no. 9, pp. 1933-1937, 2015.

\bibitem{Seyfioglu274} 
 Seyfioglu, M. Gurbuz, S.i Ozbayoglu, et al., ``Deep learning of micro-Doppler features for aided and unaided gait recognition,'' pp. 1125-1130, 2017.

\bibitem{Fioranelli275} 
 F. Fioranelli, M. Ritchie, S. Z. Gurbuz, et al., ``Feature diversity for optimized human micro-doppler classification using multistatic Radar,'' \emph{IEEE Transactions on Aerospace and Electronic Systems}, vol. 53, no. 2, pp. 640-654, 2017.

\bibitem{Zhang276} 
 J. Zhang, J. Tao, Z. Shi, et al., ``Doppler-radar based hand gesture recognition system using convolutional neural networks,'' in \emph{Proc. International Conference on Communications}, 2017, pp.  1096-1113.

\bibitem{Ren277} 
 J. Ren, X. Jiang, `` Regularized 2-D complex-log spectral analysis and subspace reliability analysis of micro-doppler signature for UAV detection,'' \emph{Pattern Recognition}, pp. 225-237, 2017.

\bibitem{Sherif278} 
 S. Abdulatif, F. Aziz, K. Armanious, ``A study of human body characteristics effect on micro-doppler-based person identification using deep learning,'' \emph{ArXiv}: 1811.07173, 2019.

\bibitem{Seyfioglu279} 
 M. S. Seyfioglu, B. Erol, S. Z. Gurbuz, et al.,  ``DNN transfer learning from diversified micro-doppler for motion classification,''  \emph{IEEE Transactions on Aerospace and Electronic Systems}, vol. 55, no. 5, pp. 2164-2180, 2019.

\bibitem{Seyfioglu280} 
 M. S. Seyfioglu, A. M. Ozbayoglu, S. Z. Gurbuz, et al.,  ``Deep convolutional autoencoder for radar-based classification of similar aided and unaided human activities,''  \emph{IEEE Transactions on Aerospace and Electronic Systems}, vol. 54, no. 4, pp. 1709-1723, 2018.

\bibitem{Seyfio281} 
 M. S. Seyfiofilu, S. Z. G¨¹rb¨¹z, ``Deep neural network initialization methods for micro-doppler classification with low training sample support,'' \emph{IEEE Geoscience and Remote Sensing Letters}, vol. 14, no. 12, pp. 2462-2466, Dec. 2017.

\bibitem{Parashar282} 
 K. N. Parashar, M. C. Oveneke, M. Rykunov, et al., ``Micro-doppler feature extraction using convolutional auto-encoders for low latency target classification,''  in \emph{Proc. 2017 IEEE Radar Conference (RadarConf)}, Seattle, WA, 2017, pp. 1739-1744.

\bibitem{Miller283} 
 A. W. Miller, C. Clemente, A. Robinson, et al., ``Micro-doppler based target classification using multi-feature integration,''  in \emph{Proc. 2013 IET Intelligent Signal Processing Conference (ISP 2013)}, London, 2013, pp. 1-6.

\bibitem{Klarenbeek284} 
 G. Klarenbeek, R. I. A. Harmanny, L. Cifola, ``Multi-target human gait classification using LSTM recurrent neural networks applied to micro-Doppler,'' in \emph{Proc. 2017 European Radar Conference (EURAD)}, Nuremberg, 2017, pp. 167-170.

\bibitem{Kim285} 
 B. K. Kim, H. Kang, S. Park, ``Drone classification using convolutional neural networks with merged doppler images,'' \emph{IEEE Geoscience and Remote Sensing Letters}, vol. 14, no. 1, pp. 38-42, Jan. 2017.

\bibitem{Tahmoush286} 
 D. Tahmoush, ``Review of micro-Doppler signatures,'' \emph{ IET Radar Sonar and Navigation}, vol. 9, no. 9, pp. 1140-1146, 2015.

\bibitem{Chen287} 
 V. C. Chen, ``Doppler signatures of radar backscattering from objects with micro-motions,'' \emph{IET Signal Processing}, vol. 2, no. 3, pp. 291-300,  2008.

\bibitem{Patel288} 
 J. S. Patel, F. Fioranelli, D. Anderson, et al., ``Review of radar classification and RCS characterisation techniques for small UAVs or drones,'' \emph{IET Radar Sonar and Navigation}, vol. 12, no. 9, pp. 911-919, 2018.

\bibitem{Ritchie289} 
 M. Ritchie, F. Fioranelli, H. Borrion, et al., ``Multistatic micro-doppler radar feature extraction for classification of unloaded/loaded micro-drones,'' \emph{IET Radar Sonar and Navigation}, vol. 11, no. 1, pp. 116-124, 2017.

\bibitem{Rahman290} 
 S. Rahman, D. A. Robertson, ''Radar micro-Doppler signatures of drones and birds at K-band and W-band,'' \emph{Scientific Reports}, vol. 8, no. 1, 2018.

\bibitem{Shi291} 
 X. Shi, F. Zhou, X. Bai, et al.,  ``Deceptive jamming for tracked vehicles based on micro-Doppler signatures,''  \emph{IET Radar Sonar and Navigation}, vol. 12, no. 8, pp. 844-852, 2018.

\bibitem{Thayaparan292} 
 T. Thayaparan, S. Abrol, E. Riseborough, et al., ``Analysis of radar micro-Doppler signatures from experimental helicopter and human data,''  \emph{IET Radar Sonar and Navigation}, vol. 1, no. 4, pp. 289-299, 2007.

\bibitem{Jokanovi293} 
 B. Jokanovic, M. Amin, ``Fall detection using deep learning in range-doppler radars,''  \emph{IEEE Transactions on Aerospace and Electronic Systems}, vol. 54, no. 1, pp. 180-189, Feb. 2018.

\bibitem{Wang294} 
 S. Wang, J. Song, J. Lien, et al., ``Interacting with soli: Exploring fine-grained dynamic gesture recognition in the radio-frequency spectrum,'' in \emph{Proc. User Interface Software and Technology}, 2016, pp. 851-860.

\bibitem{Li295} 
 Y. Li, Z. Peng, R. Pal, et al., ``Potential active shooter detection based on radar micro-doppler and range-doppler analysis using artificial neural network,''  \emph{IEEE Sensors Journal}, vol. 19, no. 3, pp. 1052-1063, 2019.

\bibitem{interference_classification} 
 R. Zhang, S. Cao, ``Support vector machines for classification of automotive radar interference,'' in \emph{Proc. 2018 IEEE Radar Conference (RadarConf18)}, Oklahoma City, OK, 2018, pp. 0366-0371. Doi: 10.1109/RADAR.2018.8378586.

\bibitem{jammer_classification} 
 Ferre R M, La Fuente A D, Lohan E S, et al., ``Jammer classification in GNSS bands via machine learning algorithms,'' \emph{Sensors}, vol. 19, no. 22, 2019.

\bibitem{dense_false_targets_jamming_recognition} 
 Z. Hao, W. Yu, W. Chen, ``Recognition method of dense false targets jamming based on time-frequency atomic decomposition,''  \emph{The Journal of Engineering}, vol. 2019, no. 20, pp. 6354-6358, Oct. 2019. Doi: 10.1049/joe.2019.0147.

\bibitem{jammer_classification_NN} 
 A. Mendoza, A. Soto, B. C. Flores, ``Classification of radar jammer FM signals using a neural network,'' in \emph{Pro. of SPIE}, 2017. https://doi.org/10.1117/12.2262059

\bibitem{satellite_interference_recognition} 
 Y. Yang, L. Zhu, ``An efficient way for satellite interference signal recognition via incremental learning,'' in \emph{Proc. International Symposium on Networks Computers and Communications}, 2019, pp. 1-5.

\bibitem{radio_interference_recognition} 
 M. Kong, J. Liu, Z. Zhang, et al., ``Radio ground-to-air interference signals recognition based on support vector machine,'' in \emph{Proc. International Conference on Digital Signal Processing}, 2015, pp. 987-990.

\bibitem{compound_jamming_recognition} 
 F. Ruo-Ran, ``Compound jamming signal recognition based on neural networks,'' in \emph{Proc. 2016 Sixth International Conference on Instrumentation and Measurement, Computer, Communication and Control (IMCCC)}, Harbin, 2016, pp. 737-740. Doi: 10.1109/IMCCC.2016.163.

\bibitem{jamming_classification_CNN} 
 Z. Wu, Y. Zhao, Z. Yin, et al., ``Jamming signals classification using convolutional neural network,'' in \emph{Proc. 2017 IEEE International Symposium on Signal Processing and Information Technology (ISSPIT)}, Bilbao, 2017, pp. 062-067. Doi: 10.1109/ISSPIT.2017.8388320.

\bibitem{jamming_recognition_CNN} 
 Q. Liu, W. Zhang, ``Deep learning and recognition of radar jamming based on CNN,'' in \emph{Proc. 2019 12th International Symposium on Computational Intelligence and Design (ISCID)}, Hangzhou, China, 2019, pp. 208-212. Doi: 10.1109/ISCID.2019.00054.

\bibitem{jamming_recognition_DL} 
 Y. Cai, K. Shi, F. Song, et al., ``Jamming pattern recognition using spectrum waterfall: A deep learning method,'' in \emph{Proc. 2019 IEEE 5th International Conference on Computer and Communications (ICCC)}, Chengdu, China, 2019, pp. 2113-2117. Doi: 10.1109/ICCC47050.2019.9064207.

\bibitem{Chongqing82} 
 ''Research on electronic jamming identification based on CNN,'' in \emph{Proc. 2019 IEEE interference on signal, data and signal processing}, Chongqing, 2019.

\bibitem{jamming_identification_DL} 
 L. Tingpeng, W. Manxi, P. Danhua, et al., ``Identification of jamming factors in electronic information system based on deep learning,'' in \emph{Proc. 2018 IEEE 18th International Conference on Communication Technology (ICCT)}, Chongqing, 2018, pp. 1426-1430, doi: 10.1109/ICCT.2018.8600065

\bibitem{active_jamming_recognition_CNN} 
 Y. Wang, B. Sun, N. Wang, ``Recognition of radar active-jamming through convolutional neural networks,'' \emph{The Journal of Engineering}, vol. 2019, no. 21, pp. 7695-7697, Nov. 2019. Doi: 10.1049/joe.2019.0659.

\bibitem{SAR_deception_jamming_recognition_CNN} 
 X. Tang, X. Zhang, J. Shi, et al., ``SAR deception jamming target recognition based on the shadow feature,''  in \emph{Proc. 2017 25th European Signal Processing Conference (EUSIPCO)}, Kos, 2017, pp. 2491-2495. Doi: 10.23919/EUSIPCO.2017.8081659.

\bibitem{jamming_recognition_MLCNN} 
 M. Zhu, Y. Li, Z. Pan, et al., ``Automatic modulation recognition of compound signals using a deep multi-label classifier: A case study with radar jamming signals,'' \emph{Signal Processing}, 2020.

\bibitem{jamming_prediction_DNN_LSTM} 
 Lee G, Jo J, Park C H, et al, ``Jamming prediction for radar signals using machine learning methods,'' \emph{ Security and Communication Networks}, pp. 1-9, 2020.

\bibitem{interference_recognition_AE_RNN} 
 Q. Wu, Z. Sun, X. Zhou, ``Interference detection and recognition based on signal reconstruction using recurrent neural network,'' in \emph{Proc. 2019 IEEE Globecom Workshops (GC Wkshps)}, Waikoloa, HI, USA, 2019, pp. 1-6. Doi: 10.1109/GCWkshps45667.2019.9024542.

\bibitem{DQN_anti_jamming_frequency_hopping} 
 L. Kang, J. Bo, L. Hongwei, et al.,  ``Reinforcement learning based anti-jamming frequency hopping strategies design for cognitive radar,'' in \emph{Proc. International Conference on Signal Processing}, 2018.

\bibitem{spatial_anti_jamming_game_RL} 
 C. Han, L. Huo, X. Tong, et al., ``Spatial anti-jamming scheme for internet of satellites based on the deep reinforcement learning and Stackelberg game,''  \emph{IEEE Transactions on Vehicular Technology}, vol. 69, no. 5, pp. 5331-5342, May 2020. Doi: 10.1109/TVT.2020.2982672.

\bibitem{RL_LSTM_anti_mutual_interference_automotive_radars} 
 P. Liu, Y. Liu, T. Huang, et al., ``Decentralized automotive radar spectrum allocation to avoid mutual interference using reinforcement learning,'' \emph{ArXiv}: Signal Processing, 2020.

\bibitem{GRU_anti_mutual_interference_automotive_radars} 
 J. Mun, H. Kim, J. Lee, ``A deep learning approach for automotive radar interference mitigation,'' in \emph{Proc. 2018 IEEE 88th Vehicular Technology Conference (VTC-Fall)}, Chicago, IL, USA, 2018, pp. 1-5. Doi: 10.1109/VTCFall.2018.8690848.

\bibitem{Wang296} 
 L. Wang, S. Fortunati, M. Greco, et al., ``Reinforcement learning-based waveform optimization for MIMO multi-target detection,''  in \emph{Proc. Asilomar Conference on Signals, Systems and Computers}, 2018, pp. 1329-1333.

\bibitem{end_to_end_learning_waveform_detection} 
 J. Wei, M. H. Alexander, S. Osvaldo, ``End-to-end learning of waveform generation and detection for radar system,'' \emph{ArXiv}: 1912.00802v1, 2019.

\bibitem{Naparstek299} 
 O. Naparstek, K. Cohen, ``Deep multi-user reinforcement learning for distributed dynamic spectrum access,''  \emph{IEEE Transactions on Wireless Communications}, vol. 18, no. 1, pp. 310-323, 2019.

\bibitem{Faganello300} 
 L. R. Faganello, R. Kunst, C. B. Both, et al., ``Improving reinforcement learning algorithms for dynamic spectrum allocation in cognitive sensor networks,'' in \emph{Proc. Wireless Communications and Networking Conference}, 2013, pp. 35-40.

\bibitem{You301} 
 S. You, M. Diao, L. Gao, et al.,  ``Deep reinforcement learning for target searching in cognitive electronic warfare,''  \emph{IEEE Access}, pp. 37432-37447, 2019.

\bibitem{Metcalf302} 
 J. G. Metcalf, S. D. Blunt, B. Himed, et al., ``A machine learning approach to cognitive radar detection,'' in \emph{Proc. IEEE radar conference}, 2015, pp. 1405-1411.

\bibitem{Elbir303} 
 A. M. Elbir, K. V. Mishra, Y. C. Eldar, et al., ``Cognitive radar antenna selection via deep learning,'' \emph{ IET Radar Sonar and Navigation}, vol. 13, no. 6, pp. 871-880, 2019.

\bibitem{Liu304} 
 Z. Liu, D. K. Ho, X. Xu, et al., ``Moving target indication using deep convolutional neural network,'' \emph{IEEE Access}, pp. 65651-65660, 2018.

\bibitem{sea_clutter_recognition} 
 X. B. Liu, Y. A. Xu, H. Ding, et al., ``High-dimensional feature extraction of sea clutter and target signal for intelligent maritime monitoring network,'' \emph{Computer Communications}, vol. 147, pp. 76-84, 2019.

\bibitem{NN_Neyman_Pearson} 
 M. P. Jaraboamores, M. Rosazurera, R. Gilpita, et al., ``Study of two error functions to approximate the Neyman¨CPearson detector using supervised learning machines,'' \emph{IEEE Transactions on Signal Processing}, vol. 57, no. 11, pp. 4175-4181, 2009.

\bibitem{few-shot-learning-survey} 
 Y. Q. Wang, Q. M. Yao. ``Few-shot Learning: A Survey,'' \emph{ArXiv}: 1904.05046v1, 2020.

\bibitem{few-shot-learning-optimization} 
 S. Ravi, H. Larochelle, ``Optimization as a model for few-shot learning,'' in \emph{Proc. International Conference on Learning Representations (ICLR)}, 2017.

\bibitem{zero-shot-learning-1} 
 R. Socher, M. Ganjoo, C. D. Manning, et al., ``Zero-shot learning through cross-modal transfer,'' in \emph{Proc. Neural Information Processing Systems}, 2013, PP. 935-943.

\bibitem{zero-shot-learning-2} 
 Z. M. Ding., H. D. Zhao, Y. Fu. ``Zero-Shot Learning,'' 2019. doi:10.1007/978-3-030-00734-8-6.

\bibitem{NAS_survey} 
 T. Elsken, J. H. Metzen, F. Hutter, et al.,  ``Neural architecture search: A survey,'' \emph{Journal of Machine Learning Research}, vol. 20, no. 55, pp. 1-21, 2019.

\bibitem{model_compression_survey} 
 Y. Cheng, D. Wang, P. Zhou, et al.,  ``A survey of model compression and acceleration for deep neural networks,''  \emph{ArXiv: Learning}, 2017.

\bibitem{black_box} 
 D. Castelvecchi, ``Can we open the black box of AI,'' \emph{Nature}, vol. 538, no. 7623, pp. 20-23,  2016.

\bibitem{XAI} 
 A. Adadi, M. Berrada, ``Peeking inside the black-box: A survey on explainable artificial intelligence (XAI),'' \emph{IEEE Access}, pp. 52138-52160, 2018.

\bibitem{medical_XAI} 
 E. Tjoa, C. Guan, ``A survey on explainable artificial intelligence (XAI): Towards medical XAI,'' \emph{ArXiv: Learning}, 2019.

\bibitem{LIME} 
 M. T. Ribeiro, S. Singh, C. Guestrin, et al.,  ``Why should I trust you?: Explaining the predictions of any classifier,''  in \emph{Proc. Knowledge Discovery and Data Mining}, 2016, pp. 1135-1144.

\bibitem{global_additive_explanation} 
 S. Tan, R. Caruana, G. Hooker, et al., ``Learning global additive explanations for neural nets using model distillation,'' \emph{ArXiv: Machine Learning}, 2018.

\bibitem{R-CNN_Unfolding_Latent_Structures} 
 T. Wu, W. Sun, X. Li, et al., ``Towards interpretable R-CNN by unfolding latent structures,''  \emph{ArXiv: Computer Vision and Pattern Recognition}, 2017.

\bibitem{visualization_1} 
 M. D. Zeiler, R. Fergus,''  ``Visualizing and understanding convolutional networks,'' in \emph{Proc. European Conference on Computer Vision}, 2014, pp. 818-833.

\bibitem{visualization_2} 
 J. Yosinski, J. Clune, A. Nguyen, et al.,  ``Understanding neural networks through deep visualization,''  \emph{ArXiv: Computer Vision and Pattern Recognition}, 2015.

\bibitem{visualization_3} 
 L. M. Zintgraf, T. S. Cohen, T. Adel, et al.,  ``Visualizing deep neural network decisions: Prediction difference analysis,'' in \emph{ Proc. International Conference on Learning Representations (ICLR)}, 2017.

\bibitem{attributes_analysis_1} 
 P. Kindermans, K. T. Schutt, M. Alber, et al.,  ``Learning how to explain neural networks: PatternNet and PatternAttribution,'' in \emph{Proc. International Conference on Learning Representations (ICLR)}, 2018.

\bibitem{attributes_analysis_2} 
 M. Ancona, E. Ceolini, C. Oztireli, et al.,  ``Towards better understanding of gradient-based attribution methods for Deep Neural Networks,'' in \emph{Proc. International Conference on Learning Representations (ICLR)}, 2018, pp. 0-0.

\bibitem{interpretable_model_1} 
 B. Letham, C. Rudin, T. H. Mccormick, et al., ``Interpretable classifiers using rules and Bayesian analysis: Building a better stroke prediction model,'' \emph{ The Annals of Applied Statistics}, vol. 9, no. 3, pp. 1350-1371, 2015.

\bibitem{interpretable_model_2} 
 V. Joel, S. Agus, et al., ``Explainable neural networks based on additive index models,'' \emph{ 	 ArXiv}: 1806.01933, 2018.

\bibitem{interpretable_model_3} 
 A. M. David, J. Tommi,  ``Towards robust interpretability with self-explaining neural networks,'' in \emph{Proc. Neural Information Processing Systems (NIPS)}, 2018, pp. 7786-7795.

\bibitem{regularized_explanation_1} 
 A. S. Maruan, D. Avinava, P. X. Eric, ``Contextual explanation networks,'' \emph{ ArXiv}: 1705.10301, 2018.

\bibitem{regularized_explanation_2} 
 Q. S. Zhang, Y. N. Wu, S. C. Zhu, ``Interpretable convolutional neural networks,'' \emph{ArXiv}: 1710.00935, 2018.

\bibitem{disentangled_1} 
 T. D. Kulkarni, W. F. Whitney, P. Kohli, et al., ``Deep convolutional inverse graphics network,'' in \emph{Proc. Neural Information Processing Systems (NIPS)}, 2015, pp. 2539-2547.

\bibitem{disentangled_2} 
 K. Hyunjik, M. Andry,  ``Disentangling by factorising,'' \emph{ArXiv}: 1802.05983v3, 2019.

\bibitem{attention_mechanism_1} 
 M. David, T. Philip, et al., ``Transparency by design: Closing the gap between performance and interpretability in visual reasoning,'' \emph{ArXiv}: 1803.05268, 2018.

\bibitem{attention_mechanism_2} 
 Xiaoran X., Songpeng Z., and et al. Modeling Attention Flow on Graphs. ArXiv:1811.00497, 2018.

\bibitem{information_bottleneck_1} 
 N. Tishby, N. Zaslavsky, ``Deep learning and the information bottleneck principle,'' in \emph{Proc. Information Theory Workshop}, 2015, pp. 1-5.

\bibitem{information_bottleneck_2} 
 R. Shwartzziv, N. Tishby, ``Opening the black box of deep neural networks via information,'' \emph{ ArXiv: Learning}, 2017.

\bibitem{physics_1} 
 Q. Li, B. Wang, M. Melucci, et al.,  ``CNM: An interpretable complex-valued network for matching,'' \emph{ArXiv: Computation and Language}, 2019.

\bibitem{physics_2} 
 Y. P. Lu, Z. H. Li, et al., ``Understanding and improving transformer from a multi-particle dynamic system point of view,'' \emph{Arxiv}: 1906 02762, 2019.

\bibitem{human_cognitive_1} 
 P. Vincke, ``Analysis of multicriteria decision aid in europe,'' \emph{European Journal of Operational Research}, vol. 25, no. 2, pp. 160-168, 1986.

\bibitem{human_cognitive_2} 
 M.Z. Guo, Q.P. Zhang, X.W Liao, et al., ``An interpretable machine learning framework for modelling human decision behavior,'' \emph{ArXiv}: 1906.01233, 2019.

\bibitem{human_cognitive_3} 
 B. M. Lake, R. Salakhutdinov, J. B. Tenenbaum, et al., ``Human-level concept learning through probabilistic program induction,'' \emph{Science}, vol. 350, no. 6266, pp. 1332-1338, 2015.

\bibitem{symbolism_connectionism_1} 
 S. Murray, N. Kyriacos, C. Antonia, et al., ``An explicitly relational neural network architecture,'' \emph{ArXiv}: 1905.10307v1, 2019.

\bibitem{Neural_symbolic} 
 J. Y. Mao, C. Gan, K. Pushmeet, et al., ``The neuro-symbolic concept learner: Interpreting scenes, words, and sentences from natural supervision,'' in \emph{Proc. International Conference on Learning Representation (ICLR)}, 2019.

\bibitem{casual_inference} 
 D. Goldfarb, S. C. Evans, ``Causal inference via conditional Kolmogorov complexity using MDL Binning,'' \emph{ArXiv: Learning}, 2019.

\bibitem{representation_learning} 
  Y. Bengio, ``Deep learning of representations: Looking forward,'' in \emph{Proc. International Conference on Statistical Language and Speech Processing} ,Springer, 2013, pp. 1-37.

\bibitem{symbolic_AI} 
 M. Garnelo, M. Shanahan, ``Reconciling deep learning with symbolic artificial intelligence: representing objects and relations,'' in \emph{Proc. Current Opinion in Behavioral Sciences 29}, 2019, pp. 17¨C23.

\bibitem{consciousness_prior} 
 Y. Bengio, ``The consciousness prior,''  \emph{ArXiv}: 1709.08568, 2017.

\bibitem{EW} 
 G.N.Rao, C.V.S.Sastry, N.Divakar, ``Trends in electronic warfare,'' \emph{ IETE Tech. Rev.}, vol. 20, no. 2, pp. 139¨C150, Feb. 2003.

\bibitem{cognitive_radar_1} 
 S. Haykin, ``Cognitive radar: a way of the future,'' \emph{IEEE Signal Processing Magazine}, vol. 23, no. 1, pp. 30-40, 2006.

\bibitem{cognitive_radar_2} 
 S. Haykin, Y. Xue, P. Setoodeh, ``Cognitive radar: Step toward bridging the gap between neuroscience and engineering,'' in \emph{Proc. of the IEEE}, vol. 100, no. 11, pp. 3102-3130, Nov. 2012. Doi: 10.1109/JPROC.2012.2203089.

\bibitem{cognitive_radar_3} 
 J. R. Guerci, ``Cognitive radar: A knowledge-aided fully adaptive approach,'' in \emph{Proc. 2010 IEEE Radar Conference}, Washington, DC, 2010, pp. 1365-1370. Doi: 10.1109/RADAR.2010.5494403.

\bibitem{optimal_waveform_design} 
 W. Huleihel, J. Tabrikian, R. Shavit, et al., ``Optimal adaptive waveform design for cognitive MIMO Radar,'' \emph{IEEE Transactions on Signal Processing}, vol. 61, no. 20, pp. 5075-5089, 2013.

\bibitem{learning_branching} 
 A. Lodi, G. Zarpellon, ``On learning and branching: A survey,'' \emph{Top}, vol. 25, no. 2, pp. 207-236, 2017.

\bibitem{learning_branch} 
 N. Balcan, T. Dick, T. Sandholm, et al., ``Learning to Branch,'' in \emph{Proc. International Conference on Machine Learning (ICML)}, 2018, pp. 344-353.

\bibitem{RL_integer_programming} 
 Y. Tang, S. Agrawal, Y. Faenza, et al., ``Reinforcement learning for integer programming: Learning to cut,''  \emph{ArXiv: Learning}, 2019.

\bibitem{combinatorial_optimization} 
 Y. Bengio, A. Lodi, A. Prouvost, et al., ``Machine learning for combinatorial optimization: A methodological tour d'Horizon,'' \emph{ArXiv: Learning}, 2018.

\bibitem{combinatorial_optimization_graphs} 
 H. J. Dai, B. K. Elias, Z. Yuyu, et al., ``Learning combinatorial optimization algorithms over graphs,'' \emph{ArXiv}: 1704.01665v4, 2018.

\bibitem{RL_vehicle_routing} 
 M. Nazari, A. Oroojlooy, L. V. Snyder, et al., ``Reinforcement learning for solving the vehicle routing problem,'' in \emph{Proc. Neural Information Processing Systems (NIPS)}, 2018, PP. 9861-9871.

\bibitem{pointer_networks} 
 O. Vinyals, M. Fortunato, N. Jaitly, et al., ``Pointer networks,'' in \emph{Proc. Neural Information Processing Systems (NIPS)}, 2015, pp. 2692-2700.


\end{thebibliography}
\end{document}